\documentstyle[12pt,epsfig,psfig]{article}

\newcommand{\ci}{\cite}

\def\setb@se#1{\baselineskip=#1 \normalbaselineskip=#1}
\setb@se{12pt}
\newcommand{\be}{\begin{equation}}
\newcommand{\ee}{\end{equation}}
\newcommand{\bea}{\begin{eqnarray}}
\newcommand{\eea}{\end{eqnarray}}
\def \R {{\rm R}}
\def \F {{\rm F}}
\def \ov {\over}\def \const {{\rm const}}
\def \la {\label}
\def \ci{\cite}
\newcommand{\rf}[1]{(\ref{#1})}
\def \a {\alpha}

\def \foot {\footnote}

\newcommand{\g}{{\bf g}}
\newcommand{\x}{{\bf x}}

\newcommand{\T}{\mbox{\bf T}}

\def\TC{\hfil$\displaystyle{##}$\hfil}

\def\seqalign#1#2{\vcenter{\openup1\jot
  \halign{\strut #1\cr #2 \cr}}}

\def\comment#1{}
\def\fixit#1{}

\def\lsim{\mathrel{\mathstrut\smash{\ooalign{\raise2.5pt\hbox{$<$}\cr\lower2.5pt\hbox{$\sim$}}}}}
\def\gsim{\mathrel{\mathstrut\smash{\ooalign{\raise2.5pt\hbox{$>$}\cr\lower2.5pt\hbox{$\sim$}}}}}

\def\mop#1{\mathop{\rm #1}\nolimits}
\def\tr{\mop{tr}}

\def\lbldef#1#2{\expandafter\gdef\csname #1\endcsname {#2}}

\def\eno#1{(\ref{#1})}

\begin{document}

\begin{titlepage}

\begin{center}
\hfill hep-th/0108205 \\
\hfill CALT-68-2329 \\
\hfill CITUSC/01-016 \\
\hfill  OHSTPY-HEP-T-01-010\\
\hfill  FSU TPI 01/05 \\

\vfil\vfil

{\Large \bf
Non-abelian 4-d black holes, wrapped 5-branes, \\ \vskip .3 cm
and their dual descriptions
}

\vskip .7 cm

\vskip 1 cm

{\large S.S. Gubser,$^{a,}$\footnote{On leave from Princeton University.}
  A.A. Tseytlin$^{b,}$\footnote{Also at
Imperial College, London and
Lebedev
Institute, Moscow.}  and M.S. Volkov$^{c,}$\footnote{After 
1st september: LMPT, Universite de Tours, 
Parc de Grandmont, 37200 Tours, France. }}  \\

\end{center}

 \centerline{\it ${}^a$ Lauritsen  Laboratory of Physics, 452-48
 Caltech,  }
\centerline{\it  Pasadena, CA 91125,    USA}

\vskip 0.4 cm

\centerline{\it ${}^b$ Department of Physics, The Ohio State University,}
\centerline{\it 174 West 18th Avenue, Columbus, OH 43210-1106, USA}

\vskip 0.4 cm

\centerline{\it  ${}^c$ Institute for Theoretical Physics, Friedrich
Schiller
University of Jena,}
\centerline{\it  Max-Wien Platz 1, D-07743 Jena,
Germany }

\vskip 0.5 cm

\begin{abstract}
 We study extremal and non-extremal generalizations of the regular
non-abelian monopole solution of \cite{Chamseddine:1997nm},
interpreted in \cite{Maldacena:2000yy} as 5-branes wrapped on a
shrinking $S^2$.  Naively, the low energy dynamics is pure ${\cal
N}=1$ supersymmetric Yang-Mills.  However, our results suggest that
the scale of confinement and chiral symmetry breaking in the
Yang-Mills theory actually coincides with the Hagedorn temperature of
the little string theory.  We find solutions with regular horizons and
arbitrarily high Hawking temperature.  Chiral symmetry is restored at
high energy density, corresponding to large black holes.  But the
entropy of the black hole solutions decreases as one proceeds to
higher temperatures, indicating that there is a thermodynamic
instability and that the canonical ensemble is ill-defined.  For
certain limits of the black hole solutions, we exhibit explicit
non-linear sigma models involving a linear dilaton.  In other limits
we find extremal non-BPS solutions which may have some relevance to
string cosmology.

\end{abstract}

\end{titlepage}

\setcounter{page}{1}
\renewcommand{\thefootnote}{\arabic{footnote}}
\setcounter{footnote}{0}

\section{Introduction}

One of the main motivations of
the AdS/CFT correspondence and its generalizations
\cite{Maldacena:1998re,Gubser:1998bc,Witten:1998qj} (see
\cite{Aharony:2000ti} for a review) is to give an account of
confinement based on string theory \ci{POL}.  Since the duality is most
naturally formulated for strongly coupled gauge theories, this goal
might not seem too distant; and indeed, there have been many attempts,
starting with \cite{Witten:1998zw}, to give a qualitatively correct
description of confinement based on semi-classical reasoning on the
supergravity side.

A particularly natural venue for making an explicit connection between
string theory and gauge theory is pure ${\cal N}=1$ super-Yang-Mills
model.
This theory exhibits chiral symmetry breaking and confinement, but
supersymmetry gives enough control to make a number of exact
statements (see, e.g.,  \ci{SV} for a review).  In particular, for gauge group $SU(N)$, there is a
$Z_{2N}$ chiral $R$-symmetry (acting as a complex phase on the
gauginos) which is the remnant of the $U(1)_R$ of the classical theory
after instanton effects are taken into account.  A choice of vacuum
breaks this further to $Z_2$ through a gaugino condensate,
  $ \langle \tr \lambda\lambda \rangle =
    e^{2\pi i k \over N} \Lambda^3 ,$
 where $k=1,\ldots,N$ labels the vacua, and $\Lambda$ is the
dynamically generated scale.  For high enough temperatures, the full
$Z_{2N}$ chiral symmetry should be restored, and we should have
$\langle \tr \lambda\lambda \rangle = 0$.

The original motivation for this paper was to study the chiral symmetry
breaking transition of ${\cal N}=1$ super-Yang-Mills theory via a
supergravity dual.

In the recent literature, there
are two particularly notable attempts
to provide supergravity duals of ${\cal N}=1$ super-Yang-Mills theory
\cite{Klebanov:2000hb,Maldacena:2000yy}.\footnote{In these papers
 the supergravity backgrounds have non-trivial dependence on the
radial coordinate (``energy scale") only.  An earlier approach, based
on a massive deformation of ${\cal N}=4$, has been studied in
\cite{Girardello:2000bd,Gubser:2000nd,Pilch:2000fu,PS}.  The
ten-dimensional supergravity solutions are more complicated in this
approach because there is angular as well as radial dependence.
Studying finite temperature in these backgrounds is difficult; see
however \cite{Freedman:2001xb}.}
 In \cite{Klebanov:2000hb} the geometry is
a warped product of ${\bf R}^{3,1}$ and the deformed conifold, which
is supersymmetric \cite{Grana:2001jj,Gubser:2000vg}, and can be
thought of as the result of wrapping $M$ D5-branes on the $S^2$ of the
conifold's base \ci{GU} and then turning on other fields
\ci{Klebanov:1999rd} to keep the dilaton fixed \ci{ Klebanov:2000nc}.
The $S^2$ shrinks, but the three-form Ramond-Ramond (R-R) flux from
the D5-branes remains; also there is a R-R five-form corresponding to
an indefinite number of D3-branes which grows with energy scale.  The
gauge theory dual involves a ``duality cascade'' of $SU(N) \times
SU(N+M)$ gauge theories with ${\cal N}=1$ supersymmetry, where $N$
also grows with energy scale.  At low energies, only pure $SU(M)$
gauge theory remains.  In
\cite{Buchel:2001ch,Buchel:2001gw,Gubser:2001ri} an understanding of
chiral symmetry restoration at high temperature was reached: black
holes were shown to exist which corresponded to thermal states in the
gauge theory with exactly zero gaugino condensate.  Unfortunately, the
supergravity equations that determine these black holes are formidable
coupled differential equations, and the best that could be done
\cite{Gubser:2001ri} was to solve them in a high-temperature
expansion.  This leaves open the nature of the chiral symmetry
breaking phase transition.

The current paper focuses on the other approach
\cite{Maldacena:2000yy}, which was based on re-interpretation of a
supergravity solution previously found in
\cite{Chamseddine:1997nm,Chamseddine:1998mc}.  Here the R-R five-form
field is turned off altogether, and only the D5-branes remain.  The
S-dual NS5-brane version of this geometry (with the R-R two-form
replaced by the NS-NS two-form) falls \cite{PT,Buchel:2001qi} into the
general category described in \cite{Strominger:1986uh}.  The gauge
theory interpretation is that one starts with little string theory
\ci{AH} on the six-dimensional D5-brane worldvolume and compactifies
on $S^2$ to obtain four-dimensional supersymmetric Yang-Mills theory
(for a discussion of some properties of this theory see \ci{SL}).

The approaches of \cite{Klebanov:2000hb,Maldacena:2000yy} thus
provide   different UV completions\footnote{We use the term ``UV
completions" loosely here since ${\cal N}=1$ is already renormalizable
and asymptotically free, so it doesn't strictly require any additional
fields in the ultraviolet.} of pure ${\cal N}=1$ super-Yang-Mills
theory which can be studied in string theory via extensions of the
AdS/CFT correspondence.

\subsection{Summary of results}
\label{Summary}

It may seem that the approach of \cite{Maldacena:2000yy} should be
simpler than the duality cascade of \ci{Klebanov:2000hb}.  Indeed, it
is technically simpler on the supergravity side, and we shall obtain
results on non-BPS solutions which are considerably more detailed
than the ones available for the duality cascade.  However, our results
suggest that the Hagedorn temperature of the little string theory
either coincides or nearly coincides with the critical temperature for
chiral symmetry breaking, so that the super-Yang-Mills modes are not
cleanly decoupled from massive modes in its parent theory.\footnote{We
are grateful to I. Klebanov for a useful discussion of this point.}
This is a particularly sharp manifestation of a persistent problem
observed in supergravity duals of confining gauge theories:
generically there is not a clean separation of scales between
higher-dimensional modes and gauge theory phenomena.  A general
argument that this should be so is that for supergravity to be valid,
the 't~Hooft coupling should be large, so if the extra matter fields
freeze out at a scale $\Lambda$, then the scale of confinement is
roughly $e^{-{c_0/(Ng_{\rm YM}^2}) } \Lambda$, where $c_0$ is some constant of
order $1$.  We may suspect that the ``AdS-QCD" enterprise teaches us at
least as much about the UV completions (in our case, little string
theory on $S^2$) as it does about the low-energy confining gauge
theories.

Besides the intrinsic interest of little string theories, there are
two reasons why the example of 5-branes on a two-sphere deserves
further study.  First, this system dual does exhibit chiral symmetry
breaking in its supersymmetric ground state, and (as we shall see)
possesses chiral-symmetry restored states at high energy density; so we
have a reasonable shot at describing the interesting chiral symmetry
breaking phase transition.  Second, it is possible to quantize
D1-branes in the background under consideration, using (in S-dual
language) nothing more than non-linear sigma model techniques.  This
is not quite ideal: ``weaving together'' planar graphs for the gauge
bosons $A_\mu$ leads to worldsheets for {\it fundamental} strings,
whereas D1-brane worldsheets are related to the dual magnetic
variables, and external magnetic charges are screened rather than
confined.  Still, it is a real novelty to be in possession of string
backgrounds for a confining gauge theory which do not require R-R
fields: from this S-dual point of view we have fundamental strings
moving in $S^2$-wrapped NS5-brane background.

In \cite{Buchel:2001qi} a first attempt was made to construct  a
non-extremal black hole
generalization of the supersymmetric solution of
\cite{Maldacena:2000yy,Chamseddine:1997nm,Chamseddine:1998mc}.
Here we shall present  a systematic study of such solutions,
which extends
the work in  \cite{Buchel:2001qi} in  several directions.
 Rather than working in ten dimensions, it is useful to go back to
four by integrating over the $S^3$ threaded by the three-form flux and
also dropping the spatial ${\bf R}^3$ factor (which is possible as
long as we are only interested in questions about translation
invariant quantities in a thermodynamic limit).  The 4-d framework
allows us to be guided by intuition about structure and properties of
familiar black-hole solutions.\footnote{In practice, since we are
interested in static spherically symmetric solutions, we will end up,
as in \ci{GM,PT,Buchel:2001gw,Gubser:2001ri}, with 1-d effective
action for the radial evolution of the unknown functions in the metric
and matter field ansatz.}  Indeed, the BPS solution arose from lifting
a non-abelian gravitating monopole in four-dimensional ${\cal N}=4$
gauged supergravity back up to ten dimensions.  This monopole
\cite{Chamseddine:1998mc} is one of the few analytically known
classical supergravity solutions involving both non-abelian gauge
fields and gravity.  For a review of such solutions, both analytic and
numerical, see \cite{Volkov:1998cc}.

Our approach will be to consider black hole solutions with asymptotics
similar to the gravitating monopole solution of
\cite{Chamseddine:1998mc}.  For the most part our non-BPS solutions
will be numerical.  As we shall explain, {\it unbroken} chiral
symmetry is equivalent to having only {\it abelian} gauge fields in
the supergravity solution: the non-abelian gauge fields yield an order
parameter for the transition.  
There is a critical value 
(depending on the normalization of the dilaton)
of the
entropy of a black hole
solution below which non-abelian gauge fields must appear.  At this
critical value, a long throat develops in the geometry which is, in
the string frame if we are describing NS5-branes on $S^2$ (or in
D1-brane frame if we are describing D5-branes), the two-dimensional
dilaton black hole geometry times ${\bf R}^3 \times \tilde T^{1,1}$.
Here the $\tilde T^{1,1}$ space \ci{ PZ} has the same symmetries and
topology as the familiar $T^{1,1}$ base of the standard conifold
\ci{Candelas:1990js}.  Its metric is only slightly different.  The
throat solution at the critical value of the entropy is available
analytically, and we are also able to provide a worldsheet sigma model
description of it as well as a description of how it is deformed as
one departs from the critical point.

One might hope to map this ``critical point'' in the space of
supergravity solutions to a second order chiral symmetry breaking
transition in the ${\cal N}=1$ gauge theory.  This does not work out
because the temperature of the critical point is actually {\it higher}
than the Hagedorn temperature $T_c$ of the little string theory, which can
be read off as the limiting temperature of black holes far from
extremality.  Rather, it seems that the black hole solutions we find
help characterize little string theory on $S^2$ {\it above} its
Hagedorn transition.  It is not possible, in a near-horizon limit, to
proceed to $T>T_c$ in classical, non-extremal, flat NS5-solutions.
However,  it seems that wrapping the NS5 on an $S^2$ changes the story
and allows us to characterize higher temperature states without
resorting to string theory corrections, as in \cite{Kutasov:2001jp}.
The specific heat is negative for the black holes we find, so that the
entropy decreases as the temperature rises.  This is reminiscent of
speculations that at very high energies, string theory has very few
degrees of freedom.  The thermodynamic instability that negative
specific implies seems likely to be reflected in tachyonic modes of
the black hole solutions
\cite{Gubser:2000ec,Gubser:2000mm,Reall:2001ag,Rangamani:2001ir},
similar to the local Gregory-Laflamme instability.  We postpone a
detailed investigation of this point, focusing instead on
translationally invariant questions.

If all our black hole solutions describe effects in little string
theory, then what, one may ask, describes the chiral symmetry breaking
transition in field theory?  There are no black hole solutions whose
Hawking temperature is less than the Hagedorn temperature of little
string theory.  Thus, semiclassically, the solution that may be
expected to dominate the path integral at lower temperatures is the
original vacuum solution of
\cite{Maldacena:2000yy,Chamseddine:1997nm,Chamseddine:1998mc},
periodically identified in Euclidean time.  This solution does have
broken chiral symmetry.  
There are
no globally regular 
solutions without horizons that have unbroken chiral symmetry.
Thus the transition which restores chiral symmetry occurs precisely
when one reaches the Hagedorn temperature and can form the abelian
black holes.  At this point it is only a question whether such black
holes are entropically favored over the periodized vacuum solution.
They in fact are,\fixit{This point is sort of moot!} so we may
provisionally conclude that chiral symmetry restoration and
deconfinement occur simultaneously, at the Hagedorn temperature of the
little string theory, and that the transition is first
order.\footnote{It is possible that spatially non-uniform black hole
solutions may have a lower minimum Hawking temperature, in which case
our conclusions would be somewhat modified.  It is almost certain that
spatially non-uniform solutions play a role in describing the high
temperature phase, since the specific heat is negative there.}  These
results are in line with the familiar conclusion \cite{Witten:1998zw}
that solutions with regular horizons describe a deconfined phase,
while horizonless solutions describe a confined phase.\footnote{The
conclusions of this paragraph were arrived at in discussions with
I.~Klebanov.}  We will revisit this issue in
section~\ref{HawkingPage}: as we shall see, some refinement is
necessary on account of the thermodynamic instability of the black
hole solutions.


\subsection{Organization of the paper}


 In section~\ref{TenFour} we shall describe the class of
ten-dimensional backgrounds we are going to consider.  These IIB
backgrounds involve only the metric, the dilaton, and a three form
field strength, which by S-duality may be taken to be the R-R field
strength or the NS one.  The ansatz will be translationally invariant
in three spatial direction as well as in the time direction, but
generally the Lorentz group $SO(3,1)$ will be broken to $SO(3)$ by
non-extremality (that is, finite temperature).  The six extra
dimensions comprise a radial direction $r$ and a transverse compact
5-d space with $S^2 \times S^3$ topology and $SU(2) \times SU(2)$
isometry. The resulting background may be interpreted
\ci{Maldacena:2000yy} as a special kind of 3-brane representing D5 (or
NS5) branes wrapped over a shrinking $S^2$.  Our general ansatz for
the supergravity fields will be parametrized by 9 functions of the
radial coordinate $r$, and we will derive the effective 1-d action for
them that reproduces the full set of supergravity equations in this
case.  We shall then consider a subset of backgrounds with only 3
independent functions which corresponds to the solution of
\ci{Chamseddine:1997nm,Chamseddine:1998mc,Maldacena:2000yy}.

In section~\ref{DFour} we shall obtain the equivalent set of equations
from the $D=4$ perspective \ci{Chamseddine:1997nm,Chamseddine:1998mc}:
by looking for non-abelian black-hole type solutions of the bosonic
$SU(2) \times U(1)$ sector of ${\cal N}=4$ gauged supergravity (which
can be obtained by compactifying $D=10$ supergravity on $S^3 \times
T^3$).  We shall explain the translation between the $D=10$ and $D=4$
descriptions.

In section~\ref{ZeroT} we shall study the extremal (or
``zero-temperature") solutions of this system -- solutions which have
$SO(3,1)$ Lorentz invariance.  They are obtained when a
non-extremality parameter $\alpha$ is set equal to zero.  We shall
first consider a subset of BPS solutions (section 4.1) which solve a
first-order system following from a superpotential and preserve ${\cal
N}=1$, $D=4$ supersymmetry.  The family of these BPS solutions is
parametrized by one essential parameter $c$; solutions with generic
values of $0\leq c\leq \infty$ are singular non-Abelian backgrounds,
while the boundary points of the family corresponding to $c=0$ and
$c=\infty$ are, respectively, the regular non-Abelian and singular
Abelian solutions of \ci{Chamseddine:1997nm,Chamseddine:1998mc}.  It
is the regular non-Abelian solution that was interpreted in
\ci{Maldacena:2000yy} as supergravity dual of ${\cal N}=1$
supersymmetric gauge theory.

Non-BPS (supersymmetry-breaking) extremal solutions will be described
in section 4.2.  We shall start with two ``fixed-point" abelian
solutions, one of which has
 a remarkably simple world-sheet description in terms of special kind
of $SU(2) \times SU(2)\over U(1)$ gauged WZW model \ci{PZ} and thus is
expected to be an exact string solution to all orders in $\alpha'$.
We shall then describe a class of regular non-extremal solutions
(depending on one parameter $b$) by analyzing asymptotics at $r=0$ and
$r=\infty$ and interpolating between them numerically. Presumably,
these solutions may be interpreted as ``excited states" of the regular
BPS solitonic background, similar to higher excitation modes of BPS
monopoles. They may be related to supersymmetry-breaking deformations
of ${\cal N}=1$ supersymmetric gauge theory dual to the regular BPS
background.

In section~\ref{NonZeroT} we shall turn to non-extremal solutions
($\alpha\not=0$) with regular black hole horizons.  We shall determine
their short-distance behavior, which depends on the two essential
parameters $(\R_h,w_h)$, the second of which may be interpreted as the
U(1)
chiral symmetry breaking parameter. The global form of the solutions
is found by numerical integration.  We shall then compute the
corresponding Hawking temperature as a function of the two horizon
parameters.  As we will explain, there is a minimal non-zero value of
the temperature, $T_c=1/4\pi$, which is achieved in the limit of large black
holes.  For $\R_h<\infty$ one has $T>T_c$, and the minimal value of
$T$ for a fixed $\R_h$ is achieved for the Abelian solution,
suggesting restoration of chiral symmetry on the gauge theory side.
The limit $T\to\infty$ will lead to globally regular non-Abelian
solutions, which break the chiral symmetry.

In section~6 we compute the energy and free energy of the black holes
we have found.  Remarkably, of the two-parameter family of black hole
solutions, only a discrete series of one-parameter families has finite
energy.  Non-abelian black holes exist only with energy less than a
certain threshold; abelian black holes exist only with energy greater
than a different threshold---lower than the first, so that there is a
range of energies where both abelian and non-abelian solutions are
possible.

In section~7 we shall address the question of chiral symmetry
restoration at temperatures higher than the Hagedorn temperature.  We
compare the free energy of a black hole solution with the free energy
of the globally regular BPS solution with the same periodicity in
Euclidean time at infinity.  The thermodynamic instability of the
black holes makes it difficult to discuss Hawking-Page transitions
meaningfully; however we describe conditions under which black holes
would be expected to form.

Section~8 contains a summary of different solutions we obtained and a
discussion of possible application of excited monopole solutions in
string cosmology context.

\medskip

While this paper was in preparation
there appeared   another
 discussion  \ci{BUCH} of  a possible relation between
 non-extremal NS5 on $S^2$ background and  issues of little string
 thermodynamics.  There is some overlap with our
section~\ref{NonZeroT}, to the extent that \ci{BUCH} also reached the
conclusion that the specific heat is negative.  We also make contact
briefly with the analysis of \ci{BUCH} in section~\ref{HawkingPage}.

\section{Ten-dimensional description of 5-branes on $S^2$}
\label{TenFour}

We shall  study solutions in
the following subsector of the  type IIB supergravity action:
\be
S_{10}=\frac14\,\int d^{10}x\,\sqrt{-\g}\, \left(
R
-\frac{1}{2}\,(\partial\Phi)^2
-\frac{1}{12}\,e^{-\Phi}\,H_3^2
-\frac{1}{12}\,e^{\Phi}\,F_3^2
\right)
\ .                            \label{0:NS}
\ee
Here $H_3=dB_2=\frac16\,H_{MNS}\,dx^M\wedge dx^N\wedge dx^S$ and
$F_3=dC_2=\frac16\,F_{MNS}\,dx^M\wedge dx^N\wedge dx^S$.
The line elements in the Einstein frame (used in the above action)
and in the string frame
are related  by
$ds_{10E}^2={\rm e}^{-\Phi/2}ds_{10S}^2=\g_{MN}dx^M dx^N$.
 We shall be studying solutions with either $F_3$ or $H_3$ being zero,
so this is a consistent truncation of the type IIB
theory.\footnote{Since the solutions we shall be discussing will have
only metric, dilaton and one three-form non-trivial, they can be
embedded into ${\cal N}=1$ $D=10$ supergravity.}  These two cases,
i.e.{} the NS-NS and R-R backgrounds, are related by $S$-duality: if
$(\g_{MN},\Phi,H_3,F_3=0)$ is a solution of the field equations, then
interchanging $H_3\leftrightarrow F_3$ and changing $\Phi\to-\Phi$
gives another solution with the same Einstein-frame metric $\g_{MN}$
(but the string frame metric changes).  In what follows we shall
mostly consider the R-R version of the solutions.

We shall be considering 3-brane-type solutions with 1+3 ``parallel"
directions ($t,{\rm x}^n$) and 6 transverse directions
$(r,\theta_1,\phi_1,\psi,\theta_2,\phi_2)$ representing a manifold
with topology ${\bf R} \times S^2\times S^3$ and metric similar to conifold
metrics \ci{ Candelas:1990js,PAN}.  We shall assume that
the metric and matter fields have non-trivial dependence on the radial
direction $r$ only, while all angular dependence will be fixed by
global symmetries.

Let $(\theta_1,\phi_1)$ be the standard coordinates on $S^2$,
and $(\psi,\theta_2,\phi_2)$ be the Euler angles on $S^3$.
We choose the 1-form basis on $S^2$ as $(e_1,e_2)$,
\be                                  \label{0:1}
e_1=d\theta_1,\ \ \ \ \ e_2=-\sin\theta_1 d\phi_1\, ,
\ \ \ \ \ \ e_3=-\cos\theta_1 d\phi_1\, ,
\ee
where $e_3$ is the spin connection,
and the invariant 1-forms on $S^3$ as
\be                             \label{0:2}
\epsilon_1=\cos\psi d\theta_2+\sin\psi\sin\theta_2 d\phi_2,\
\epsilon_2=-\sin\psi d\theta_2+\cos\psi\sin\theta_2 d\phi_2,\
\epsilon_3=d\psi+\cos\theta_2 d\phi_2\, .
\ee
These forms satisfy  the Maurer-Cartan equation
$d\epsilon_a+\frac12\,\epsilon_{abc}\,\epsilon_b\wedge\epsilon_c=0$.
Let $r$ be the transverse to the brane radial coordinate, while
$t$ and ${\rm x}^n$ are the time and three longitudinal coordinates.

We shall consider metrics of the following  form
\be                            \label{0:3}
ds^2_{10E} =  -Y_1\,dt^2+Y_2\,d{\rm x}^n d{\rm x}^n
+ Y_3\, dr^2 + Y_4\,(e_1^2+e_2^2)
+ Y_5\,(\tilde{\epsilon}_1^2+\tilde{\epsilon}_2^2)
+ Y_6\,\tilde{\epsilon}_3^2 \, ,
\ee
where
\be                               \label{0:4}
\tilde{\epsilon}_1\equiv \epsilon_1-Y_7\,e_1,\ \ \ \ \ \
\tilde{\epsilon}_2\equiv \epsilon_2-Y_7\,e_2,\ \ \ \ \ \
\tilde{\epsilon}_3\equiv \epsilon_3-e_3,\ \ \
\ee
and $Y_i=Y_i(r)$ are seven functions of
 the {\it radial} coordinate $r$ only.

Strings in such metric may describe  confining
 gauge theories   \ci{POL}, provided
 $Y_1$ and $Y_2$ have finite limit for $r\to 0$.
 That means one has  finite fundamental  string tension in the
  IR limit in dual
 gauge theory.

In the ``extremal" case of $Y_1=Y_2$ one has  Lorentz invariance
  in 1+3 dimensional part,
 while non-extremal black-hole type solutions should have
 $Y_1/Y_2\not=$const.  The regular horizon case should then represent finite
 temperature gauge theory in a deconfined state.\footnote{An alternative
 option for a finite temperature state is
  $Y_1/Y_2=$const with $t$ replaced by periodic euclidean time.}

  This general class of metrics  includes \ci{PT,PAN}
as special cases all
 3-brane-on-conifold metrics recently studied in the literature.
{}For example, the subclass with $Y_4=Y_5, Y_7=0$ contains
metrics whose transverse 6-space is the standard
Ricci-flat conifold,
$ds_6^2=dr^2+r^2(dM_5)^2$,
where the base $M_5=T^{1,1}=SU(2)\times SU(2)/U(1)$  has
\be
dM_5^2=\frac16(d\theta_1^2+\sin^2\theta_1d\phi_1^2)
+ \frac16(d\theta_2^2+\sin^2\theta_2d\phi_2^2)
+\frac19(d\psi+\cos\theta_1 d\phi_1+\cos\theta_2 d\phi_2)^2
\ .   \label{stan}
\ee
Resolved conifold corresponds to $Y_4\not=Y_5, Y_7=0$,
and deformed conifold has $Y_7\not=0$.

For  $Y_7=0$  the metric has additional $U(1)$ symmetry
  under  $\psi\to\psi+ \psi_0 $, which should
correspond to  chiral symmetry on the gauge field  theory side
\cite{Klebanov:2000hb,Maldacena:2000yy}.
If $Y_7(r)\to 0$ for $r\to\infty$, this may be interpreted as a
supergravity
manifestation of
chiral symmetry restoration in the  high energy (UV)  limit.
As we shall see below,  the
symmetry under $\psi\to\psi+\psi_0$  may be  restored also for $Y_7=1$
and  $Y_5=Y_6$.

In addition to the metric, we shall make the following ansatz
for the closed R-R 3-form $F_3$ \ci{PT} ($Y'\equiv { d Y \over dr}$)
\bea                      \label{0:555}
  F_3=P\bigg[ \,\tilde{\epsilon}_3\wedge\left\{
{\epsilon}_1\wedge {\epsilon}_2
+ {e}_1\wedge{e}_2  - Y_8 ({\epsilon}_1\wedge{e}_2-
{\epsilon}_2\wedge{e}_1)\right\} \nonumber  \\
+\ \, Y'_8 dr\wedge({\epsilon}_1\wedge{e}_1+
{\epsilon}_2\wedge{e}_2)\bigg] \, ,
\eea
or, in terms of $\tilde \epsilon_{1},\tilde \epsilon_2$,
\bea                      \label{0:5}
F_3=P\bigg[\,\tilde{\epsilon}_3\wedge\left\{
\tilde{\epsilon}_1\wedge\tilde{\epsilon}_2
+(Y_7^2-2Y_7Y_8+1){e}_1\wedge{e}_2+
(Y_7-Y_8)(\tilde{\epsilon}_1\wedge{e}_2-
\tilde{\epsilon}_2\wedge{e}_1)\right\} \nonumber  \\
+\ \, Y'_8  dr\wedge(\tilde {\epsilon}_1\wedge{e}_1+
\tilde {\epsilon}_2\wedge{e}_2)\bigg]\, .
\eea
Here $P$ is a constant  which may be interpreted
as a  charge of D5-brane  wrapped on  $S^2$.  Note that
 $dF_3=0$ for any function
$Y_8=Y_8(r)$.
Finally, we shall assume that the  dilaton may be also non-constant:
 $\Phi\equiv Y_9(r)$.

The global symmetries of our background allow
one to derive all supergravity equations from a single
1-d effective action for functions $Y_i$.
Inserting the above ansatz for the metric and the matter fields
into the action (\ref{0:NS}), integrating over all coordinates except $r$
and dropping the surface term (and the overall volume factor)
gives the effective one-dimensional action
$S_1=\int dr \ L $, where
\be                 \label{0:6}
L= \sum_{i,k} G_{ik}(Y) Y_i'Y_k'-U(Y)\equiv T-U.
\ee
The action has the residual reparametrization invariance
$r\to\tilde{r}(r)$ unbroken by our ansatz.
Expressing the $Y_i$'s in terms of 9 other functions
$x,y,z,l,q,p,w,\tilde{w},\Phi$
$$
Y_1={\rm e}^{2z-6x},\ \ \ \ \
Y_2={\rm e}^{2z+2x},\ \ \ \ \
Y_3={\rm e}^{10y-2z+2l} ,\ \  $$ $$
 Y_4={\rm e}^{2y-2z+2p+2q} ,\ \ \ \
Y_5={\rm e}^{2y-2z+2p-2q}, \ \ \ \
Y_6={\rm e}^{2y-2z-8p}, $$
\be Y_7=w , \ \ \ \ \ \  \ Y_8=\tilde{w},\ \ \ \ \ \ \  Y_9=\Phi, \la{defo}
\ee
to make $G_{ik}$ diagonal,  one finds  (equivalent action
was given in \ci{PT})
\bea                    \label{0:7}
T&=&{\rm e}^{-l}\left(5y'^2-3x'^2-2z'^2-5p'^2-q'^2
-\frac14\,{\rm e}^{-4q}w'^2-\frac{1}{4}P^2 \,{\rm e}^{\Phi+4z-4y-4p}
\tilde{w}'^2-\frac18\Phi'^2 \right)\, , \nonumber \\
U&=&\frac18\,{\rm e}^{l}
\bigg[{\rm e}^{8y}\left\{{\rm e}^{-12p}\,
[{\rm e}^{4q}+{\rm e}^{-4q}(w^2-1)^2+2w^2(1-{\rm e}^{10p-2q})^2]
-8{\rm e}^{-2p}\cosh\ 2q \right\}  \nonumber   \\
&&\ \ \ \ \ \  +\ {P^2}\,{\rm e}^{\Phi+4z+4y+4p}
\left\{{\rm e}^{4q}+{\rm e}^{-4q}(w^2-2w\tilde{w} +1 )^2
+2(w-\tilde{w})^2\right\}\bigg].
\eea
Here  $l$, which has  no kinetic term,
 is a  pure gauge degree of freedom reflecting remaining
 reparametrization  invariance
(${\rm e}^{l}$ plays the role  of an einbein).
Varying with respect to $l$ one can then set it to any value as  a
reparametrization gauge.
In the gauge
$$l=0 \ , $$
the  equation of motion for $l$ takes the form of
the ``zero-energy" constraint
 $T+U=0$.
 Another variable with a  simple equation of motion is  the
 function  $x(r)$: it is a ``modulus" of the 1-d action as it does not enter
 the potential.
In the gauge $l=0$ we get
\be x''=0,\ \ \ \ \ \  {\rm  i.e.}\ \ \ \ \
x= - \frac{1}{4}\alpha\ r  \ , \ \ \ \ \ \
\alpha=\const\geq 0
\ . \la{xex}
\ee
The constant $\alpha$ is the ``non-extremality'' parameter
(the choice of its sign is of course a convention):
note that
$ Y_1/Y_2 = e^{-8x}$  so that $\alpha\not=0$ corresponds to
 breaking of the $SO(1,3)$  Lorentz symmetry in the
 parallel directions in the 10-d  metric.

 As  is clear from the action (1),(11),  the charge
$P$ can be absorbed  into  a constant part  of the
 dilaton, and so  we shall assume below that $P=1$.

We shall be interested in the special subclass of solutions
with
$$w=\tilde{w} \ , $$
 which corresponds to  the class of solutions including that of
\cite{Chamseddine:1997nm,Chamseddine:1998mc,Maldacena:2000yy}.
The consistency with the other equations then
requires that
$$q=5p \ , \ \ \ \ \ \    \ \ \
\Phi+4z-4y+16p=0 \ , $$
in which case the
equation of motion for $z$ can be integrated to give
$$
z= \frac12 y - 2 p  + \frac14 \gamma r \ , \ \ \
\ {\rm i.e. }\ \ \ \   z=   \frac14
\Phi+ \frac12   \gamma r    \ ,
      \ \ \ \ \  \ \  \gamma =\const\ , $$
where $\gamma$ is another integration
constant.

The  functions  in the ``parallel" part of the metric are then
$$Y_1=\exp[\frac12\,\Phi+(\gamma+\frac32\alpha)r]\ , \ \ \ \ \
Y_2=\exp[\frac12\,\Phi+(\gamma-\frac12\alpha)r]\ .  $$
Assuming that $\gamma+\frac32\alpha>0$, the point $r=-\infty$ is the
event horizon (as we will see below, $\Phi$ is {\it finite}
 at the horizon).
 To have {\it regular} horizon, we must require
 that the scale  of the flat 3-space factor  $Y_2$ is
{\it finite}
 at the horizon.\footnote{Equivalently,
 after compactifying on 3 parallel directions,
 $Y_2$ becomes  a scalar in 7-d theory, and,
 in view of the ``no-hair theorem'' intuition,   one would
  expect that
 7-d black hole will have a regular
  horizon  only if that scalar does not have a charge at
   infinity.}
 This
  gives the condition
$$\gamma=\frac12\,\alpha \ .  $$
Introducing finally
$$ s\equiv 2y+ 2 p  \ , \ \ \ \ \ \ \ \ \   g\equiv 2q=10 p  \ , $$
the metric becomes
\be                   \label{0:8}
ds_{10E}^2={\rm e}^{\Phi/2}\left[
-{\rm e}^{2\alpha r}dt^2+d{\rm x}^n d{\rm x}^n+
{\rm e}^{4s}dr^2+{\rm e}^{2g}(e_1^2+e_2^2)+
\tilde{\epsilon}_1^2+\tilde{\epsilon}_2^2+
\tilde{\epsilon}_3^2\right]\, ,
\ee
where
$$\Phi=s- g- \frac12 \alpha\ r\ ,$$
 while the 3-form is given by (\ref{0:5})
with $Y_7=Y_8=\tilde{w}=w$.

We are finally left with  only {\it three}
 independent
functions
 $s$, $g$, and $w$, whose dynamics is determined by the Lagrangian
\be                               \label{0:9}
\hat L=s'^2-\frac12\,g'^2-\frac12\,{\rm e}^{-2g}w'^2-
\frac14{\rm e}^{4s}\left[ {\rm e}^{-4g}(w^2-1)^2
-2{\rm e}^{-2g}-1\right]
\equiv \hat T- \hat U\, . \la{lll}
\ee
The only effect of the  integration constant $\alpha $
is to modify the zero-energy constraint,
\be\la{kkk}
 \hat  T+\hat U =  \frac14 \alpha^2\ . \ee

\section{$D=4$ description: non-Abelian  black holes \\ in  gauged
${\cal N}=4$ supergravity}
\label{DFour}

Before we proceed to analyzing the equations of motion for the
Lagrangian (\ref{0:9}), let us  re-derive these equations
using the $D=4$ approach.
This is motivated by the fact that the solution
of  \cite{Chamseddine:1997nm,Chamseddine:1998mc,Maldacena:2000yy}
was originally obtained  in the context of
the $D=4$  supergravity \cite{Chamseddine:1997nm},
and then  was uplifted to $D=10$ \cite{Chamseddine:1998mc}.
It turns out that the subclass of
$D=10$ solutions determined by (\ref{0:8}), (\ref{0:9}) can be  obtained
in a similar way -- by uplifting the $D=4$ solutions. It will be convenient
in what follows to use both the $D=10$ and $D=4$ descriptions, and we shall
now establish  the precise correspondence between the two.

Let us consider the bosonic part of the action of the
four-dimensional ${\cal N}=4$ half-gauged%
\footnote{The full SU(2)$\times$SU(2) FS model
contains two independent SU(2) gauge fields
\cite{Freedman:1978ra}. The half-gauged model is
obtained by setting the second field together with its
coupling constant to zero.
The coupling constant for the first gauge field in (\ref{FS}) is set
to $\sqrt{2}$,
while in \cite{Chamseddine:1997nm,Chamseddine:1998mc} it was set to one.
The full FS model can be obtained from the ${\cal N}=1$, $D=10$  supergravity
 by  dimensional reduction
on $S^3\times S^3$ \cite{Chamseddine:1998mc}.}
SU(2)$\times$[U(1)]$^3$ supergravity
of Freedman and Schwarz (FS) \cite{Freedman:1978ra}:
\bea
S_4=\int d^4x  \,  \sqrt{-\g}\,   \left(\frac{1}{4}\,R \right.
 &-&
\frac12\,\partial_\mu\Phi \,\partial^\mu\Phi
-\frac{1}{2}\,{\rm e}^{-4\Phi}\,
\partial_\mu{\bf a}\,\partial^\mu{\bf a}  \nonumber \\
&-&
\frac18\,{\rm e}^{2\Phi}
\F^{a}_{\ \,\mu\nu}\F^{a \mu\nu}
- \left.\frac14\,{\bf a}
\ast\! \F^{a}_{\ \,\mu\nu}\F^{a\mu\nu}
+\frac14\, {\rm e}^{-2\Phi}\right)
\, . \label{FS}
\eea
Apart from the gravitational field $\g_{\mu\nu}$,
the model contains the axion ${\bf a}$, the dilaton
$\Phi$, and the non-Abelian SU(2) gauge field $A^{a}_\mu$
with
$\F^{a}_{\ \mu\nu}=
\partial_{\mu}A^{a}_{\nu}-\partial_{\nu}A^{a}_{\mu}
+\varepsilon_{abc}A^{b}_{\mu}A^{c}_{\nu}$.
The dual field tensor is
$\ast\! \F^{a}_{\mu\nu}=
\frac12\sqrt{-\g}\varepsilon_{\mu\nu\lambda\rho}\F^{a\lambda\rho}$,
where $\varepsilon^{0123}=1$.
As was shown in \cite{Chamseddine:1998mc}, this model
can be obtained via dimensional reduction
of the  $D=10$  supergravity (${\cal N}=1$ truncation of  (\ref{0:NS}))
 on $S^3\times T^3 $
 (the normalizations of the kinetic terms agree
 after taking into account  that the radius of the internal manifold is
 $\Phi$-dependent).
As a result, any
on-shell configuration in the FS model,
$(\g_{\mu\nu},A^{a}_{\mu}, \Phi, {\bf a})$,
can be uplifted to $D=10$ to become a solution
of ten-dimensional equations of motion for the action (\ref{0:NS}).
The uplifted fields are obtained as follows.
The $D=10$ metric in the Einstein frame is given by
\be        \label{E}
d{s}^2_{10E} ={\rm e}^{\Phi/2}\left(
{\rm e}^{-2\,\Phi}\g_{\mu\nu}dx^\mu dx^\nu
+
\Theta^{a}\Theta^a+d{\rm x}^n d{\rm x}^n \right)\, ,
\ee
where ($a,b,c=1,2,3$)
$$\Theta^{a}\equiv \epsilon^a-A^{a}\ , \ \ \ \ \ \ \ \
A^{a}=A^{a}_\mu dx^\mu$$ while
$\epsilon^a$
are the invariant 1-forms on $S^3$.
The R-R 3-form
is given by
\be           \label{ns}
F_3=
\Theta^{1}\wedge\Theta^{2}\wedge\Theta^{3}
- \Theta^{a}\wedge \F^{a}
- 2 {\rm e}^{4\Phi}\ast\! d{\bf a} \, ,
\ee
Here $\F^{a}=\frac12\,\F^{a}_{\mu\nu}
dx^\mu\wedge dx^\nu$, and the asterisk stands for the four-dimensional
Hodge dual,
$\ast(d{\bf a})=\frac16\sqrt{-\g}\,
\varepsilon_{\mu\nu\rho\delta}\, \partial^\mu{\bf a}\, dx^\nu\wedge
dx^\rho\wedge dx^\delta$, while $H_3=0$.
The $D=10$ dilaton is given by
$\Phi+\ln 4$.%
\footnote{Since they differ by a constant shift, and since shifting
the dilaton is a symmetry,
we denote both the 4d and 10d dilaton by the same letter $\Phi$.}
If the four-dimensional configuration
is supersymmetric, then its $D=10$ analog preserves the same
amount of supersymmetry.

This correspondence between $D=4$ and $D=10$ backgrounds
 may be  useful for constructing
solutions in $D=10$,
provided one has some insight into how to
solve the 4-dimensional problem.
 In general, however, it is not easy to
solve the equations for the action (\ref{FS}), unless some simplifying
assumptions are made.
Let us assume that $\partial/\partial x^0$
is the hypersurface-orthogonal Killing vector.
In this case the most general 4-metric can be represented as
\be                                                 \label{a2+}
ds^2_4 =\g_{\mu\nu}dx^\mu dx^\nu= -{\rm e}^{2\Phi+2X}dt^2+
{\rm e}^{2\Phi-2X} h_{ik}(x)\,dx^{i} dx^{k}\, .
\ee
We shall also assume that temporal component of the gauge
field vanishes, $A_0=0$. This implies that the field is purely
magnetic, so that $\ast\! \F^{a}_{\ \,\mu\nu}\F^{a\mu\nu}=0$,
and one can therefore consistently set the axion to zero.
We are now left with the 3-metric $h_{ik}$, the gauge field
$A^a_i$, and  two scalars $X$ and $\Phi$.
The equations of motion for (\ref{FS})
imply that $X$ is a harmonic function,
\be
\tilde{\nabla}_i\tilde{\nabla}^i X=0,
\ee
where $\tilde{\nabla}_i\tilde{\nabla}^i$ is the covariant Laplacian with
respect to the 3-metric ${\rm e}^{-2X} h_{ik}$.
Since a  harmonic function is necessarily unbounded,
solutions with non-constant $X$ are singular, or possibly have
event horizons.
Using (\ref{E}),(\ref{ns}), any
on-shell configuration $(h_{ik},A^a_i,\Phi,X)$ gives rise to the
solution in $D=10$:
\bea        \label{EE}
d{s}^2_{10E}&=&{\rm e}^{\Phi/2}
\left[-{\rm e}^{2X}dt^2+d{\rm x}^n d{\rm x}^n+
{\rm e}^{-2X} h_{ik}dx^i dx^k
+\Theta^a\Theta^a\right]\, , \nonumber  \\
F_3&=&
\Theta^{1}\wedge\Theta^{2}\wedge\Theta^{3}
-\Theta^{a}\wedge
\F^{a} \,.
\eea
Although this could,  in principle,  give new solutions in $D=10$,
the  equations of motion for the general static fields $(h_{ik},A^a_i,\phi,X)$
are still rather complicated.

For this reason we
now make a further simplifying assumption by demanding that
the $D=4$ system is {\it spherically symmetric}. In this case the
most general 4-metric can be chosen in the form
 \be                                                 \label{a2}
ds^2_4={\rm e}^{2\Phi}\left[-{\rm e}^{2X}dt^2+
{\rm e}^{-2X+2\lambda}dr^2+{\rm e}^{2g}(d\theta ^{2}
+\sin^{2}\theta \,d\phi ^{2})\right],
\ee
where $\Phi$, $X$, $\lambda$, $g$ are functions of the radial coordinate
$r$.
The components $A^a$ of the
spherically symmetric, purely magnetic gauge field
can be read off from
\begin{equation}               \label{aa2}
\T_a A^{a}=w\ (\T_1\,d\theta  -\T_2\,\sin \theta \,d\phi
) -\T_3\,\cos \theta \,d\phi .
\end{equation}
Here $w=w(r)$ and $\T_a=\frac12 \tau_a$ are constant SU(2) generators
($\tau_a$ being Pauli matrices).
The corresponding gauge field tensor is
\be                            \label{aaa2}
\T_a \F^a=dw\wedge(\T_1 d\theta-\T_2\sin\theta d\phi)
-\T_3(w^2-1)\sin\theta d\theta\wedge d\phi\, .
\ee
If $w(r)=0$ then the gauge field is of the Abelian Dirac magnetic
monopole type. If $w(r)=\pm 1$, then $\F^a=0$, which implies that
the gauge field $A^a$ is pure gauge and,  therefore, can
 be gauged away.
Below we shall use the fact that the choice $w=\pm 1$
corresponds, in fact,  to the  vanishing gauge field.

In order to derive the 4d equations of motion, it is convenient to
redefine the variables as
\be
\lambda=X+2s+l,~~~~~~~~~~~~~\Phi=s-g-\frac12\,X\, .
\ee
Since $X$ is a harmonic function, its equation of motion is
\be
(X'{\rm e}^{-l})'=0\, ,
\ee
which gives
\be                \label{X}
X=X_0+\alpha\int{\rm e}^l dr\, ,
\ee
where $X_0$ and $\alpha$ are integration constants.
Inserting the ansatz (\ref{a2}), (\ref{aa2}) into the
action (\ref{FS}), integrating and dropping the surface term,
the result is $S=4\pi\int dt\int dr L$,
where (cf. (\ref{0:9}))
\be           \label{1aa}
 L={\rm e}^{-l}\left(s'^2-\frac12\,{\rm e}^{-2g}w'^2
-\frac12\,g'^2\right)-
\frac14\,{\rm e}^{4s+l}\left[{\rm e}^{-4g}(w^2-1)^2
-2{\rm e}^{-2g}-1\right] + \frac{1}{4}\alpha^2\,{\rm e}^{l}.
\ee
Varying this effective Lagrangian gives the system of radial
equations
\bea                \label{e1}
({\rm e}^{-l}s')'&=&\frac12\,{\rm e}^{4s+l}\left(-{\rm e}^{-4g}(w^2-1)^2
+2{\rm e}^{-2g}+1\right), \\
({\rm e}^{-l-2g}w')'&=&{\rm e}^{4s-4g+l}(w^2-1)w, \label{e2}\\
({\rm e}^{-l}g')'&=&{\rm e}^{4s+l}\left(-{\rm e}^{-4g}(w^2-1)^2
+{\rm e}^{-2g}\right), \label{e3} \\
-4s'^2+2{\rm e}^{-2g}w'^2
+2g'^2&=&
{\rm e}^{4s+2l}\left({\rm e}^{-4g}(w^2-1)^2
-2{\rm e}^{-2g}-1\right)- \alpha^2{\rm e}^{2l},  \label{e4} \\
X'&=&\alpha {\rm e}^l\, .  \label{e5}
\eea
The same radial equations can be obtained by inserting the ansatz
 (\ref{a2}), (\ref{aa2}) into the general equations for the action
(\ref{FS}). Notice that the integration constant $\alpha$ enters
only the last two equations.
Since the equations are invariant under
$l\to l+l_0$, $s\to s-l_0/2$, $\alpha\to\alpha{\rm e}^{-\l_0}$, the
actual
value of $\alpha$ is irrelevant, what matters is whether
$\alpha$
vanishes or not.
Eq.(\ref{e4}), which is the ``zero energy
condition,'' is in fact the initial value constraint.
It is sufficient to impose it on the initial (boundary) values
of solutions of the
independent equations (\ref{e1})--(\ref{e3}).
The constraint generates
reparameterizations $r\to\tilde{r}(r)$, which is the residual gauge
freedom of the ansatz (\ref{a2}), (\ref{aa2}). One can fix the gauge
by imposing a gauge condition on the fields
$(s,l,g,w)$. For example, one can impose the gauge condition $l=0$,
in which case the equation for $X$ can be integrated,
$X=X_0+\alpha r$.

In the  $l=0$ gauge the Lagrangian (\ref{1aa})  coincides with
the one  (\ref{0:9}) obtained within $D=10$ approach.
Let us also compare the uplifted fields with those
given by Eqs.(\ref{0:5}),(\ref{0:8})
(identifying $\theta=\theta_1, \ \phi=\phi_1$).
Using the notation of Eq.(\ref{0:1}) one has
$A^1=we_1$, $A^2=we_2$, $A^3=e_3$, also $\F^1=dw\wedge e_1$,
$\F^2=dw\wedge e_2$, $\F^3=(w^2-1)e_1\wedge e_2$.
The 1-forms $\Theta^a$ are then the
same as $\tilde{\epsilon}_a$ in (\ref{0:4}):
$$\Theta^1=\tilde{\epsilon}_1=\epsilon_1-w\,e_1, \ \ \
\Theta^2=\tilde{\epsilon}_2=\epsilon_2-w\,e_2, \ \ \ \
\Theta^3=\tilde{\epsilon}_3=\epsilon_3-e_3. $$
Then the metric  (\ref{EE}) takes the form
\bea        \label{EEa}
d{s}^2_{10E}& =&{\rm e}^{\Phi/2}
\left[-{\rm e}^{2X}dt^2+d{\rm x}^n d{\rm x}^n+
{\rm e}^{4s+2l}dr^2+{\rm e}^{2g}(d\theta ^{2}
+\sin^{2}\theta \,d\varphi ^{2})
+\tilde{\epsilon}_c\tilde{\epsilon}_c\right]\, ,\nonumber  \\
F_3&=&\,\tilde{\epsilon}_3\wedge\left[
\tilde{\epsilon}_1\wedge\tilde{\epsilon}_2
+(1-w^2){e}_1\wedge{e}_2\right]
+\, w'dr\wedge({\epsilon}_1\wedge{e}_1+
{\epsilon}_2\wedge{e}_2)\, .
\eea
Setting again $l=0$, in which case
$X=\alpha r$ (with $X_0=0$), these expressions
are exactly the same as  in (\ref{0:5}), (\ref{0:8}).

Summarizing, the  four-dimensional solutions in the static, spherically
symmetric, purely magnetic sector of the half-gauged FS model are
equivalent to the ``3-brane" backgrounds
  of  Eqs.(\ref{0:5}),(\ref{0:8}).
In what follows we shall study solutions for gravitating Yang-Mills fields
in four dimensions described by Eqs.(\ref{e1})--(\ref{e5}), using (\ref{EEa})
in order to construct their ten-dimensional 3-brane analogs.

Before  starting  to solve the equations of motion, let us rewrite
them in another gauge, i.e. choice of the radial coordinate $r$.
 While the gauge $l=0$ is sometimes  useful,
  in this gauge a finite
vicinity of $r=0$ is  mapped
 into an infinite region at spatial infinity, which may
cause difficulties in numerical analysis.
For that reason,
we shall often use instead the gauge where
\be
\lambda=0\ ,   \ \ \ \ \ \
{\rm i.e.}    \ \ \ \ \ \    l= - 2s- X \ .
\la{gll}
\ee
Introducing    the functions
$$\nu \equiv{\rm e}^{2X}\ , \ \ \ \   \ \ \ \   \R \equiv{\rm e}^g, $$
the metric becomes
\be                                                 \label{00a2}
ds^2_4={\rm e}^{2\Phi}(-\nu\, dt^2+
\nu^{-1} {dr^2} +\R^2d\Omega^2).
\ee
Introducing also another function
 $$ Z  \equiv \Phi' \ , $$
the equations  (\ref{e1})--(\ref{e5})  take
the following form  in this gauge
\bea
\R''+\frac{3w'^2-\R'^2}{\R} +\frac{\R^2+1}{\nu \R}
-\frac{\nu'}{\nu}(\R'+2\R Z) -4\R Z^2-6Z\R'=
0\ ,&&\label{1a}
\\ Z'+4Z^2+\frac{\R'^2-2w'^2}{\R^2}
-\frac{\R^2+1}{\nu \R^2}+\frac{\nu'}{\R\nu}(\R'+2Z\R)
+6\frac{Z\R'}{\R}=0\ , &&\label{1b}
\\ w''+(2Z+\frac{\nu'}{\nu})w'-\frac{(w^2-1)w}{\nu \R^2}=0\ ,&& \label{1c} \\
2\R^2Z^2+4\R Z\R'+\R'^2
+\R\frac{\nu'}{\nu}(\R'+\R Z) -w'^2
+\frac{(w^2-1)^2}{2\nu \R^2}
-\frac{\R^4+2\R^2}{2\nu \R^2}=0\ ,&& \label{1d}\\
\nu'-\frac{2\alpha}{\R^2} \,{\rm e}^{-2\Phi}=0\ , &&\label{1e} \\
\Phi'-Z=0\ . &&
\label{1f} \eea
The transformation (with constant $d$)
\be              \label{scaling}
r\to{\rm e}^{2 d}r,\ \ \
\Phi\to\Phi+ d,\ \ \ \nu\to{\rm e}^{-4d}\,\nu ,\ \ \
w\to w,\ \ \ \R\to \R,\ \ \
Z\to{\rm e}^{2d}Z,\ \ \
\ee
maps one solution  $\{w(r),\Phi(r),\R(r),\nu(r) \}$
into another solution
$\{w({\rm e}^{2 d}r),\Phi({\rm e}^{2 d}r) +d,\R({\rm e}^{2 d}r),
{\rm e}^{-4d} \nu({\rm e}^{2d}r) \}$.
Note that in this gauge
the  constant $\alpha^2$ term is absent in the
constraint (\ref{1d}) but is present instead in the equation for $\nu$
in   (\ref{1e}).

Another obvious symmetry of the equations is
($C$=const)
\be              \label{scaling1}
\Phi\to\Phi+ C\ ,\ \ \  \  \ \ \ \
\alpha\to{\rm e}^{2C}\alpha\, ,
\ee
with  all other functions  remaining unchanged.  Since $\alpha$ appears only in combination with
${\rm e}^{-2\Phi}$, it can be  set, when it is non-zero,
 to some fixed value
by a constant shift of $\Phi$.

Finally, there is the symmetry with respect to translations,
when argument of all functions is replaced as
\be              \label{scaling2}
r\to r+r_0\, .
\ee

\section{Extremal  solutions}\label{ZeroT}


Let us  now study solutions of the  above system
of equations.
There are two distinct cases:
  $\alpha=0$ and $\alpha\neq 0$,
where $\alpha$ is the integration constant in \rf{xex} or
 (\ref{e5}). In the first, ``extremal," case
 $$ \alpha=0 \ , $$
 the $D=10$
 metric has  $SO(1,3)$ Lorentz symmetry in the 3-brane directions.
 In the $D=4$ description
  one has $X'=0$, so
 that $\nu={\rm e}^{2X}=\const$. In view of the scaling symmetry
(\ref{scaling}) (or simply rescaling $t$ and $r$)
one can assume,
without loss of generality,  that
$\nu=1$.
Then the  $D=4$  metric (\ref{00a2})  becomes
\be                                                 \label{0a2}
ds^2_4={\rm e}^{2\Phi}(-dt^2+
dr^2+\R^2d\Omega^2).
\ee
Written in the string frame, i.e.  without the
${\rm e}^{2\Phi}$ factor, the $t-r$ part of the metric
is thus flat.
 The resulting solutions are  either globally
regular (i.e. geodesically-complete)
 or have naked singularities. There is  a special
subset of  BPS
solutions  preserving part of supersymmetry.

For $\alpha\neq 0$  the 4-d
 metric function $\nu={\rm e}^{2X}$ is non-trivial,
and we get black-hole type
solutions that may  have  a (regular)  event horizon.
Such finite temperature solutions will
 be considered in the next section.

\subsection{BPS solutions}
The system of second-order equations following from
\rf{lll}, \rf{kkk} or \rf{1aa}  in the case of
 $\alpha=0$  admits a special subset of
  solutions  which satisfy the first-order system of equations, following
  from a
   superpotential $W$.
As in many other similar  cases,
such BPS solutions preserve part of supersymmetry
(see, e.g., \ci{town}).\footnote{The
existence of superpotential
  is related to a possibility to embed
   the  effective 1-d system
\rf{0:6} into a globally-supersymmetric action.
 This, in turn,  is related to
the fact that we consider solutions of a bosonic system
 that can be embedded
into locally-supersymmetric supergravity,
as well as to  special properties of the  ansatz.
 Though highly plausible,
in general, the existence of a BPS solution (i.e.
a solution of 1-st order
system) may
 not automatically imply that it
  will be preserving part of supersymmetry.
  }

In fact, in  the present case, the corresponding  first-order system
 was originally derived in \cite{Chamseddine:1997nm}
   from the conditions for  unbroken supersymmetry, i.e.
for the  existence of non-trivial  Killing spinors.
In \ci{PT} the  same  system was obtained  by first
finding the
 superpotential  for the action  \rf{lll}.
 Since the existence of residual supersymmetry
 was already  checked in \cite{Chamseddine:1997nm}
 (with independent arguments given also in \ci{Maldacena:2000yy,PT})
 below we shall follow this more transparent superpotential
 approach.

Let us  write the
Lagrangian (\ref{1aa}) with $\alpha=0$  in the form \rf{0:6}
\be                    \label{lag}
L=G_{ik}(y){ dy^i\over  dr} { dy^k\over  dr} -U(y)\ , \
\ \ \ \ \ \  \ \ \ \  y^i= (s,w,g) \ ,
\ee
where
$G_{ik}={\rm e}^{-l} {\rm diag}(1,-\frac12{\rm e}^{-2g},
-  \frac12   )$.  Direct inspection shows
that the  potential $U$ can be  represented as
\be
U=-G^{ik}\frac{\partial W}{\partial y^i}\frac{\partial W}{\partial y^k}\, ,
\ee
where the superpotential $W$  is  \ci{PT}
\be
W=\pm\frac14{\rm e}^{2s}\sqrt{
{\rm e}^{-4g}(w^2-1)^2     + 2{\rm e}^{-2g}\,(w^2+1)             +1}\, \ .
\ee
As a result, the Lagrangian  (\ref{lag})  can be written as
\be
L=G_{ik}
\left({ dy^i\over  dr}-G^{ij}\frac{\partial W}{\partial y^j}\right)
\left({ dy^k\over  dr}-G^{kn}\frac{\partial W}{\partial y^n}\right)+2W' \, ,
\ee
and this, in turn,  implies that solutions of
the first order equations
\be                   \label{BOG}
  { dy^i\over  dr}  =G^{ik}\frac{\partial W}{\partial y^k}\ , \la{bog}
\ee
solve also the second-order system.

 \def \x {u}
 \def \y {v}

Writing down the explicit form of the ``Bogomol'nyi equations"
 \rf{bog},
one finds  that the equations for $g'$ and $w'$ contain only
$g$ and $w$, and  thus,
taking their ratio, gives  one  first-order
equation
$dg\over dw$=$f(g,w)$.
 Introducing
 $$\x=w^2 \ , \ \ \ \ \ \ \  \y={\rm e}^{2g}\ ,$$
this equation reads
\be
\x(\y+\x-1)\frac{d\y}{d\x}+(\x+1)\y+(\x-1)^2=0.
\ee
Remarkably,  the substitution \cite{Chamseddine:1997nm}
\be
 (\x,\y(\x)) \ \to \ (\rho,\xi(\rho))\ : \ \ \ \ \ \ \ \
 \x=\rho^2{\rm e}^{\xi(\rho)},\ \ \ \ \ \ \
\y(\x)=-\rho\frac{d\xi(\rho)}{d\rho}-     \x        -1
\ee
reduces the problem to  the simple  Liouville equation
\be
\frac{d^2\xi(\rho)}{d\rho^2}=2{\rm e}^{\xi(\rho)}\, .
\ee
 As a result, one finds the following analytic form of the
general solution of the first-order  equations \rf{bog}:
in the  gauge \rf{gll}
(i.e. $l=-2s$)
 the  functions  in the
 gauge field \rf{aa2} and in  the  4-d metric (\ref{0a2})  are
$$
w(r)=\frac{r+r_0}{\sinh(r+r_0 + c)}\, ,\ \ \ \ \ \ \ \ \ \ \
{\rm e}^{2g(r)}=2(r+r_0)\coth(r+r_0 + c)-w^2(r)-1\, ,   $$
\be             \label{BPSa}
 \Phi= s-g \ , \ \ \ \ \ \ \ \
 {\rm e}^{2[\Phi(r)-\Phi_0]}={\rm e}^{-g(r)}\sinh(r+r_0 + c)\, .
\ee
Here $r_0$, $c$, and $\Phi_0$ are the three integration constants
for the three equations.
Different choices of $\Phi_0$ correspond
 to global rescalings of the
solution, while $r_0$ can be absorbed  by  shifting
 $r\to r-r_0$.

 The parameter $c$ (which without loss of generality may be assumed to be
 non-negative)
  is essential, as
different values of $c$ lead to qualitatively  different solutions.
Setting $c=r_0=0$ we obtain the  globally {\it regular} solution,
\be             \label{BPS}
c=0: \ \ \ \ \
w=\frac{r}{\sinh r}\, ,\ \ \ \
{\rm e}^{2g }=2r\coth r-w^2 -1\, ,\ \ \
{\rm e}^{2(\Phi-\Phi_0)}={\rm e}^{-g}\sinh r\, .
\ee
Since $w\not=0$, the  corresponding 4-d gauge field \rf{aa2}  is
non-Abelian.
The  $r\to 0$  asymptotics of this solution is
\be
w = 1 - {r^2\ov 6} + O(r^3) \ , \ \ \ \ \
 {\rm e}^{2g }= r^2  - {r^4\ov 9} + O(r^6) \ , \ \ \ \ \
{\rm e}^{2(\Phi-\Phi_0)}=1 + { 2r^2  \ov 9 } + O(r^4) \ ,
\ee
while the $r\to \infty$ asymptotics  is given by eq. \rf{chir}
below.
Since the dilaton (string coupling) grows for $r \to \infty$, for
large $r$ (i.e. in the UV) one is to switch \ci{Maldacena:2000yy}
from the R-R  background (describing the IR region of the dual theory)
to the  S-dual NS-NS one
 with the same Einstein-frame
metric  \rf{EEa} and the dilaton
${\rm e}^{2(\Phi_{NS} +  \Phi_0)}={{\rm e}^{g}\ov \sinh r}
\to_{r \to \infty}  \sqrt{r} {\rm e}^{-r}$.

For $c\neq 0$ solutions have a curvature singularity at the point,
where ${\rm e}^{2g}$ vanishes, and the parameter $r_0$ can be
chosen so that
${\rm e}^{2g}\geq 0$ for $r\geq 0$.\footnote{The existence of a
1-parameter family of   BPS solutions which are singular
for non-zero value of the parameter
is similar to what happens in the case of fractional D3-branes
on conifolds \ci{PAN}.}
For finite values of $c$  these singular solutions
have  non-Abelian gauge field,
 while in the limit $c\to\infty$ we get $w=0$, i.e.
the gauge field becomes Abelian,
\be                    \label{chir}
w=0\ ,\ \ \ \ \ \ \ {\rm e}^{2g}=2r\ ,\ \ \ \ \ \ \
{\rm e}^{2(\Phi-\Phi'_0)}=\frac{1}{\sqrt{r}}\,{\rm e}^{r}\  .
\ee
We have  set $r_0=1/2$ and shifted $\Phi_0$  by an infinite constant
($ - c/2$)
to put solution into  this form. Note that \rf{chir}
represents the large $r$ asymptotics of the  family of
BPS solutions
\rf{BPSa}.

We conclude that, as  $c$ is changed
 from zero to infinity,   the family of BPS solutions
connects  the
regular non-Abelian
solution (\ref{BPS}) with the Abelian solution (\ref{chir}).
All these BPS solutions
preserve  ${\cal N}=1$, $D=4$  supersymmetry.

\subsection{Non-BPS solutions}

Let us now consider other solutions of the second-order equations
(\ref{e1})--(\ref{e5}) or (\ref{1a})--(\ref{1f}) which {\it do not}
satisfy \rf{bog}, and thus do not preserve supersymmetry.  First note
that the ``Higgs" form of the potential for the gauge-field function
$w$ in \rf{lll},(\ref{1aa}) implies that the equation (\ref{e3}) for
$w$ admits two simple ``fixed-point" solutions, $w=\pm 1$, and $w=0$.
More general non-BPS solutions will not have a simple analytic form (a
standard situation for non-BPS monopoles in gauge theories) and will
be analyzed by a combination of short- and long-distance expansions
and numerical interpolation.

\subsubsection{Vanishing gauge field ($w=\pm 1$)}

Let us set  $w=\pm 1$.  In the $l=-2s$ gauge, the
 field equations (\ref{e1})--(\ref{e5}) reduce to
\be                 \label{w=1}
s''+2s'^2-\frac12={\rm e}^{-2g}\ ,\ \ \ \ \ \ \ \ \ \ \
g''+2s'g'={\rm e}^{-2g}\ , \ee $$
-4s'^2+2 g'^2+2{\rm e}^{-2g}+1=0\, .
$$
As was explained above, for $w=\pm 1$ the gauge field can be
gauged away, $A^a=0$. As a result, there is no
mixing
between the $S^2$ and $S^3$ angles  ($\tilde \epsilon_a= \epsilon_a$)
 in the uplifted $D=10$ background
\be                \label{10dw=1}
d{s}^2_{10E} ={\rm e}^{\Phi/2}\left[
-dt^2+d{\rm x}^n d{\rm x}^n+
dr^2+{\rm e}^{2g}(e_1^2+e_2^2)
+\epsilon_1^2+\epsilon_2^2+\epsilon_3^2\right], \ \ \ \ee
$$ \Phi=s-g\ , \ \ \ \ \ \ \ \ \ \ \
F_3=\epsilon^1\wedge\epsilon^2\wedge\epsilon^3\ . $$
The compact angular part of this  is a direct product of $S^2\times S^3$,
i.e. the
 symmetry of this solution is enhanced as compared to all
other solutions with $w\neq \pm 1$:
it is invariant under $SU(2) \times SU(2) \times SU(2)$.

Using the third equation in (\ref{w=1}) to eliminate ${\rm e}^{-2g}$
from the first two,  and introducing $\y=g'$ and $\x=s'$,
the system reduces to
\be
\y'=2\x^2-2\x\y-\y^2-\frac12\, ,\ \ \ \ \ \ \ \x'=-\y^2\, ,
\ee
which gives
\be               \label{zerofield}
\frac{d\y}{d\x}=1+\frac{2\x}{\y}+\frac{1-4\x^2}{2\y^2}\, .
\ee
The numerical solution of this equation
 will be described below.

\subsubsection{Special Abelian  solution ($w=0,\ g=0$) and
its  NS-NS  coset sigma model counterpart}

For  $w=0$, the
gauge field (\ref{aa2}) is of the Abelian
Dirac magnetic monopole type.
For $l=-2s$, $\alpha=0$  the  equations (\ref{e1})--(\ref{e5})
reduce to
\be                 \label{w=0}
s''+2s'^2={\rm e}^{-2g}-
\frac12\,{\rm e}^{-4g} + \frac12\ ,\ \ \ \ \ \ \ \ \ \
g''+2s'g'={\rm e}^{-2g}
-{\rm e}^{-4g}\ ,\ee
$$
-4s'^2+2 g'^2+2{\rm e}^{-2g}+1-{\rm e}^{-4g}=0 \ .
$$
This system does not seem to  have a simple general solution,
but there are two important special solutions.

One special   solution is already
known -- the Abelian BPS configuration  (\ref{chir}).
There is another
simple  but non-BPS solution representing background
 with  $g=0$, i.e. with constant radius of $S^2$.

Indeed, $g=0$  solves the second equation in \rf{w=0}, and then
the resulting solution is
\be                 \label{tube}
w=0,\ \ \ \ \   g=0 \ , \ \ \ \ \ \    s  =
\frac{r}{\sqrt{2}}+s_0\ ,\ \ \ {\rm i.e.}
\ \ \   \ \
\R={\rm e}^g=1,\ \ \ \Phi=s-g=\Phi_0  +  \frac{r}{\sqrt{2}}\, .
\ee
The 4-geometry (\ref{0a2}) is thus the  direct product of
$R^2$ and unit $S^2$.
This solution will be important in what
follows, as  it will play the role of an attracting fixed point
 for a class of globally regular non-BPS  solutions.

One may wonder if this non-supersymmetric solution is stable.
In fact, the  instability  of the $w=0,g=0$
solution is suggested by the ``Higgs"
 form of the potential
for $w$ in  the 1-d action  \rf{lll},(\ref{1aa}).
Indeed, using the fact that our background   is
static, and that the metric has 2-d Lorentz symmetry in the $(t,r)$ plane,
it is straightforward to generalize  the equations
\rf{w=0} to the case of time $t$ and $r$ dependent perturbations  near
the solution \rf{tube} (note that linear $s$ or linear  dilaton
provides a spatial friction term):
\be
 - \partial^2_t  \delta w  + \delta w''+\sqrt{2}\delta w'+\delta w=0,\ \ \
 - \partial^2_t \delta \R +
  \delta \R''+\sqrt{2}\delta \R'-2\delta \R=0,\ \ \
\delta Z=-\delta \R'\, . \la{pert}
\ee
 $  w $  has ``tachyonic" mass term, and thus its
perturbations may grow with time, just as in the standard
$(w^2 -1)^2$ scalar  potential case.\footnote{This argument does not
contradict the expected
stability of the  $w=0$  Abelian  BPS
(supersymmetric)
solution \rf{chir}:   there $g$
 is non-trivial and $w$,  $g$  and $\Phi$  perturbations  mix.}
Ignoring time dependence,
 the  four basic solutions  of  \rf{pert}  are
\be             \label{lambdas}
\delta w=\exp(-\frac{1\pm i}{\sqrt{2}}\,r)\, ,\ \ \ \ \ \ \ \
\delta \R=\exp(-\frac{1\pm \sqrt{5}}{\sqrt{2}}\,r).
\ee
Because of the spatial friction term related to linear dilaton,
 $\delta w$ tends to zero for large $r$,  oscillating
infinitely many times as it decreases.

\medskip

The $D=10$ form  of this solution (written in S-dual form with $F_3$
replaced by the NS-NS 3-form $H_3$) has
very simple form: in  the  the string frame
  the  background is the
direct product  of  flat  $R^{1,3}$,
radial $r$-direction with linear dilaton,
and angular 5-space $M^5$ supported by $H_3$ flux.
Explicitly (restoring the dependence
on  the 3-form charge $P$ and changing the sign of the dilaton)
\be        \label{EEtube}
d{s}^2_{10\ {\rm NS-NS}}=
P\,\left(-dt^2+d{\rm x}^n d{\rm x}^n+dr^2+ dM^2_5 \right) \ ,
\ \ \ \ \ \   dM^2_5= e_1^2+e_2^2+
{\epsilon}_1^2+
{\epsilon}_2^2+
\tilde{\epsilon}_3^2\, ,  \ee
\be \la{hhd}
\Phi_{\rm NS-NS} = -\Phi  =- \Phi_0 - \frac{r}{\sqrt{2}}\ , \ \ \ \ \ \ \ \ \ \ \ \
H_3 = P\,\tilde{\epsilon}_3\wedge\left(
{\epsilon}_1\wedge{\epsilon}_2
+{e}_1\wedge{e}_2\right) \ .
\ee
This NS-NS background  may be interpreted
as a  near-throat region of NS5-brane wrapped
over the transverse $S^2$ in a special  way that
breaks all supersymmetries.
As in other  NS5 brane cases (like the regular BPS solution
\rf{BPS}, this NS-NS description is valid for  $r \gg  0$
when the coupling is small, while for small
$r$ one needs to consider the S-dual background \ci{itz}.

 Like the throat region of the standard
NS5-brane \ci{HAR}
described by  $R^{1,6}\times S^3$ or
 $SU(2)$ WZW model with linear dilaton,
 this  model
has a remarkably simple  {\it world-sheet
conformal sigma model}  interpretation.

Indeed,  the $M^5$ metric
 \be \la{tet}
dM_5^2=  d\theta_1^2+\sin^2\theta_1d\phi_1^2
+ d\theta_2^2+\sin^2\theta_2 d\phi_2^2 +
       (d\psi+\cos\theta_1 d\phi_1+\cos\theta_2 d\phi_2)^2
\ee
is of the same $\frac{SU(2)\times SU(2)}{U(1)}$ coset
 form  as the $T^{1,1}$ metric \rf{stan},
but now the relative coefficients of the $U(1)$ and $S^2$ factors
are equal since this is not an Einstein space but
 rather a solution
of the 5-d Einstein equations with the  ${H}_3$ stress tensor term.
We shall call this space $\tilde T^{1,1}$.

The 3-form
\be
H_3=P\,(d\psi+\cos\theta_1 d\phi_1 +\cos\theta_2 d\phi_2)\wedge
(\sin\theta_2 d\theta_2\wedge d\phi_2-\sin\theta_1 d\theta_1\wedge d\phi_1),
\ee
has potential ($H_3=dB_2$)
\be\la{bee}
B_2=P \left[ (\cos\theta_1 d\phi_1-
\cos\theta_2 d\phi_2) \wedge d\psi
+\cos\theta_1\cos\theta_2 d\phi_1\wedge d\phi_2\right]\ .
\ee
Combining the  $\tilde T^{1,1}$ metric \rf{tet}
 with this antisymmetric 2-tensor we get  the same  D=5  NS-NS
 background that was  discovered recently \cite{PZ}
  as a  simplest  representative
  in a special class of
   $\frac{G\times G'}{H} = \frac{SU(2)\times SU(2)}{U(1)} $
 coset   sigma models introduced
in \cite{Guadagnini:1987ty}.
As was  checked in \ci{PZ},
 the corresponding bosonic
sigma model  is conformally invariant in the one- and two-loop
approximation (3-loop approximation in the world-sheet
 supersymmetric case),
 and there are good reasons to believe that (in a proper scheme)
these backgrounds are exact NS-NS string solutions
to all orders in $\alpha'$.

The string world-sheet action of this
$\frac{SU(2)\times SU(2)}{U(1)}$ coset model is
obtained as follows. Let $(\psi_1,\theta_1,\phi_1)$ and
$(\psi_2,\theta_2,\phi_2)$ be the Euler angles that
parametrize the two SU(2) group
manifolds. Taking the  sum of the  two SU(2) WZW models
with  equal  levels $k=P\in Z$ and  adding
the current-current  interaction term  \cite{Guadagnini:1987ty}
with the same  coefficient $P$  one finds  \cite{PZ}
\bea                        \label{WZW}
I=\frac{P}{4\pi}\int d^2\sigma
 \bigg[
\partial_\mu\theta_1\partial^\mu\theta_1 +
\partial_\mu\phi_1\partial^\mu\phi_1+
\partial_\mu\psi_1\partial^\mu\psi_1+
\cos\theta_1\partial_\mu\phi_1\partial_\nu\psi_1\,
(\eta^{\mu\nu}+\epsilon^{\mu\nu}) &&
 \nonumber \\
+\  \partial_\mu\theta_2\partial^\mu\theta_2 +
\partial_\mu\phi_2\partial^\mu\phi_2+
\partial_\mu\psi_2\partial^\mu\psi_2+
\cos\theta_2\partial_\mu\phi_2\partial_\nu\psi_2\,
(\eta^{\mu\nu}+\epsilon^{\mu\nu})  && \nonumber\\
+\  (\cos\theta_1\partial_\mu\phi_1+\partial_\mu\psi_1)
(\cos\theta_2\partial_\nu\phi_2+\partial_\nu\psi_2)\,
(\eta^{\mu\nu}+\epsilon^{\mu\nu})\bigg]. &&
\eea
The  U(1) gauge invariance of this action
allows one to set $\psi_2=0$ as a gauge choice.
 Denoting then $\psi\equiv\psi_1$,
the coset model  (\ref{WZW}) becomes the same as the
string  sigma model corresponding to  the D=5
target space \rf{tet},\rf{bee}.

The exact central charge of (world-sheet  supersymmetric version
of) this model   is
\be                                  \label{cc}
c=2\times\frac{3k'}{k'+2}-1=5-\frac{12}{k}\ , \ \ \ \  \ \ \
k'=k-2\ ,\ \ \ \ ~~~k=P\, .
\ee
As in the case of the NS5 throat model,
 the central charge deficit of this coset model
 is canceled by the  linear dilaton in \rf{hhd}.
Indeed,  the central charge  (dilaton $\beta$-function) equation
\be                         \label{beta}
\bar \beta^\Phi=\frac14\,(D-10)+\alpha'\left[
-\frac12\nabla^2\Phi+(\partial\Phi)^2-\frac{1}{24}H_3^2\right]
+O(\alpha'^4)\,
\ee
vanishes for the $D=10$ background  (\ref{EEtube})
(here  $D=10, \ (\partial\Phi)^2 = { 1\over 2}=
\frac{1}{24}H_3^2 = { 1\over 6}R$).

It is possible to  check directly (e.g., following the discussion in \ci{PT})
that this solution breaks all
 supersymmetries (all such coset models in \ci{PZ}
 were claimed to be non-supersymmetric).
  It
 may have a relation to some  non-supersymmetric
 deformation of D=6  little string model compactified  on $S^2$.
Returning back to the S-dual  R-R background supported by the
 3-form $F_3$, one may write down the
corresponding string-frame metric as  ($g_s= {\rm e}^{\Phi_0}$)
\be
ds_{10\ {\rm R-R}}^2=  g_s P {\rm e}^{ {1 \over\sqrt 2} r }(-dt^2+d{\rm x}^n d{\rm x}^n+dr^2+dM_5^2)
 \equiv  d\rho^2 + { 1 \ov 2} \rho^2
  ( -dt^2+d{\rm x}^n d{\rm x}^n + dM_5^2)\ .
\ee
One may
speculate that string theory in this
simple background may be  dual  to a
   non-supersymmetric deformation of
 ${\cal N}=1$ supersymmetric  theory
discussed in    \ci{Maldacena:2000yy}.\footnote{While
the string coupling
${\rm e}^{\Phi} = g_s  {\rm e}^{ {1 \over\sqrt 2} r } $
decreases for small $r$,  as in the near-horizon
D5 brane case \ci{itz}
the curvature grows indefinitely
 at $r \to -\infty$  and thus  the supergravity
approximation  breaks down there. There is also the usual
problem of non-decoupling (at supergravity level)
of KK modes corresponding to $M_5$ space
since its scale is naturally of the  order of the string  scale.}

This NS-NS (or R-R)   solution admits a trivial
non-extremal generalization (to be discussed below):
one is simply to replace the $(t,r)$ part of the
metric and the dilaton by  the 2-d dilatonic black hole background
\ci{BH}.

\subsubsection{Globally regular solutions}

Consider now general extremal non-BPS solutions
of the second order field equations
with   {\it non-constant} $w$.
For $\alpha=0$ (i.e.  $\nu=\const=1$) the
independent
field equations
(\ref{1a})--(\ref{1f})
reduce to
\bea            \label{1q}
&&\R''+\frac{3w'^2-\R'^2+1}{\R}-4\R Z^2+\R-6Z\R'=0 ,\nonumber \\
&&Z'+4Z^2+\frac{\R'^2-2w'^2-1}{\R^2}+6\frac{Z\R'}{\R}=1 ,\nonumber \\
&&w''+2Zw=\frac{(w^2-1)w}{\R^2},\ \ \ \ \ \ \ \  Z\equiv\Phi'\ ,
\eea
plus the constraint
\be          \label{2q}
2\R^2Z^2+4\R Z\R'+\R'^2-1-w'^2+\frac{(w^2-1)^2}{2\R^2}-\frac12\R^2=0.
\ee
We  will  be interested in
solutions that are {\it globally regular}. This means that either the curvature
is everywhere bounded or it takes an infinite geodesic time to reach
the region with unbounded curvature -- the spacetime
manifold is geodesically complete.
First of all, we shall consider solutions that have a
{\it regular origin}, which is the point $r=r_0$ where $\R$
vanishes
but the
curvature is bounded. One can set $r_0=0$.
The manifold cannot be analytically continued towards
negative $r$ in this case, and so one can assume
without loss of generality that $r\geq 0$.\footnote{Not all globally
 regular solutions considered below will
have a regular origin, and so the restriction
$r\geq 0$ will not
always apply.}
The inspection of the field equations shows that
 such solutions form a one-parameter  family,
  with the following small $r$   Taylor expansion:
  \begin{equation}\seqalign{\span\TC}{
w=1-b r^2+O(r^4), \ \ \ \ \ \ \
Z=\Phi'=2(b^2+\frac{1}{12}) r+O(r^3) ,  \cr\noalign{\vskip1\jot}
  \R={\rm e}^{g}=r-(b^2+\frac{1}{36})r^3+O(r^5),\ \ \
\Phi=\Phi(0)+(b^2+\frac{1}{12}) r^2+O(r^4)   \,. } \label{6}
  \end{equation}
Here $b$ and $\Phi(0)$ are free parameters. The value
$$b={1\over 6}$$
corresponds to the
regular BPS solution (\ref{BPS}),
while for $b\neq { 1 \ov 6} $ we obtain its regular, non-BPS
deformations.
Expansions (\ref{6}) determine only  local
solutions for small $r$, and the next step is to extend these
solutions  to finite values of $r$.
Our strategy  will be  to numerically integrate
Eqs. (\ref{1q}) in the interval $r\in[0,\infty)$
using (\ref{6}) as the boundary conditions
at $r=0$. Since the constraint (\ref{2q}) is fulfilled by the
initial values (\ref{6}), it holds for all $r$.

Let us  discuss the boundary conditions at $r=\infty$.
Having in mind  future  applications, let us  consider the
general equations (\ref{1a})--(\ref{1f}) with $\alpha\neq 0$.
Assuming that $\R\to\infty$ for large $r$, we
find the following series solutions
in the vicinity
of $r=\infty$:\footnote{These expressions
  apply only  to solutions for which $\R$ is unbounded.
Similar expansions exist for solutions  where $\R$ is bounded.}
\bea                   \label{inf}
\R&=&\sqrt{2x}-\frac{\Upsilon^2}{\sqrt{2}x^{3/2} }\,
(1-\frac14\cdot \frac{3\Upsilon^2-10}{x}+ \ldots)
+\sqrt{2}{\cal P}x{\rm e}^{-x}(1+\frac2x+\ldots)+O({\rm e}^{-2x})
  \, , \nonumber \\
\Phi&=&\Phi_\infty +\frac{x}{2}
-\frac{1}{4}\ln x
+\frac{5\Upsilon^2}{16x^2}\,(1-\frac25\cdot \frac{2\Upsilon^2-7}{x}+\ldots)
-{\cal P}\sqrt{x}{\rm e}^{-x}(1+\frac1x+\ldots)+O({\rm e}^{-2x})
 \, ,\nonumber  \\
w&=&\frac{\Upsilon}{\sqrt{x}}\,(1+\frac{1-\frac12\Upsilon^2}{x}
+\ldots)+
{\cal C}x{\rm e}^{-x}(1+\ldots)+O({\rm e}^{-2x})\, ,\nonumber  \\
\nu&=&\frac{1}{\mu^2}-\frac{\alpha}{\sqrt{x}}\,
{\rm e}^{-x-2\Phi_\infty}(1+\ldots)+O({\rm e}^{-2x})\, , \ \ \ \ \ \ \ \ \
x\equiv\mu(r+r_\infty) \ .
\eea
Here $\mu$, $r_\infty$,
${\cal P}$, $\Phi_\infty$, $\Upsilon$, ${\cal C}$ are six integration
constants. Notice that 6 is the maximal number of integration
constants a solution can have: Eqs.(\ref{1a})--(\ref{1f})
can be reformulated as a system of 7 first order equations plus
one constraint. As a result, (\ref{inf}) determines
asymptotics of a
{\sl generic} solution for which $\R\to\infty$ for $r\to\infty$.
There are
also solutions for which $\R$ is bounded for large $r$.
It is worth noting that solutions with asymptotics (\ref{inf})
are geodesically complete for large $r$, and moreover
all curvature invariants determined by
(\ref{inf}) vanish for $r\to\infty$.

The parameter $\mu$ (which may be interpreted as the dilaton charge at
infinity)
 reflects the scaling symmetry (\ref{scaling}) of the
equations.
Comparing with (\ref{BPS}), (\ref{chir}),
we conclude that for large $r$ the solutions generically
have the same asymptotics as the BPS solutions (\ref{BPS}),
up to a rescaling and a shift, plus the  polynomial terms
proportional to $\Upsilon$, and 
plus also the exponentially small terms proportional to ${\cal P}$.

In  the extremal case of $\alpha=0$ the solutions for
$r\to\infty$ are then given by (\ref{inf}) with $\mu=1$
(we are assuming $\nu=\nu_0=1$).
The next step is to numerically interpolate between
the  $r\to 0$ asymptotics (\ref{6})
and these large $r$ asymptotics,
to find  the one-parameter family of regular
solutions in the whole interval $[0,\infty)$.
 It turns out that for any $b\in (0,1/2)$ the local regular
solution (\ref{6}) can be extended all the way up to the infinity
to meet the asymptotic solution (\ref{inf})  with  certain  special
$b$-dependent
values of the parameters
$\mu(b)$, $r_\infty(b)$,
${\cal P}(b)$, $\Phi_\infty(b)$, $\Upsilon(b)$, ${\cal C}(b)$
(see Figs.\ref{figUps},\ref{figfPAR}).

\begin{figure}[h]
\begin{minipage}[b]{0.45\linewidth}
  \centering\epsfig{figure=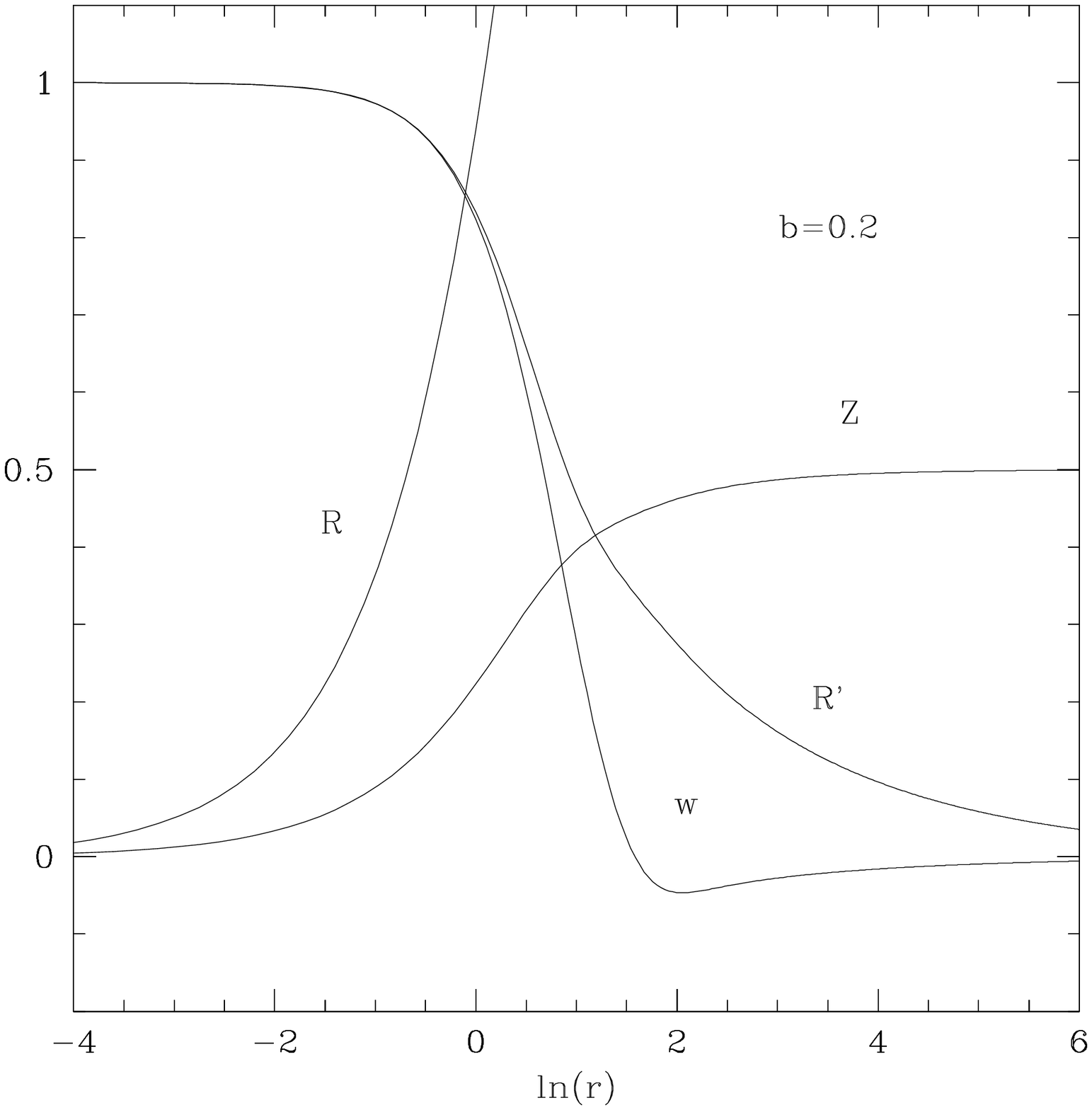,width=1.2\linewidth}
  \caption{{\fontsize{10}{12}\selectfont
Non-BPS solution for $b=0.2$.
}}
  \label{fig1}
   \end{minipage}\hspace{4 mm}
\begin{minipage}[b]{0.45\linewidth}
 \centering\epsfig{figure=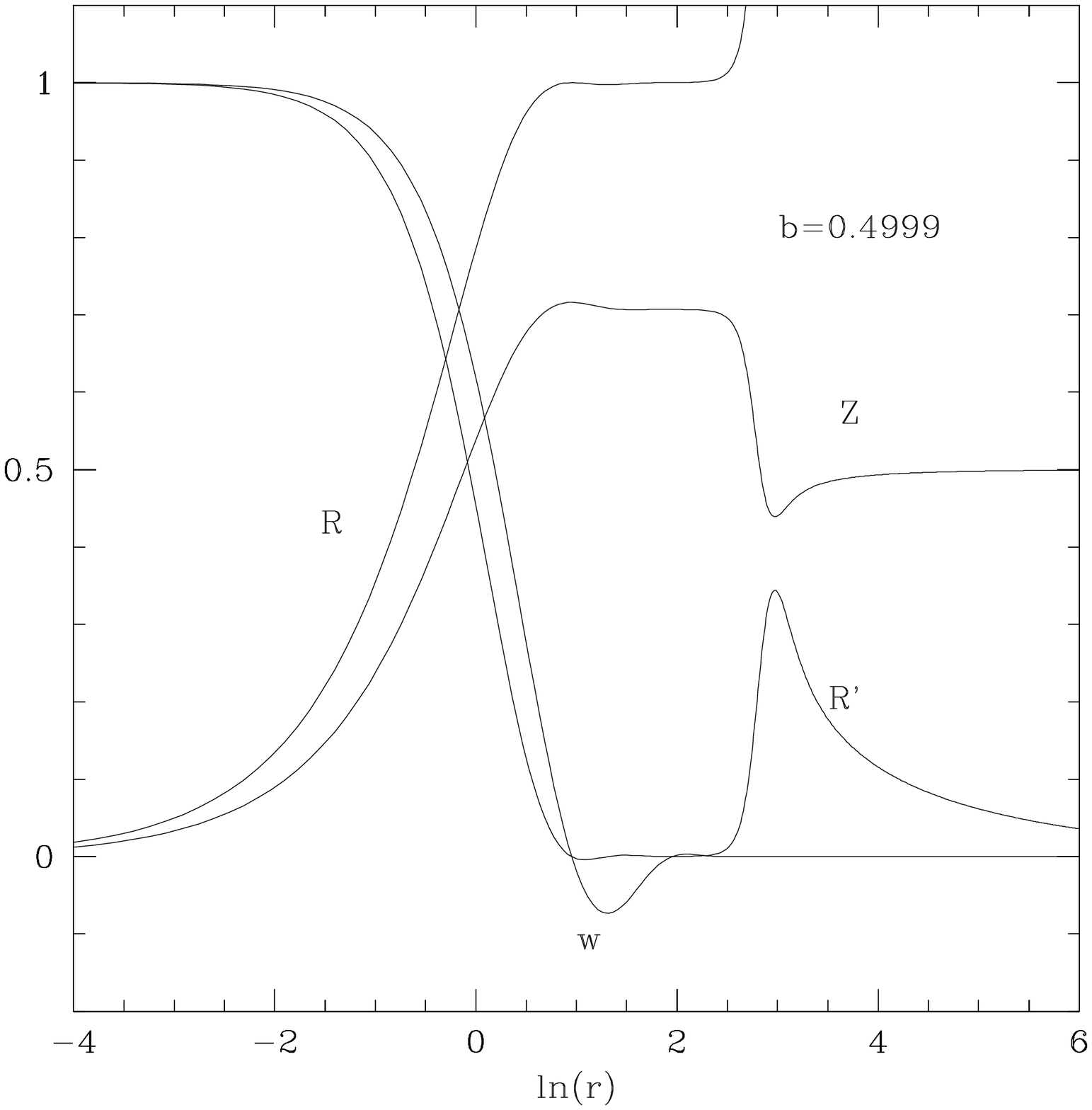,width=1.2\linewidth}
  \caption{{\fontsize{10}{12}\selectfont Non-BPS solution for
$b=0.499$.
}}
  \label{fig2}
   \end{minipage}
\end{figure}

The  behavior of the solutions is illustrated in Fig.\ref{fig1} and
Fig.\ref{fig2}.
For $0<b<1/6$ the function  $w$ is always positive, while
for $b>1/6$ it has at least one zero.
As $b$ tends to 1/2, $w$ develops more and more oscillations
around zero, while the functions  $\R$ and  $Z$ start oscillating around
their   constant values ($1$ and $1/\sqrt 2$, respectively)
 corresponding to the special Abelian solution
(\ref{tube}). Thus, one
  may  say that the solution (\ref{tube})
    acts as the large-$r$ attracting
fixed point for these regular solutions.
Specifically, among  the four independent linear fluctuation
modes (\ref{lambdas}) near this special solution  there are three modes that
are regular for large $r$. These modes parameterize the ``stable
manifold'' in the vicinity of the fixed point, and their existence
is the reason why
the nearby phase trajectories approach the fixed
point. As a result, the trajectory that starts from the
origin gets attracted by the fixed point (\ref{lambdas}) and stays
longer and longer in its vicinity as $b$ tends to 1/2. However,
for $b<1/2$, the trajectory finally gets repelled from the fixed point
due to the existence of the  fourth,
unstable,  mode  in (\ref{lambdas}), and after that
it
goes to the region where R is infinite.

\subsubsection{Limiting solutions}

A very interesting phenomenon occurs for the special case
of $b=1/2$.
For $b\to 1/2$ the trajectory approaches the fixed point
(\ref{tube}) closer and closer, and finally for $b=1/2$
the limiting trajectory
 splits into two parts. For the first, interior part the
trajectory starts from the origin at $r=0$, and in the
limit $r\to\infty$ arrives exactly at the fixed point (\ref{tube}) --
after infinitely many oscillations.
The second, exterior
part of the limiting trajectory corresponds to the solution that
interpolated between the fixed point (\ref{tube}) and infinity.

\begin{figure}[h]
\begin{minipage}[b]{0.45\linewidth}
  \centering\epsfig{figure=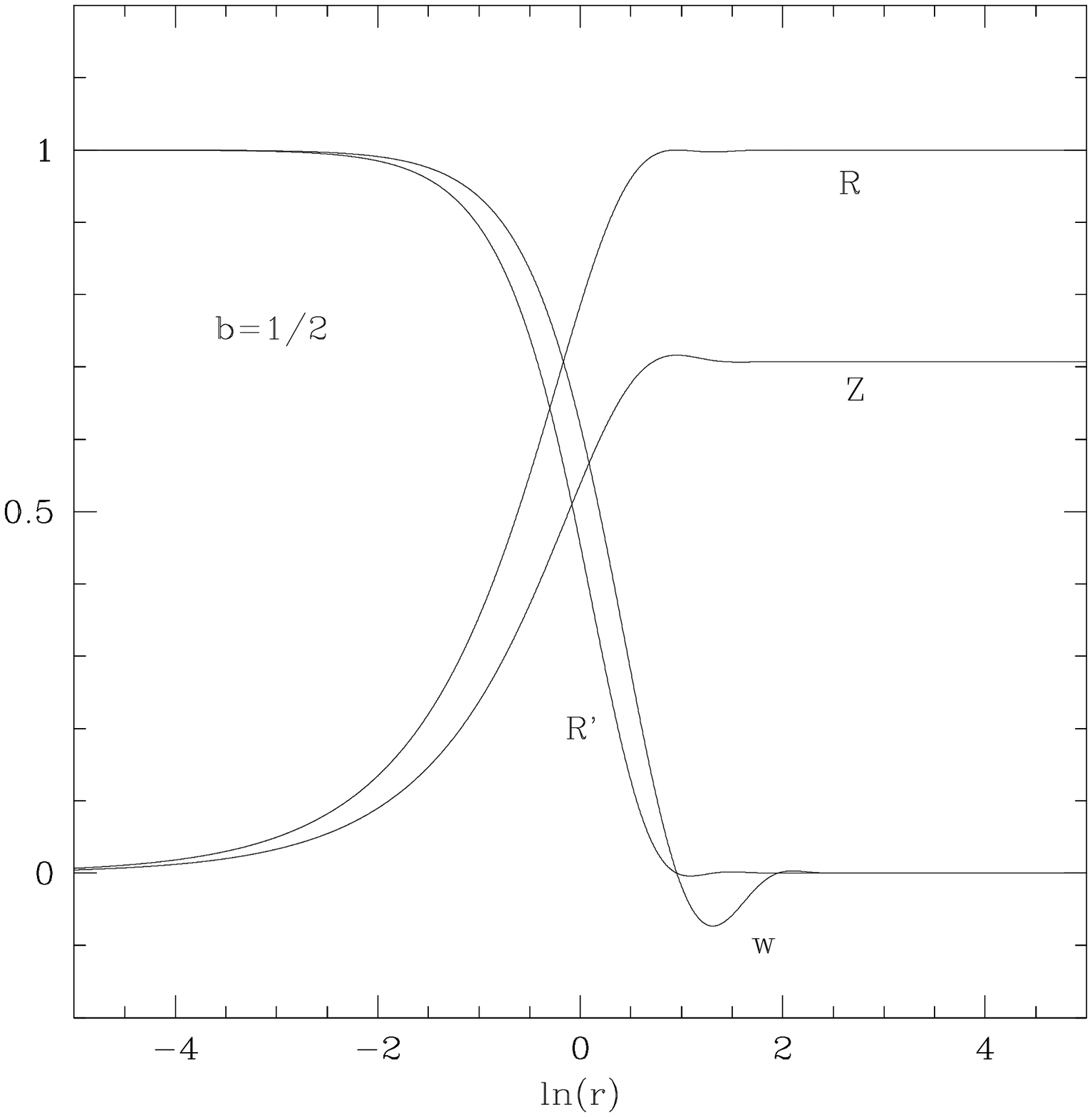,width=1.2\linewidth}
  \caption{{\fontsize{12}{12}\selectfont
Interior limiting solution
}}
  \label{fig3}
   \end{minipage}\hspace{5 mm}
\begin{minipage}[b]{0.45\linewidth}
  \centering\epsfig{figure=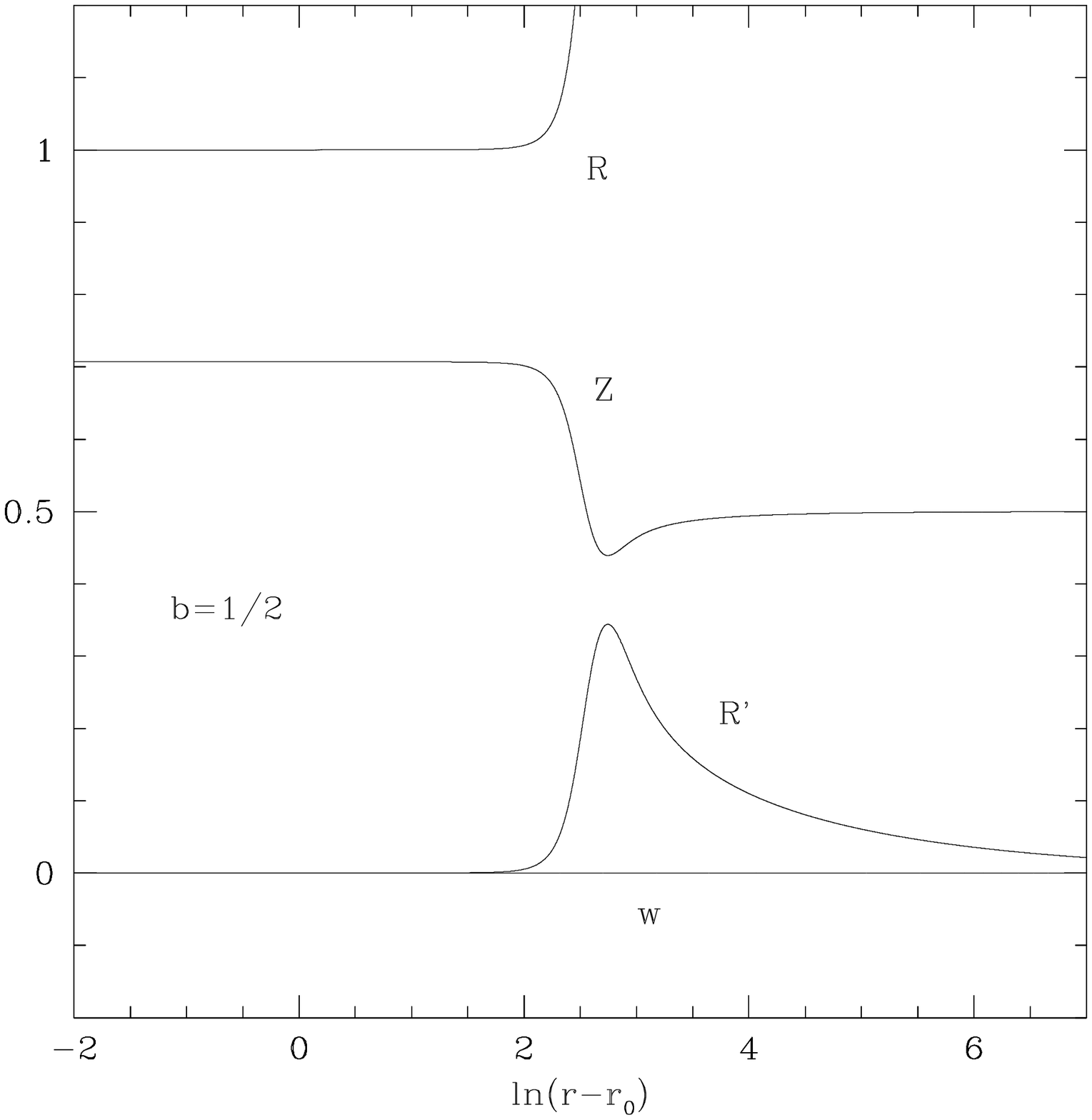,width=1.2\linewidth}
  \caption{{\fontsize{12}{12}\selectfont Exterior limiting solution
}}
  \label{fig4}
   \end{minipage}
\end{figure}

Let us construct first the interior limiting solution.
Returning back to the Lagrangian (\ref{1aa}), we introduce the
new variables $p(r)$, $f(r)$ related to $w(r)$, $g(r)$ via
\be               \label{angle1}
w=\ \cosh p \ \cos f ,\ \ \ \ \ \ \ \ \ \ \
 {\rm e}^g=\ \cosh p   \ \sin f \, .
\ee
The Lagrangian then becomes
\be           \label{angle2}
L={\rm e}^{-l}\left(s'^2-\frac{f'^2}{2\sin^2 f }
-\frac{\tanh^2 p\  \,p'^2}{2\sin^2 f }
\right)-
\frac14\,{\rm e}^{4s+l}
\left(
\frac{\tanh^4p }{\sin^4 f }-\frac{2}{\sin^2 f }\right)
-\frac{\alpha^2}{4}\,{\rm e}^{l}\, .
\ee
The advantage of such a parameterization is that, as one can
immediately see, $p(r)=0$ is a solution of the equations of motion.
This means that the field equations admit the following first integral
\be           \label{angle3}
w^2+{\rm e}^{2g}=1\, .
\ee
It turns out that for $b=1/2$ this condition
arises automatically.
Indeed,
the equation for $p(r)$ derived from  (\ref{angle2}) shows that
for $b=1/2$ the function  $p$ and all its derivatives at $r=0$
vanish. As a result, we have $p=0$, and the
Lagrangian (\ref{angle2}) becomes  simply
\be           \label{angle4}
L={\rm e}^{-l}s'^2-\frac{{\rm e}^{-l}f'^2}{2\sin^2 f }
+\frac{{\rm e}^{4s+l}}{2\sin^2 f }\  .
\ee
The field equations are then (in the gauge $l=-2s$)
\be                  \label{angle5}
s''+2s'^2=\frac{1}{\sin^2 f},\ \ \ \ \
f''+2s'f'= \ (1+f'^2)\cot f\ ,\ \ \  \  \ 2\,s'^2\sin^2f =f'^2+1\, .
\ee
The solution will be regular at the  origin if
$s=\ln r+O(r^2)$ and $f=r+O(r^3)$ for $r\to 0$.
Integrating (\ref{angle5}) with these boundary conditions
shows that $f\to\pi/2$ for large $r$. Reconstructing
$w$, $\R$, and $Z=s'-f'\cot f , $ finally gives the solution shown in Fig.\ref{fig3}.
This solution is globally regular (regular at $r=0$)
and for large $r$ it tends to the special Abelian solution (\ref{tube}).

Consider now the exterior limiting solution.
 Here $ \R$  never vanishes, so the range of $ r $ is to be taken from
 $-\infty$ to $+\infty$.
 The solution
 starts from
the special Abelian solution (\ref{tube}) at $r=-\infty$.
Eq.(\ref{lambdas}) shows that there is only one mode around
this solution which is stable for $r=-\infty$:
$\delta w=0$,
$\delta \R=\exp(-\frac{1-\sqrt{5}}{\sqrt{2}}\,r)$,
$\delta Z=-\delta \R'$. This shows that we must keep $w=0$
for all $r$, while  $\R$, $Z$ can deviate from the values
determined by the solution (\ref{tube}), so that for
$r\to -\infty$
\be
\R=1+\exp(\frac{\sqrt{5}-1}{\sqrt{2}}\,(r-r_0))+\ldots\, ,\ \ \ \
Z=\frac{1}{\sqrt{2}}+\frac{1-\sqrt{5}}{\sqrt{2}}
\exp(\frac{\sqrt{5}-1}{\sqrt{2}}\,(r-r_0))+\ldots\, .\ \ \
\ee
Here $r_0$ is an arbitrary parameter corresponding to the
possibility of global translations. Integrating the field
equations with such boundary conditions shows that for
$r\to+\infty$ the solution follows the asymptotic behavior
(\ref{inf}); see Fig.\ref{fig4}.

To recapitulate, both the interior and exterior limiting solutions
shown in Fig.\ref{fig3} and Fig.\ref{fig4} are globally regular. The interior solution
interpolates in the interval $[0,\infty)$ between the regular
origin and the special Abelian solution (\ref{tube}). The
exterior solution interpolates for $r\in(-\infty,+\infty)$
between the solution (\ref{tube}) and the asymptotic (BPS) solution
(\ref{inf}).

Summarizing this section, globally regular solutions exist for $b\in[0,1/2]$.
The solution with $b=0$ has not been
described so far: in this case $w(r)=1$,
which corresponds to the case described by Eq.(\ref{zerofield}).
The qualitative
behavior of $\R$ and $Z$ is then similar to that shown in Fig.\ref{fig1}.
If $b<0$ then  solutions are  still  regular at the origin,
 but
$w$ diverges at some finite $r$, where these solutions
develop a curvature singularity. For
 $b>1/2$ solutions have compact spatial
sections, since $\R$ develops a second zero
(in addition to the one at $r=0$)
 at some finite $r$, where the geometry
is singular. This type of behavior is somewhat similar to what is shown
in Fig.\ref{fig6} for black holes.

As we shall see below, among all globally regular solutions
described above, there is only a discrete subset of solutions
for which the energy is finite.

\section{Non-Extremal solutions:  Black holes }
\label{NonZeroT}

\subsection{Solutions with regular horizon}

We shall now turn to non-extremal solutions  that have a
non-constant function $X$ in the 10-d metric  \rf{EEa}
 or $\nu$ in the 4-d metric  \rf{00a2}, corresponding to the case of non-zero
 non-extremality parameter $\alpha$  in \rf{e5} or
 \rf{1e}.  Such solutions generalize
 the regular extreme solutions described in the previous section
to the case when an event horizon is present.
 Since $\alpha$ enters  \rf{1e} in combination
$2 \a e^{-2 \Phi}$,
it can be rescaled  by shifting $\Phi$ by a constant.
In particular, one  can set $\alpha=1/2$, which
we shall  assume in our  numerical analysis.
Since   $\nu={\rm e}^{2X}$ is  non-constant,   such non-extremal
solutions may  have a regular event horizon.
A solution has a regular event horizon
if there is a point $r=r_h$ where $\nu$ has a simple zero, while
all other
functions
 are finite and differentiable at this point.

Without loss of generality one can set $r_h=0$
(since the equations are autonomous).
The field equations then  admit, in  the  vicinity of $r=0$,
local solutions characterized by the following  Taylor
expansions:
$$
\nu= \frac{ 2\a {\rm e}^{-2\Phi_h}}{\R_h^2}\,r+O(r^2)\, ,\ \ \ \ \ \ \ \  \ \
w=w_h+
(2\a {\rm e}^{-2\Phi_h})^{-1}
  w_h(w_h^2-1)\,r+O(r^2)\, ,
$$
$$
\R=\R_h+ (2 \a {\rm e}^{-2\Phi_h})^{-1}\frac{\R_h^2-(w_h^2-1)^2}{\R_h}\,r
+O(r^2)\, ,
$$  \be                \label{ini}
\Phi=\Phi_h+
(2 \a {\rm e}^{-2\Phi_h})^{-1}\frac{\R_h^4+(w_h^2-1)^2}{2\R^2_h}\,r
+O(r^2).
\ee
The parameter $\kappa \equiv 2 \a e^{-2 \Phi_h}$
may be interpreted as a characteristic ``mass  scale''  of black hole.
The free parameters
$\Phi_h$, $\R_h$, and $w_h$
determine
the value of the dilaton at the horizon,  the ``radius'' of the
horizon, and
the value of $w$ at the horizon.
One may  check that all curvature invariants are finite at the
horizon.

We now numerically integrate  Eqs.(\ref{1a})--(\ref{1f})
towards large $r$
using (\ref{ini}) as initial values at $r=0$.
For each set of values of $\Phi_h$, $w_h$, and $\R_h$ this gives us
a black hole solution living in the interval $r\in[0,r_\ast]$, where
$r_\ast$ can be either finite or infinite.
The set of black hole solutions is therefore three dimensional and
has one dimension more as compared
to the regular solutions described in the previous section,
where we had only two parameters -- $b$ and $\Phi(0)$
in (\ref{6}). The additional parameter arising in the black hole case
determines the radius of the even horizon.

In order to qualitatively
describe these black hole solutions for different values
of $\Phi_h$, $w_h$, and $\R_h$, we first notice that choosing
different values of $\Phi_h$ leads merely to global rescalings of the
configurations. For this reason we can set
$\Phi_h=0$, since for other values of $\Phi_h$ the structure of solutions
is qualitatively similar.

\begin{figure}[h]
\begin{minipage}[b]{0.45\linewidth}
 \centering\epsfig{figure=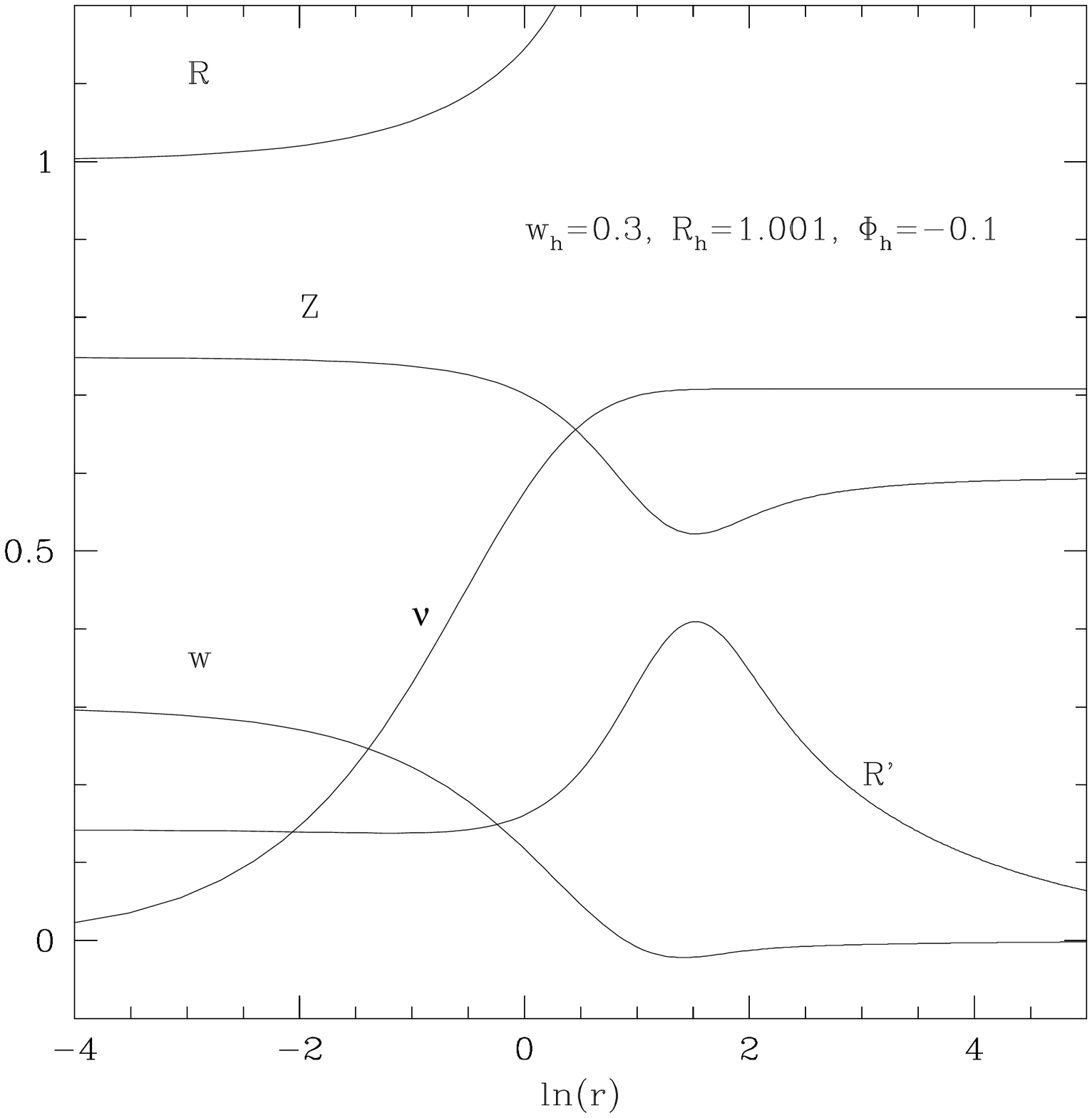,width=1.2\linewidth}
  \caption{{\fontsize{10}{12}\selectfont
Black holes with 
$\R_h>\sqrt{1-w_h^2}$.
This corresponds to Figure~\ref{figA}a. 
}}
  \label{fig5}
   \end{minipage}\hspace{4 mm}
\begin{minipage}[b]{0.45\linewidth}
 \centering\epsfig{figure=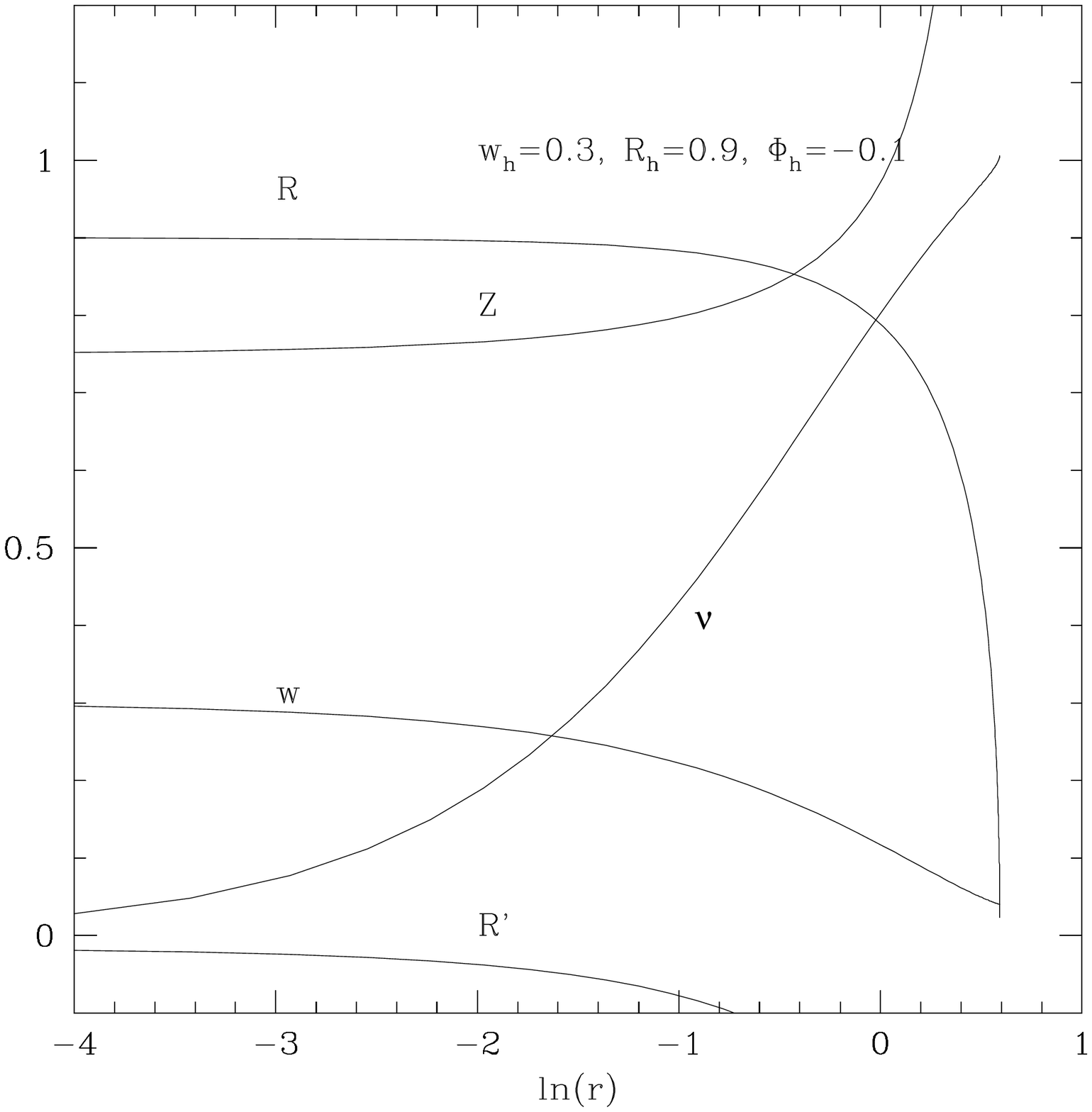,width=1.2\linewidth}
  \caption{{\fontsize{10}{12}\selectfont 
Black holes with $\R_h<\sqrt{1-w_h^2}$.
This corresponds to Figure~\ref{figA}b.
}}
  \label{fig6}
   \end{minipage}
\end{figure}

Since the equations \rf{1a}--\rf{1f}
are symmetric under $w\to -w$, one can
assume that $w_h\geq 0$, and
then one can show that $w_h$ must belong to the interval  $[0,1]$,
since otherwise $w$ diverges at some finite $r$.

Setting $w_h=0$, we will
 obtain Abelian solutions with $w=0$, while
$w_h\neq 0$  will give  non-Abelian solutions. They are qualitatively
similar, the only difference is that for Abelian solutions $w=0$
everywhere, while for non-Abelian ones $w$ starts from a finite value
at the horizon and then approaches zero for large $r$.
 As was discussed above,
configurations with $w=0$ respect the $U(1)$  symmetry
($\psi\to \psi +
\psi_0$),
so
$w_h$  may  be regarded as an
 order parameter for chiral symmetry
breaking.

The horizon  value of $\R$ -- the parameter
 $\R_h$  plays a crucial role.
For $\R_h >\sqrt{1-w_h^2}$,
the solution has the asymptotic form \eno{inf},
such that $\R\to\infty$ for $r\to\infty$.
A typical solution of this form is illustrated in
Fig.\ref{fig5}.
For $\R_h < \sqrt{1-w_h^2}$,
the event horizon is
still regular, but the asymptotics change completely.
$\R$ is no longer unbounded, but reaches a maximal value
at some finite $r$;  after that  it decreases and
finally vanishes at some $r=r_\ast$, where
there is a curvature singularity.
Such a solution is illustrated in
 Fig.\ref{fig6}.\footnote{Solutions of this
 type are sometimes called ``bags of gold.''}

\begin{figure}[h]
\begin{minipage}[b]{0.45\linewidth}
\centering\epsfig{figure=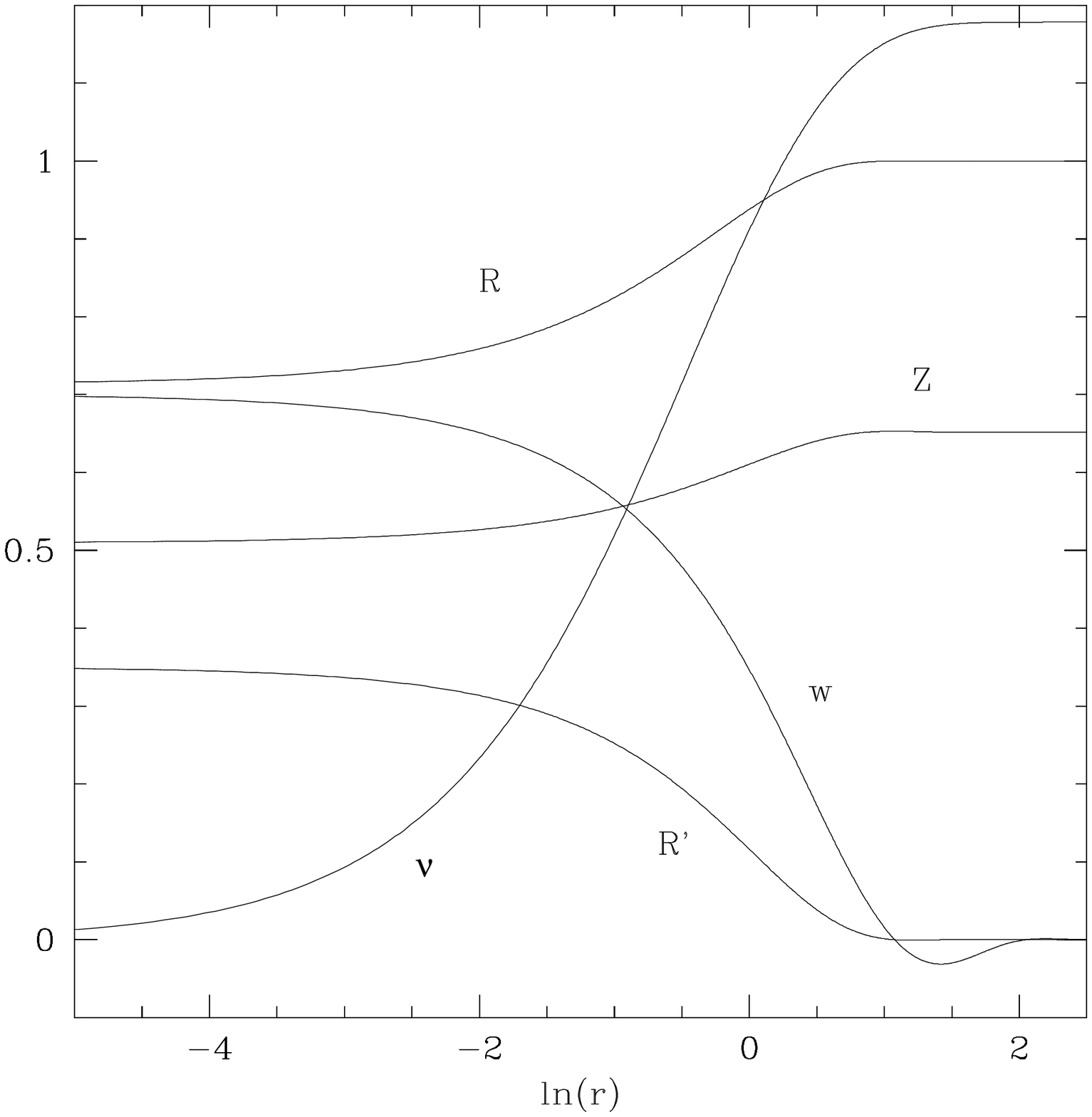,width=1.2\linewidth}
  \caption{{\fontsize{10}{12}\selectfont A typical solution with
$\R_h=\sqrt{1-w_h^2}$.  The oscillations in $w$ are matched by
oscillations in $\R$, too small to be seen in this figure.
  These
oscillations are depicted
in Figure~\ref{figA}d in magnified  form.}}
  \label{figB}
   \end{minipage}\hspace{6 mm}
\begin{minipage}[b]{0.45\linewidth}
\centering\epsfig{figure=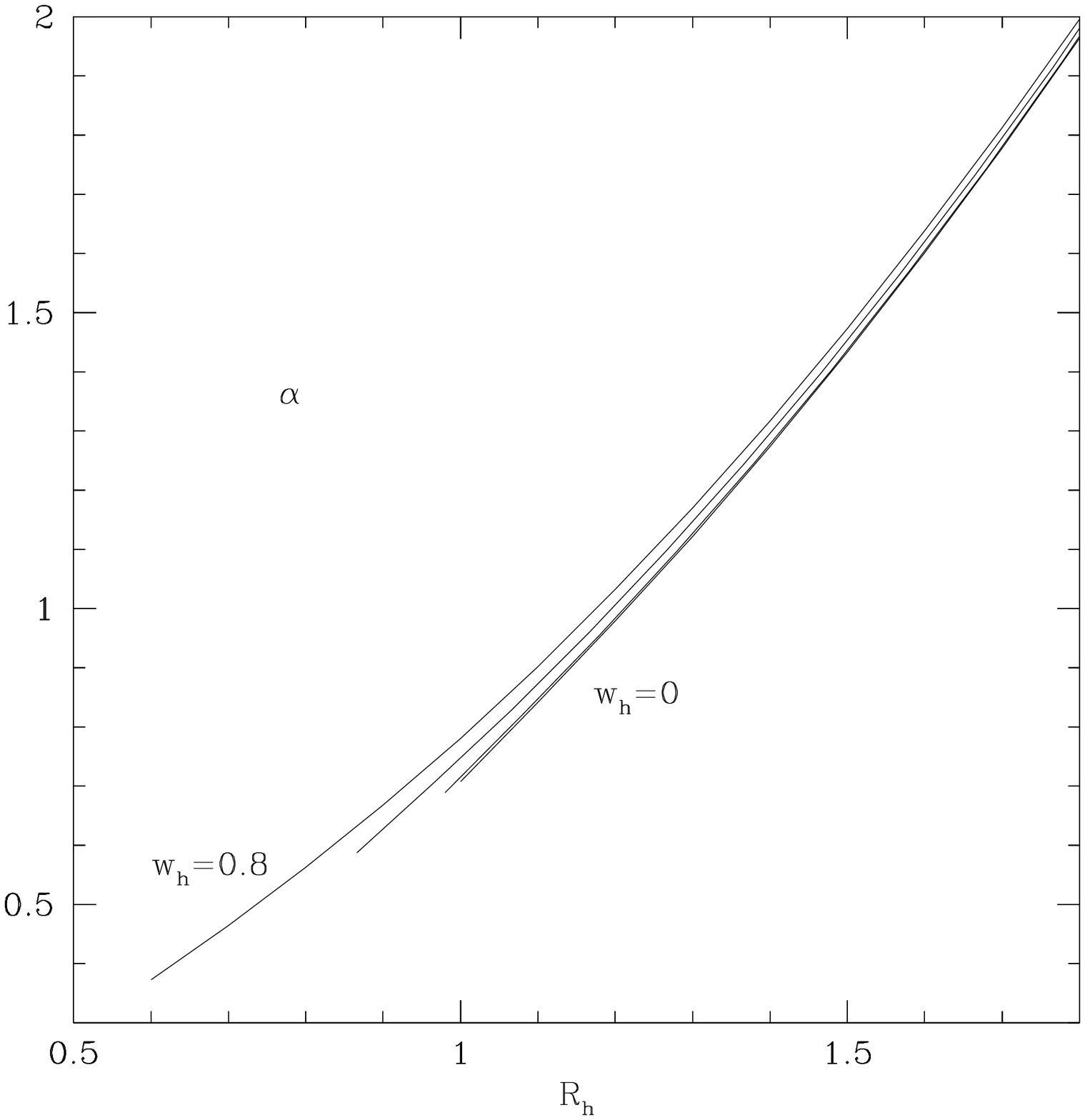,width=1.2\linewidth}
  \caption{{
  \fontsize{10}{12}\selectfont
Non-extremality $\alpha$ for black hole solutions with
R$_h >\sqrt{1-w_h^2}$ normalized such that $\Phi_h=0$,
$\nu(\infty)=1$. The region above (below) the curves corresponds
to values of $\alpha$ for solutions with $\Phi_h>0$
($\Phi_h<0$).
}}
  \label{fig_scale}
   \end{minipage}
\end{figure}

In the ``intermediate'' case, i.e.
for  $\R_h=\sqrt{1-w_h^2}$, the function  $\R$ tends,
for large $r$,
 to a constant  $\R_\infty$. The whole configuration
asymptotically approaches the
(rescaled)
special Abelian solution
(\ref{tube}), so that  $w$ oscillates,
$w \sim e^{-Z_\infty r} \sin(Z_\infty(r-r_0))$, and
$Z-Z_\infty\sim \R-\R_\infty\sim e^{-Z_\infty r}$.
 Such a solution is
illustrated in Fig.~\ref{figB}.
 For $w_h=0$ and
$\R_h=1$ the solution is easy to find  analytically
by solving  \rf{1a}--\rf{1f}:
\be
w=0,\ \ \ \ \  \R=1,\ \ \ \ \ \ Z={\rm const},\ \ \ \
\Phi= \Phi_0 +  Z r,\ \ \ \ \
\nu=\frac{1}{2Z^2}  - \frac{\a}{Z}\,{\rm e}^{-2\Phi_0 - 2 Z r}.
\label{dilaton}
\ee
For $\a=0$, choosing $Z= { 1 \ov \sqrt 2}$
we get the   extremal solution \rf{tube}.
In the  case  of $\a\not=0$
the  4-d metric \rf{00a2}
  is  simply the direct product of $S^2$ and the  2-d
 dilatonic black hole  background (with the
``cigar'' metric in euclidean signature case) \ci{BH}.

For $w_h\neq 0$ and $\R_h=\sqrt{1-w_h^2}$ the non-abelian
component of the gauge field is turned on, leading to
more general solutions which may be thought of as
finite deformations of the ``cigar''.

\begin{figure}[h]
   \centerline{\psfig{figure=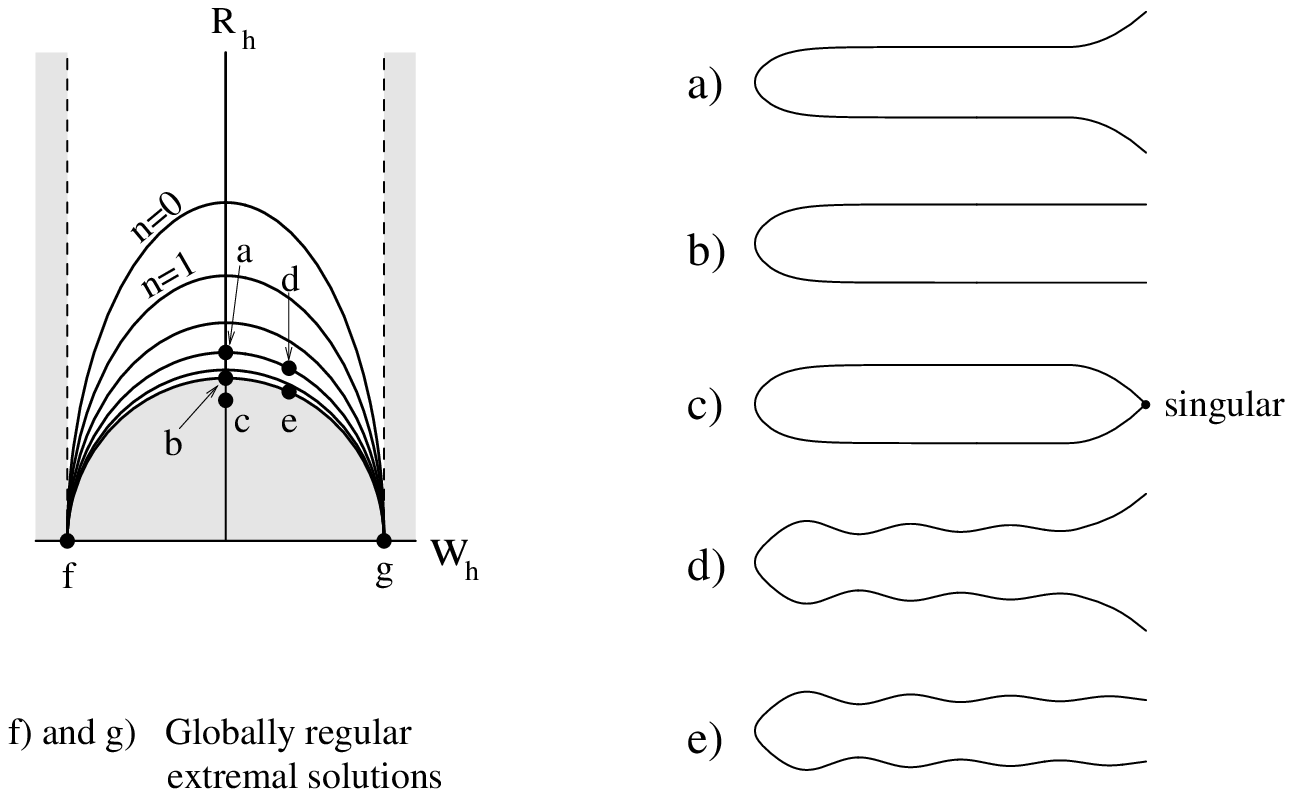,width=5.5in}}
  \caption{{\fontsize{10}{12}\selectfont A qualitatively correct
depiction of the ``phase diagram'' of black hole solutions, and of
particular solutions.  Quantitatively correct plots showing some of
the same information can be found in Figs.\ref{fig5}, \ref{fig6},
\ref{figB}, \ref{fig_scale}, \ref{figw}, and~\ref{figfw}.  Left: In
the unshaded region, solutions are asymptotic to \eno{inf}; in the
shaded region, solutions are singular at finite $r$; and on the
semi-circular border between I and II, solutions are asymptotic to the
cigar geometry.  The dark lines represent those solutions for which
$\Upsilon=0$ in \eno{inf}, which means that the asymptotics at
infinity is asymptotically close to the BPS solution.  Right: The
$(t,r)$ parts of the metrics, in Euclidean signature, are the surfaces
of revolution of the curves shown.}}
  \label{figA}
\end{figure}

The results of the previous paragraph
were discovered numerically,
although it may be  possible
to prove them directly by qualitative analysis of the system of
differential equations.
 To support the claim that for
$\R_h = \sqrt{1-w_h^2}$ the solution is asymptotic to the cigar
geometry for large $r$, recall the parametrization \eno{angle1}.
Putting $\R_h=\sqrt{1-w_h^2}$ amounts to setting
the function $p$  in    \eno{angle1}
to zero at the horizon,
and,  as we saw before, this implies $p=0$ everywhere,
so
 that $w^2 +
\R^2 = 1$.  Linearizing the analytic solution
 \eno{dilaton} around
$w=0$, one finds the claimed damped oscillatory behavior, which is actually
the same as in Eq.(\ref{lambdas}),
so this
solution is a stable attractor as one proceeds to large $r$.  It turns
out (as is confirmed by numerical  analysis)
that for all $w_h$ in the interval $(0,1)$, $\R_h = \sqrt{1-w_h^2}$
leads to this attractor at large $r$.
 A
summary  of the resulting picture
is given
in Figures~\ref{fig5}, \ref{fig6},
\ref{figB}, and~\ref{figA}.

One may regard the behavior as one crosses from $\R_h > \sqrt{1-w_h^2}$
to $\R_h < \sqrt{1-w_h^2}$ as some kind of phase transition,
with $\R_h$ being the order parameter.

Having qualitatively characterized black holes in the theory,
we would like now to choose a suitable normalization for
solutions whose asymptotic
behavior for large $r$ is given by \eno{inf}.
So far we have assumed that $\alpha=\frac12$
and $\Phi_h=0$; this choice leads to an
asymptotic value of the metric function $\nu$ which is not
generically  equal to one, $\nu(\infty)\neq 1$.
We now wish to rescale all solutions in  such a way that
\be                    \label{nu}
\nu(\infty)=1.
\ee
At the same time, we would like to keep the value of the dilaton
at the horizon fixed, since it determines the coupling
constant on the gauge theory side. Let us assume again that
$\Phi_h=0$. In order to be able to fulfill these two conditions
at the same time, it is necessary to allow for arbitrary values
of the non-extremality parameter $\alpha$. The procedure is then
as follows. Given a solution with $\Phi_h=0$ and $\alpha=\frac12$
for some $w_h$ and $\R_h>\sqrt{1-w_h^2}$,
for which $\nu$ asymptotically approaches some value $\nu(\infty)$,
we apply the scale transformation (\ref{scaling}) with
$d=\frac14\ln(\nu(\infty))$. This maps the solution to another
black hole solution for which $\nu$ asymptotically tends to one.
For this new solution we still have $\alpha=\frac12$, but
$\Phi_h$ is not longer zero but rather $\Phi_h=d$.
In order to restore the original value of $\Phi_h$
we apply the scale transformation (\ref{scaling1})  with $C=-d$.
This preserves the asymptotic value of $\nu$, but changes
the value of $\alpha=\frac12$ to  $\alpha=\frac12{\rm e}^{-2d}$.
As a result, the non-extremality parameter $\alpha$ is now fine-tuned
in such a way that we have a black hole solution with both
$\Phi_h=0$ and $\nu(\infty)=1$. In Fig.\ref{fig_scale} we show
the values of the non-extremality in such normalization
for both abelian and non-abelian black hole solutions.

In order to obtain solutions with $\nu(\infty)=1$ and for some other
value of dilaton at the horizon, we apply the scale
transformation (\ref{scaling1})
with $C=\Phi_h$, which multiplies the  vertical
coordinate of the curves
in Fig.\ref{fig_scale} by ${\rm e}^{2\Phi_h}$. It follows then that
for solution with $\Phi_h>0$
the values of $\alpha$ belong to the region
above the curves in Fig.\ref{fig_scale}, while for those with
$\Phi_h<0$, $\alpha$ is in the region below the curves.

\subsection{Hawking temperature}

\begin{figure}[h]
\begin{minipage}[b]{0.45\linewidth}
  \centering\epsfig{figure=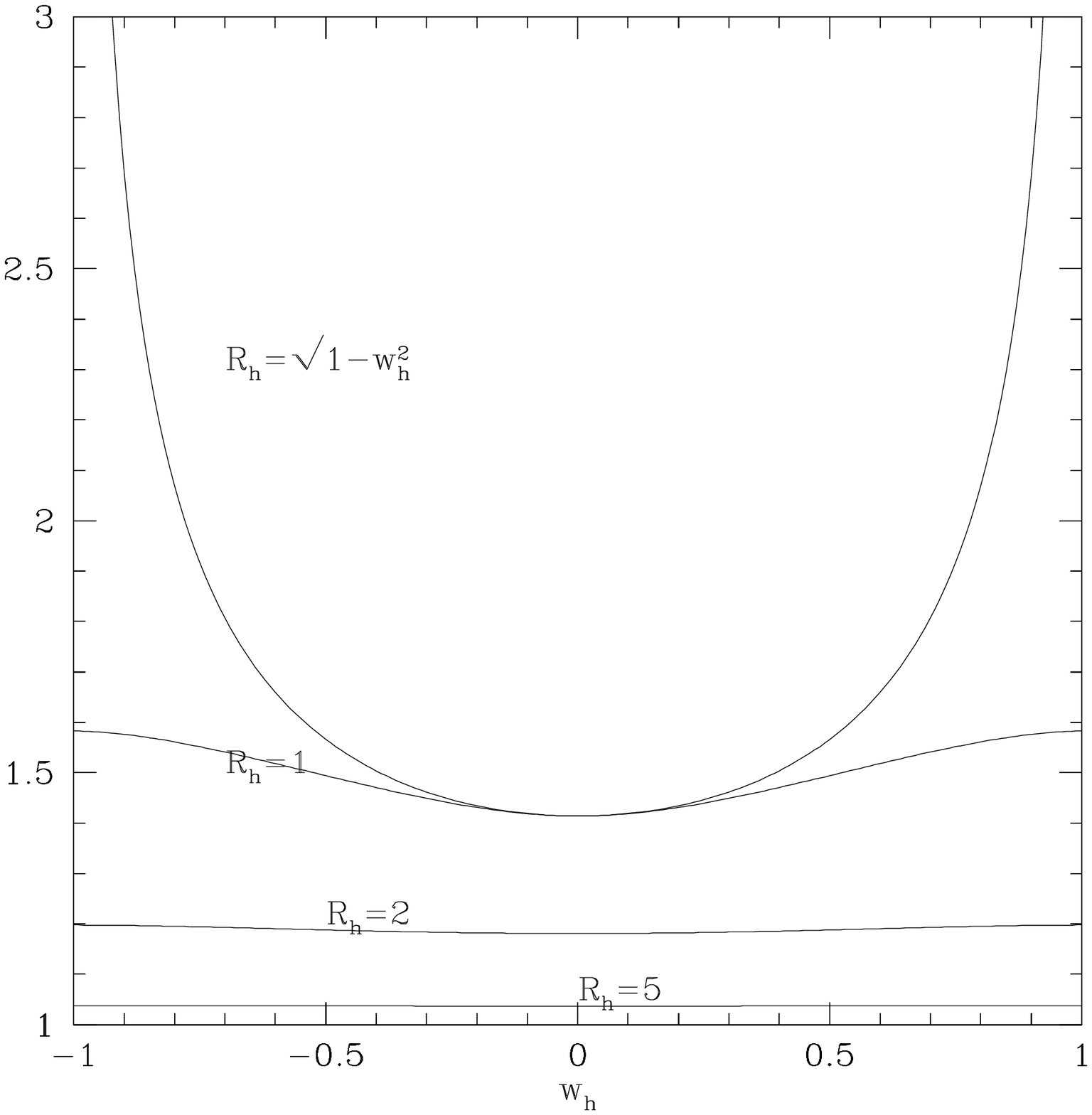,width=1.2\linewidth}
  \caption{{\fontsize{12}{12}\selectfont
$4\pi T(w_h,\R_h)$ fixed $\R_h>1$
and for $R^2+w_h^2=1$.
}}
  \label{fig8}
   \end{minipage}\hspace{5 mm}
\begin{minipage}[b]{0.45\linewidth}
 \centering\epsfig{figure=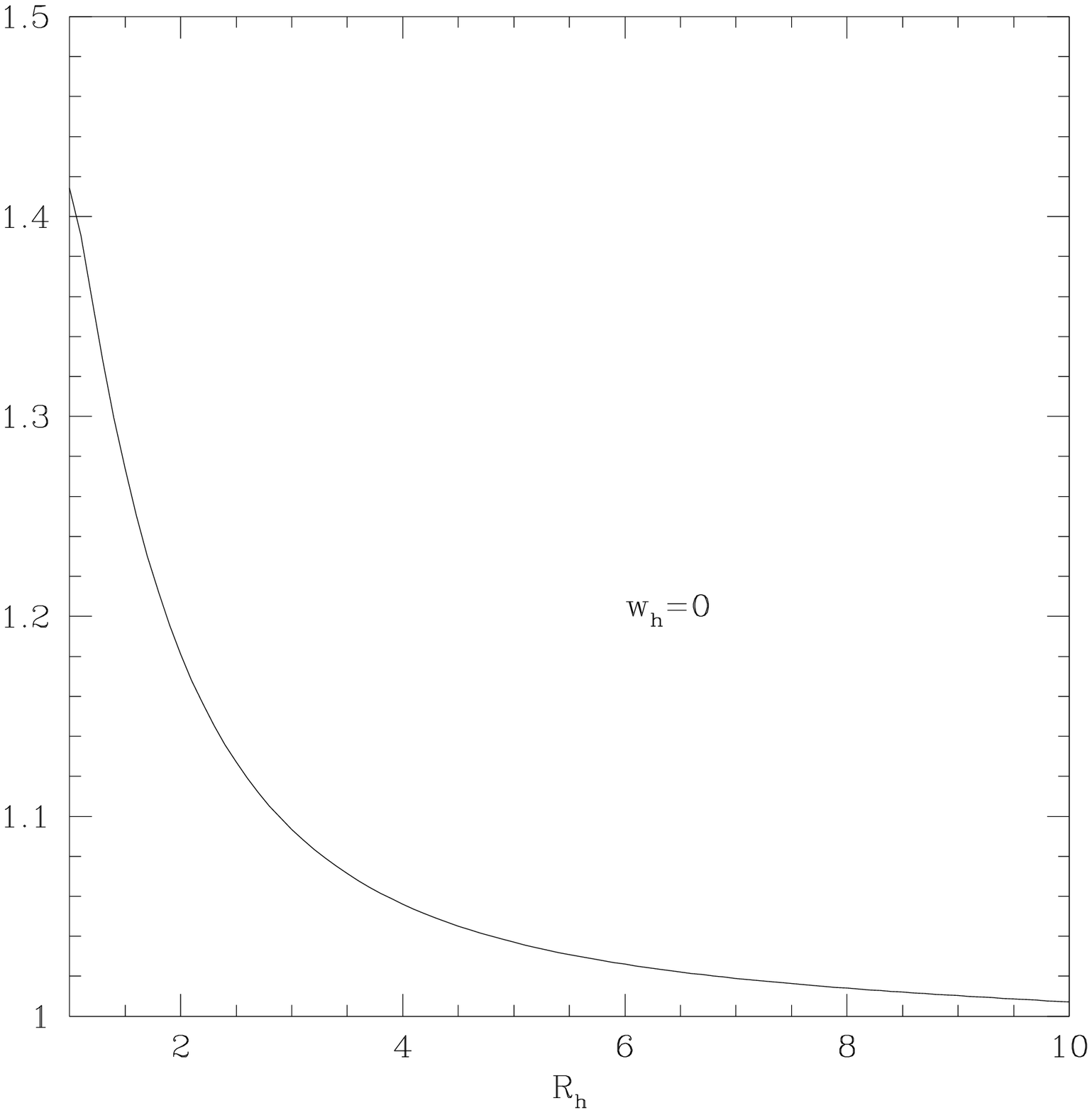,width=1.2\linewidth}
  \caption{{\fontsize{12}{12}\selectfont
$4\pi T(w_h,\R_h)$ for abelian ($w=0$) solutions.
}}
  \label{fig9}
   \end{minipage}
\end{figure}

Let us compute the Hawking temperature.
Switching to the NS-NS description and passing to the
string frame, the  10-d metric   becomes (cf.  \rf{EEa})
\be        \label{stringT}
d{s}^2_{10S} =
-\nu dt^2+d{\rm x}^n d{\rm x}^n+
{\nu}^{-1} {dr^2}+{\rm e}^{2g}(d\theta ^{2}
+\sin^{2}\theta \,d\phi ^{2})
+\tilde{\epsilon}_c\tilde{\epsilon}_c\, .
\ee
Let us examine the $(t,r)$ part of the  metric
analytically continued to the Euclidean region:
\be               \label{TT}
ds_2^2=\nu(r)  d\tau^2+{\nu}^{-1}(r)  {dr^2} \  .
\ee
Near
$r=0$ we have $\nu\sim \nu'r$, where $\nu'$ can be
read off from (\ref{ini}): $\nu'=2\alpha{\rm e}^{-2\Phi_h}/\R_h^2$.
As a result,
 $ds^2=\nu'rd\tau^2+\frac{dr^2}{\nu'r}$.
Introducing $\rho=\sqrt{4r/\nu'}$ and $\vartheta={1 \over 2} \nu'\tau$, the
metric becomes $ds^2=\rho^2d\vartheta ^2+d\rho^2$.
Since $\vartheta $ should be periodic with the period $2\pi$,
 $\tau$
should be  periodic with the period $\beta=4\pi/\nu'$, which determines
the inverse temperature.
In the normalization (\ref{nu}) the metric (\ref{TT})
is asymptotically flat, and evaluating the temperature
at infinity then gives
$T^{-1} =\beta$. If one uses some other normalization of solutions,
then the temperature at infinity
will include the additional correction factor $1/\sqrt{\nu(\infty)}$,
which finally gives
\be                                     \label{temp}
T=\frac{\alpha}{2\pi}\frac{{\rm e}^{-2\Phi_h}}{\sqrt
{\nu(\infty)}\R_h^2}\, \ .
\ee
It is worth noting that this
expression is invariant with respect to the scale
transformations (\ref{scaling}), and so it does not, in fact,
depend on value of $\nu(\infty)$. In addition, the temperature is
invariant also under (\ref{scaling1}), and this implies that
it does not depend on $\Phi_h$ as well. As a result,
the temperature depends only
on  the two essential parameters:  $T=T(w_h,\R_h)$.
Here $w_h$ and $\R_h$ must belong to the physical
region, $-1\leq w_h\leq 1$, $\sqrt{1-w_h^2}\leq \R_h$; this is the
unshaded region in Fig.\ref{figA}.
 For $w_h=0$, $\R_h=1$ we have the exact
solution (\ref{dilaton}),
for which  $T(0,1)=\sqrt{2}/4\pi$.
The numerical evaluation reveals that for a fixed $\R_h\geq 1$ the function
$T(w_h,\R_h)$ reaches its minimum for $w_h=0$ and maximum for $w_h=1$.
For $\R_h\to\infty$ the temperature tends to a constant value,
while for $\R_h\to 0$ the temperature diverges;
see Fig.\ref{fig8} and Fig.\ref{fig9}.

The limit $\R_h\to 0$ corresponds to the lower corners of
the unshaded region in Fig.\ref{figA}, and so it requires that $w_h\to\pm 1$.
Solutions obtained in this limit can be viewed as the globally
regular extremal configurations of  section 4, but containing
in addition a small
black hole in the center. In the limit $\R_h\to 0$ the size of this
black hole shrinks to zero, and outside the event horizon the
configuration tends to the globally regular solution.

Such a phenomenon is actually well known in the theory of hairy black holes
\cite{Volkov:1998cc}: gravitating solitons are often capable of
containing a small black hole inside.
The regular solutions in our case
belong to a family labeled by $b\in[0,1/2]$ (with BPS solution
corresponding to $b= { 1\ov 6}$), and which member of this family
emerges in the limit $\R_h\to 0$ depends on how the limit is taken.
For example, if we take the limit along the left or right boundary of
the unshaded region in Fig.\ref{figA}, that is keeping $w_h=\pm 1$, then the
result will be the regular solution with $b=0$, i.e.\ with zero
gauge field. If we take the limit along the circle $\R_h^2+w_h^2=1$,
then the result will the limiting solution with $b=1/2$. All other
possibilities lead to regular solutions with $0<b<1/2$.

It is important to emphasize that the black hole configurations tend to the
regular ones for $\R_h\to 0$ pointwise but not uniformly, and the limit
is actually singular -- since it is eventually accompanied by the topology
change. As a result, the temperature diverges in the limit. This is very
similar to the situation with the ordinary Schwarzschild black hole
with vanishing mass, $M\to 0$, in which case the metric tends pointwise to the flat
metric, but the temperature $T\sim 1/M\to\infty$.

 Summarizing, for all
solutions in the lowest corners of the unshaded region in Fig.\ref{figA}
the  temperature
diverges. In particular, one can show that if the parameters belong to
the  circle
$\R_h^2+w_h^2=1$, then
\be                          \label{temp1}
4\pi \lim_{w_h\to\pm 1}\sqrt{1-w_h^2} \, \
T(w_h,\sqrt{1-w_h^2})=1\, .
\ee

Let us consider now the opposite limit of {\it large}  black holes,
 having
$\R_h\to\infty$. For asymptotically flat black holes the
temperature would vanish in this limit. This  does not happen
in our case  since large black holes are sensitive to the
asymptotic structure of spacetime, while metrics under consideration
are not asymptotically flat.
In turns out that  $T(w_h,\R_h)$ decreases
for large $\R_h$, but does not vanish and tends to a finite limit
independent of $w_h$:
\be                   \label{temp2}
\lim_{\R_h\to\infty}T(w_h,\R_h)=\frac{1}{4\pi}\,.
\ee
This is  a numerical result, but one can show directly
that the limit exists. For large $\R_h$ the function
$\R\geq \R_h$ is also large,
and we can expand equations (\ref{1a})--(\ref{1f}),
keeping only the leading terms in $\R$. The gauge field
then decouples, while the resulting equations become
\bea
&&\R''-\frac{\R'^2}{\R} +\frac{\R}{\nu}
-\frac{\nu'}{\nu}(\R'+2\R Z) -4\R Z^2-6Z\R'=
0 ,\label{t1a} \\
&&Z'+4Z^2+\frac{\R'^2}{\R^2}
-\frac{1}{\nu}+\frac{\nu'}{\R\nu}(\R'+2Z\R)
+6\frac{Z\R'}{\R}=0 ,\label{t1b} \\
&&2\R^2Z^2+4\R Z\R'+\R'^2
+R\frac{\nu'}{\nu}(\R'+\R Z) -\frac{\R^2}{2\nu}=0, \label{t1c} \\
&&\nu'=2 \a {\rm e}^{-2\Phi}\ R^{-2} ,
\ \ \ \ \ \ \ \    \ \     \Phi'=Z \ .       \label{t1f}
\eea
The space of solutions of this system admits the following
symmetry transformation:
\be               \label{sc}
\R\to {k }\R,\ \ \ \ \ \
\Phi\to\Phi-\ln k,\ \  \ \ \ \
Z\to Z,\ \ \ \ \ 
\nu\to\nu\,  ,
\ee
where  $k$ is a constant   scaling parameter.
The limit $\R_h\to\infty$ can then  be understood as $k\to\infty$. 
Since the temperature (\ref{temp}) is invariant
under such rescalings, its limit for large $R_h$ exists. 
In order to explain the value $T={1\ov 4\pi}$, one has to solve
Eqs.(\ref{t1a})--(\ref{t1f}).

Summarizing: there is a minimal
non-zero value of the temperature, $T_c=  {1\ov 4\pi}$,
which is achieved for
large black holes and is the same for all
solutions. For a finite radius of the horizon
$\R_h<\infty$ one has $T>T_c$, and
there exist both Abelian and non-Abelian black holes, but
the {\it minimal} value of $T$ for a fixed $\R_h>1$ is achieved for the
{\it Abelian} solution, with $w=0$.
 The temperature of this Abelian
solution increases from $T_c$ for large $\R_h$ to
$\sqrt{2}T_c$ for $\R_h=1$. For $\R_h<1$ this Abelian
solution no longer exists and $T>\sqrt{2}T_c$.
In the limit $T\to\infty$ solutions may again become Abelian,
if the limit is taken along the boundaries of
the  unshaded region with $w_h=\pm 1$.
In this case the chiral symmetry will be restored. However, in  most
cases the limit $T\to\infty$ will lead to globally regular
non-Abelian solutions, which break the chiral symmetry.


\section{Free Energy}

Having obtained the extreme and non-extreme
non-BPS generalizations of the BPS
solutions described above, our goal is to consider their
contribution to the thermodynamics. For this we need to
compute the free energy. Passing to the Euclidean region,
such that the 4-d metric \rf{00a2} is (cf. \rf{TT})
\be                  \label{f1}
ds^2_4={\rm e}^{2\Phi}(\nu d\tau^2+\nu^{-1}{dr^2}+\R^2d\Omega^2)\ ,
\ee
with the periodic time $\tau\in[0,\beta]$, the free energy $F$ is defined
by $I=\beta F  . $
 Here  the Euclidean 4-d  action $I$  (cf.  \rf{FS})
  consists of the
volume and surface terms,
\bea         \label{f2}
I[\varphi,\Sigma]&=&\frac{1}{4\pi}\int_\Omega d^4x  \,  \sqrt{\g}\,
\left(-\frac{1}{4}\,R
 +
\frac12\,\partial_\mu\Phi \,\partial^\mu\Phi
+
\frac18\,{\rm e}^{2\Phi}
\F^{a}_{\ \,\mu\nu}\F^{a \mu\nu}
-\frac14\, {\rm e}^{-2\Phi}\right) \nonumber \\
&-&\frac{1}{8\pi}\oint_{\Sigma}Kd\Sigma\equiv I_{vol}+I_{surf}
\, ,
\eea
where $\varphi$ collectively denotes all physical fields, and
the volume integral is taken over a four-volume
$\Omega$ enclosed by a 3-boundary $\Sigma$.
The surface term is determined by the extrinsic curvature of the
boundary, $K$. If $N^\mu$ is the outward normal to
the boundary $\Sigma$, then
\be                     \label{f3}
K=\nabla_\mu N^\mu=\frac{1}{\sqrt{\g}}\partial_\mu(\sqrt{\g}N^\mu).
\ee
We assume that the boundary
$\Sigma$ is defined by the condition that $r$ is constant, whose value
is large and is taken to infinity at the end of
calculations.
The unit normal to the boundary is
$N^\mu=\sqrt{\nu}{\rm e}^{-\Phi}\delta^\mu_r$,
the 3-metric induced on the boundary is
$dl^2={\rm e}^{2\Phi}(\nu d\tau^2+\R^2d\Omega^2)$,
and
$d\Sigma=\sqrt{\nu}\R^2{\rm e}^{3\Phi}d\tau\, d\Omega_2$.

Let us consider first the volume term in the action, $I_{vol}$.
As in any theory with local diffeomorphism invariance,
the on-shell value of this term reduces to a volume integral 
of a total derivative, and so
can be expressed in terms of surface integrals.
Explicitly, using the equations of motion one obtains
\bea                 \label{f4}
I_{vol}[\varphi,\Sigma]&=&
\frac{1}{8\pi}\int_\Omega d^4x  \,  \sqrt{\g}\, \nabla_\mu\nabla^\mu \Phi
=
\frac{1}{8\pi}\int_\Omega d^4x
\partial_\mu(\sqrt{\g}\g^{\mu\nu}\partial_\nu\Phi) \nonumber \\
&=&
\frac12\,\beta \int dr (\nu \R^2{\rm e}^{2\Phi}\Phi')'=
\lim_{r\to\infty}
\frac12\,\beta (\nu \R^2{\rm e}^{2\Phi}\Phi') \, .
\eea
Here the lower integration limit  makes no contribution,
since by assumption it corresponds either to the origin of the
coordinate system for the regular solutions, in which case $\R=0$,
or to the event horizon, $\nu=0$, for the black holes.

Consider now the surface term in the action.
One has for the extrinsic curvature
\be                          \label{f6}
K=\frac{1}{R^2}\,{\rm e}^{-4\Phi}(\sqrt{\nu}\R^2{\rm e}^{3\Phi})'\, ,
\ee
which gives
\be                         \label{f7}
I_{surf}[\varphi,\Sigma]=-\frac12\,\beta\lim_{r\to\infty}\sqrt{\nu}
{\rm e}^{-\Phi}(\sqrt{\nu}\R^2{\rm e}^{3\Phi})'\, .
\ee
Adding the volume and surface terms together and using the field
equation $\R^2{\rm e}^{2\Phi}\nu'=2\alpha$, we finally obtain
\be           \label{f8}
I[\varphi,\Sigma]=-\frac12\,\beta\lim_{r\to\infty}\nu
(\R^2{\rm e}^{2\Phi})'-\frac12\,\beta\alpha\, .
\ee
This gives the on-shell value of the  action in terms of the
asymptotic values of the fields at infinity, the latter being
described by (\ref{inf}).

 Since for all solutions the dilaton
is linearly divergent at infinity, the action turns out to be
infinite. Therefore, we  need to regularize it. For this we
 subtract the value of the action
for a  reference background \cite{Hawking:1996fd},
choosing the latter to be the regular BPS solution (\ref{BPS}).
This is the natural choice, since
all solutions under consideration can be viewed as
excitations over the BPS vacuum.
For the BPS solution the metric is given by (\ref{f1}) with
$\R=\R_{\rm BPS}$, $\Phi=\Phi_{\rm BPS}$, and with $\nu=1$.
The asymptotic value of the  temperature of the 
black hole solution should be matched properly 
with the temperature of the  BPS solution, 
i.e. with the (inverse)  periodicity of its Euclidean 
time. 
To  do this  in a systematic  way, we shall 
assume that for both solutions the coordinate $\tau$ 
has the same period $ \beta$, but in addition for 
the BPS solution the time is rescaled
in such a way that an (a priori
arbitrary) constant factor $\nu_{\rm BPS}$ appears in the BPS metric, 

\be                  \label{f8b}
ds^2_4={\rm e}^{2\Phi_{\rm BPS}}(\nu_{\rm BPS}\, d\tau^2+
{dr^2}+\R_{\rm BPS}^2d\Omega^2)\ . 
\ee
In other words, 
 $\beta_{eff} = \beta \sqrt{\nu_{\rm BPS}}$  is  the effective 
temperature of the BPS solution. 

We now repeat the same calculation of $I$ 
as above, but since, in contrast to \rf{f1}, 
 $\nu_{\rm BPS}$ does not
enter the $g_{rr}$  component  of the BPS  metric  (\ref{f8b}), the result
looks slightly different.
The volume part of the action  is found to be 
\be                 \label{f8c}
I_{vol}[\varphi_{\rm BPS},\Sigma]
= \frac12\,\beta
\sqrt{\nu_{\rm BPS}} \lim_{r\to\infty}
(\R^2{\rm e}^{2\Phi}\Phi')_{\rm BPS} \, .
\ee
Since the unit normal to the boundary at $r$=const is now
$N^\mu={\rm e}^{-\Phi_{\rm BPS}}\delta^\mu_r$,  which does
not contain
$\sqrt{\nu_{\rm BPS}}$,
the surface term of the action is 
\be                         \label{f8d}
I_{surf}[\varphi_{\rm BPS},\Sigma]=-\frac12\,\beta \sqrt{\nu_{\rm BPS}} \lim_{r\to\infty} \ 
{\rm e}^{-\Phi_{\rm BPS}}(\R^2{\rm e}^{3\Phi})_{\rm BPS}'\, .
\ee
Adding the two terms together and subtracting the result from
the black hole action 
$I[\varphi,\Sigma]$ in \rf{f8}, we obtain the regularized 
value of the action:
\be                         \label{f1000}
  I \equiv I[\varphi,\Sigma]-I[\varphi_{\rm BPS},\Sigma]
=-\frac12\,\beta  \lim_{r\to\infty}
\left\{\nu(\R^2{\rm e}^{2\Phi})'-
\sqrt{\nu_{\rm BPS}}(\R^2{\rm e}^{2\Phi})'_{\rm BPS}\right\}
-\frac12\,\beta  \alpha\, .
\ee
 The free energy is then defined\foot{Alternatively, one could define
first the value of the free energy at a given large $r$ 
by dividing $I(r)$  by the local  inverse  temperature $\beta 
\sqrt{\nu(r)}$ and then take $r\to \infty$. Since the factor $\sqrt{\nu(r)}$
approaches 1 quite fast, this leads to the same 
limiting  expression for the $F$.}
in a $r\to \infty$ limit:
\be                         \label{f10}
 F\equiv \beta^{-1}  I 
=-\frac12\, \lim_{r\to\infty}
\left\{\nu(\R^2{\rm e}^{2\Phi})'-
\sqrt{\nu_{\rm BPS}}(\R^2{\rm e}^{2\Phi})'_{\rm BPS}\right\}
-\frac12\, \alpha\, .
\ee
Before the limit is taken, the matching conditions at the
boundary $\Sigma$ are to be imposed \cite{Hawking:1996fd}.
These conditions require that the 3-geometries induced on $\Sigma$
are the same for both backgrounds.  Since the boundary
is $\Sigma=S^1\times S^2$ with the induced
3-geometries
$dl^2={\rm e}^{2\Phi}(\nu d\tau^2+\R^2d\Omega^2)$ and
$dl^2={\rm e}^{2\Phi_{\rm BPS}}(\nu d\tau^2+\R_{\rm BPS}^2d\Omega^2)$,
respectively, these geometries will be the same if the following 
conditions
 \be                        \label{f10a}
{\rm e}^{\Phi}\R={\rm e}^{\Phi_{\rm BPS}}\R_{\rm BPS}\, ,\ \ \ \ \
{\rm e}^{2\Phi}\nu={\rm e}^{2\Phi_{\rm BPS}}\nu_{\rm BPS}\,
\ee
are satisfied on $\Sigma$.
In addition, 
the values of the matter fields for the two backgrounds 
should also be matched at the boundary
\cite{Hawking:1996fd}.

\subsection{Energy and entropy}

According to the analysis of
\cite{Hawking:1996fd}, for stationary spacetimes admitting
foliations by spacelike hypersurfaces $\Sigma_t$ (which is the
case for our solutions),
the regularized free energy obtained from the action as described
above can be related to the energy via the usual 
thermodynamic  equation 
\be                \label{f11}
F=E-ST\, .
\ee
Here $T=1/\beta$, $S$ is the entropy, and
$E$ is the conserved ADM energy
\be                   \label{f12}
E=-\frac{1}{8\pi}\int_{S_t^\infty}
\sqrt{|g_{00}|}\,(^2K-\, ^2K_0)\,dS_t^\infty,
\ee
where the integration is over the 2-boundary $S_t^\infty$ of the
3-surface $\Sigma_t$.
Here $^2K$ and $^2K_0$ are the extrinsic
curvatures of $S_t^\infty$ in  the geometry under
consideration and  in  the reference background geometry, respectively.
It is assumed that both geometries induce the same 2-metric on
$S_t^\infty$, and that the time coordinate is rescaled
in such a way that the $g_{00}$ metric components at $S_t^\infty$
are also the same for both 4-geometries. In addition, it is required
that the matter fields at the boundary agree or ``agree up to a sufficiently
high order''  \cite{Hawking:1996fd}.

This definition of the ADM energy is quite general, it does not
require the reference background to be asymptotically flat
\footnote{For static 4-metrics written in
Schwarzschild coordinates,
$ds^2= -A^2(r)dt^2+\frac{dr^2}{B^2(r)}+r^2d\Omega^2$,
Eq.(\ref{f12}) reduces to
$E=-\lim_{r\to\infty}rA(\sqrt{B}-\sqrt{B_0})$,
where $B_0$ refers to the reference background. For example,
for Schwarzschild-de Sitter solution with
$A=B=1-2M/r+\Lambda r^2$ and
$B_0=1+\Lambda r^2$ this gives $E=M$.},
and it agrees \cite{Hawking:1996fd}
with the definition based on the
asymptotic symmetries \cite{Abbott:1982ff}.
In particular, (\ref{f12}) can be applied to our solutions,
which are not asymptotically flat.
Let us therefore compute the energy for our solutions.
We have the three-geometry on
a hypersurface $\Sigma_t$ of constant time
$dl^2_t={\rm e}^{2\Phi}(\nu^{-1} {dr^2}+\R^2d\Omega^2)$,
while for the BPS solution this changes to
$dl^2_t={\rm e}^{2\Phi_{\rm BPS}}({dr^2}+\R_{\rm BPS}^2d\Omega^2)$.
The boundary $S_t^\infty$ of $\Sigma_t$ is a 2-sphere of constant $r$
in the limit where $r$ tends to infinity.  The 2-geometries induced on
$S_t^\infty$
are ${\rm e}^{2\Phi}\R^2d\Omega^2$ and
${\rm e}^{2\Phi_{\rm BPS}}\R_{\rm BPS}^2d\Omega^2$, respectively.
They agree if
\be                        \label{f12a}
{\rm e}^{\Phi}\R={\rm e}^{\Phi_{\rm BPS}}\R_{\rm BPS}
\ee
at $S_t^\infty$. This condition fixes the geometrical Schwarzschild
radius of the boundary.
The $g_{00}$ metric components for the two backgrounds
agree if
\be               \label{f12b}
{\rm e}^{2\Phi}\nu={\rm e}^{2\Phi_{\rm BPS}}\nu_{\rm BPS}\, .
\ee
In addition, the matter field functions $\Phi$ and $w$ should
also agree at $S_t^\infty$,
or at least the mismatch should tend to zero fast enough as $S_t^\infty$
expands to infinity.
 Notice that these matching conditions
are equivalent to those in (\ref{f10a}) required  in  the
calculation of the action.

The unit normal to $S_t^\infty$ is
$n^k=\sqrt{\nu}{\rm e}^{-\Phi}\delta^k_r$,
such that $^2K=\nabla_k n^k=
\frac{\sqrt{\nu}}{\R^2}{\rm e}^{-3\Phi}(\R^2{\rm e}^{2\Phi})'$,
while for the BPS we have
$^2K_0={\R_{\rm BPS}^2}{\rm e}^{-3\Phi_{\rm BPS}}
(\R_{\rm BPS}^2{\rm e}^{2\Phi_{\rm BPS}})'$.
Inserting this into (\ref{f12}) and taking (\ref{f12a}) and (\ref{f12b}) into
account, gives
\be                         \label{f13}
E=-\frac12\,\lim_{r\to\infty}
\left\{\nu(\R^2{\rm e}^{2\Phi})'
-\sqrt{\nu_{\rm BPS}}(\R^2{\rm e}^{2\Phi})'_{\rm BPS}\right\}\, .
\ee
This is in exact correspondence  with the first term in (\ref{f10}), and so
our calculations of the energy and free energy agree with each other
and with the general thermodynamic relation (\ref{f11}), giving
the following expression for the entropy of the solutions:
\be                              \label{f14}
S=\frac12\,\beta\alpha=\pi\R^2_h{\rm e}^{2\Phi_h}\, .
\ee
Here we have used Eq.(\ref{temp})
for the Hawking temperature
$T=1/\beta$ (assuming that  $\nu(\infty)=1$).
Since $\R_h{\rm e}^{\Phi_h}$ is the
invariant geometrical radius of the event horizon, the entropy is
equal to a quarter of the geometrical area of the event horizon.
Notice that the energy and the action do not change under
translations of $r$ (\ref{scaling2}), while under
(\ref{scaling1}), $\Phi\to\Phi+C$, $\alpha\to{\rm e}^{2C}\alpha$, both
$E$ and $I$ acquire the overall factor ${\rm e}^{2C}$.

Let us now use the above expressions  in order to evaluate
the energy and free energy.
Let us choose a non-BPS solution and shift its radial coordinate
to set $r_\infty=0$ in (\ref{inf}). The BPS solutions
actually comprise the two-parameter family. One parameter in (\ref{BPS})
is $\Phi_0$, which represents the constant part  of the dilaton.
Another parameter  accounts for the  freedom   to shift  the origin of the 
radial coordinate, $r\to r+r_0$ (see \rf{BPSa}).
 These two parameters can be fine-tuned
in order to fulfill the matching conditions.  Indeed, let us fix a large
but finite value of $r$, which specifies the position of the boundary.
Then the condition (\ref{f12b})  can be
fulfilled by the suitable choice of $\nu_{\rm BPS}$ --  so far
this parameter has not been specified. Next, one can choose $r_0$
and $\Phi_0$ such that (\ref{f12a}) is also fulfilled, and in addition
\be                       \label{f14a}
\Phi=\Phi_{\rm BPS}
\ee
at the boundary. As a result, we can exactly match the boundary
geometries and the boundary value of the dilaton
for the two solutions. The gauge field functions
$w$ and $w_{\rm BPS}$ will not, however, exactly match at the boundary,
 unless
the boundary is strictly at infinity (where 
$w$ and $w_{\rm BPS}$ are equal to zero).%
\footnote{It is usually impossible to exactly match the
matter fields at the boundary.
For example, for a Reissner-Nordstrom black hole there is always
a jump of the electric field ${\cal E}$ at the boundary $\Sigma$, since
${\cal E}\sim 1/r^2$ for the solution, while ${\cal E}=0$
for the reference background (flat space). However, the value
of this jump tends to zero as the boundary recedes to infinity
fast enough to ensure that fields at the boundary
``agree up to a sufficiently high order''.
Physically,  this condition means
that excitations over the background are sufficiently localized
for the energy to be finite.
If the boundary values of fields for the
solution and for the reference background do not
agree up to a sufficiently high order,
the excitations are too spread and
their energy will be divergent. }
If the boundary is at finite $r$, there will be some
boundary discrepancy $\Delta w=w-w_{\rm BPS}$,
which will   measure the fall-off rate
with which the non-BPS solution approaches the
BPS background. For the energy to be finite,
$\Delta w$ should tend to zero fast enough as $r\to\infty$.
Otherwise the excitations
over the BPS background will  not be  well-localized and their energy
will be infinite.

As can be seen from Eq.(\ref{inf}), all non-BPS solutions
approach the BPS asymptotic for large $r$. If
the parameter $\Upsilon$ in (\ref{inf}) vanishes, then $\Delta w\sim\exp(-r)$
and the asymptotic values are reached exponentially fast. If
$\Upsilon\neq 0$, then  the exponential fall-off is
replaced by polynomial fall-off. In terms of 
 the Schwarzschild radial coordinate
$r_s=\R{\rm e}^\Phi\sim {\rm e}^{r/2}$, the excitations with
$\Upsilon=0$ behave as $1/r_s$, while those with  $\Upsilon\neq 0$
decay only as inverse powers of $\ln r_s $.
It is instructive to compare this, say, to 
the Schwarzschild-AdS solution,  where the excitations
decay as $1/r_s$ and the energy is finite.
One can then think that all solutions with $\Upsilon\neq 0$
approach their asymptotics too slowly for the energy to be finite.
This is confirmed
by the direct calculation (see below):
matching the boundary geometries
at finite $r$ and inserting the result into (\ref{f13})
gives
$E\sim r^{-5/2}{\rm e}^r$, which is divergent as $r\to\infty$.

The conclusion is that
non-BPS excitations over the BPS background for which
$\Upsilon\neq 0$ are too much delocalized and have infinite energy.

\subsection{Solutions with finite energy}

Let us now study the special case of the solutions for which
$$\Upsilon=0 \ . $$
 As we shall see, the energy then turns out to be finite.
Non-BPS solutions with $\Upsilon=0$
exist,
one example being the Abelian black holes with $w(r)=0$.
In addition,
there are also non-Abelian solutions with $\Upsilon=0$.

\begin{figure}[h]
\begin{minipage}[b]{0.40\linewidth}
  \centering\epsfig{figure=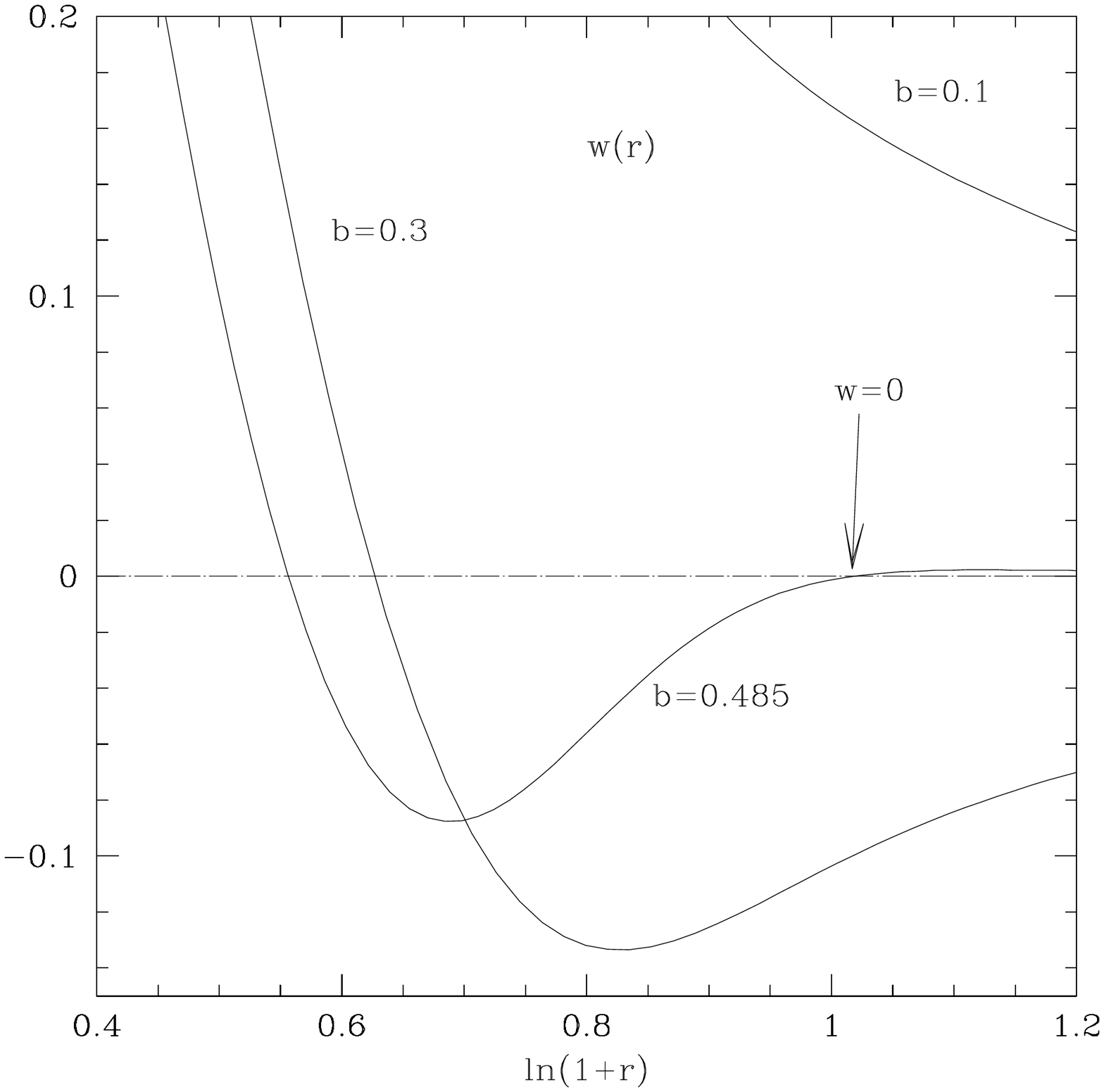,width=1.2\linewidth}
  \caption{{\fontsize{12}{12}\selectfont
$w(r)$ for the globally regular solutions. It has no nodes for
$b=0.1$; one zero for $b=0.3$; two zeroes for $b=0.485$,
and so on.
}}
  \label{figREG}
   \end{minipage}\hspace{15 mm}
\begin{minipage}[b]{0.40\linewidth}
  \centering\epsfig{figure=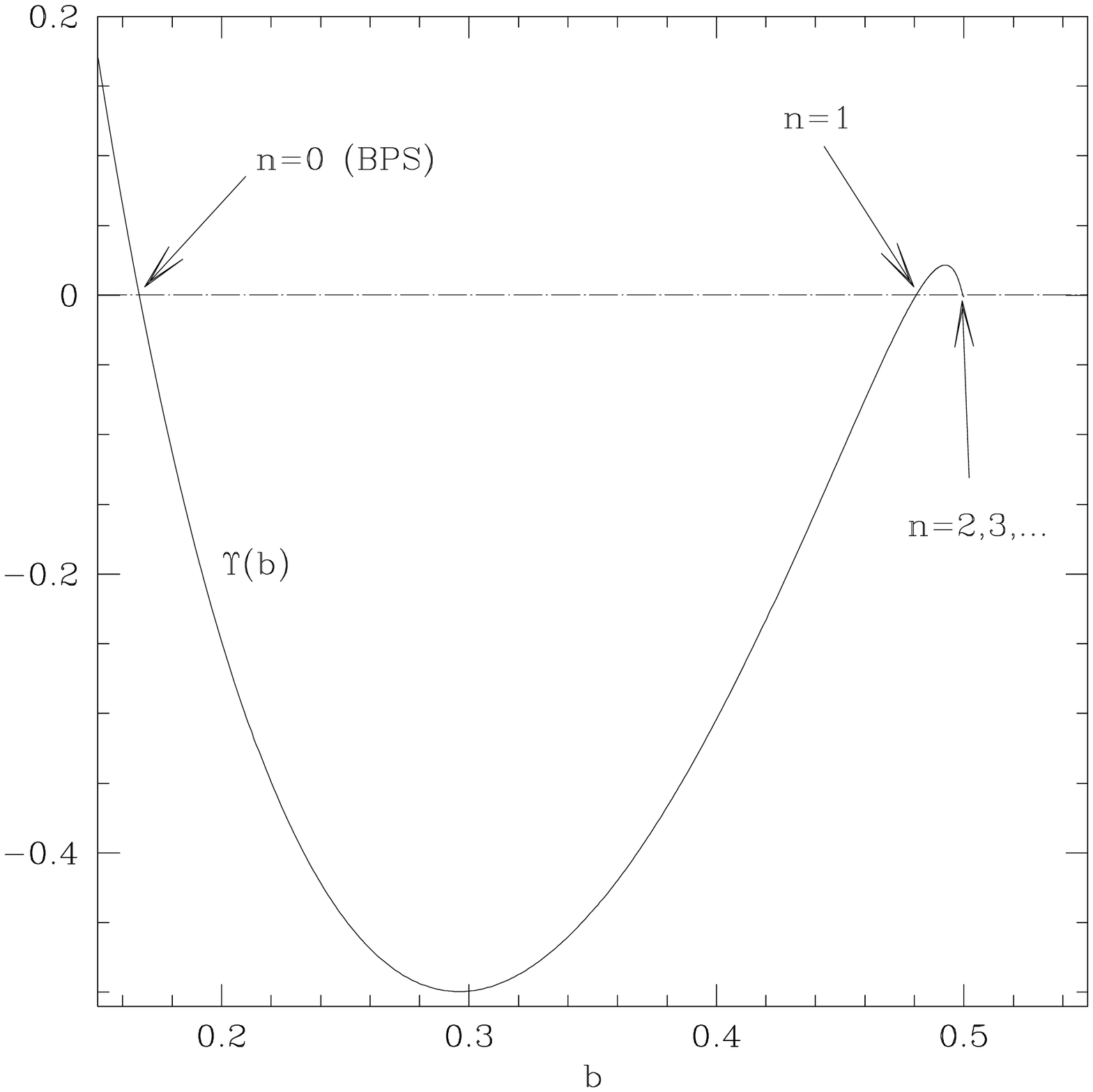,width=1.2\linewidth}
  \caption{{\fontsize{12}{12}\selectfont
$\Upsilon(b)$ for the globally regular solution.
Zeros of this function at $b=b_n$
correspond to finite energy solutions.
}}
  \label{figUps}
   \end{minipage}
\end{figure}

\begin{figure}[h]
\begin{minipage}[b]{0.40\linewidth}
 \centering\epsfig{figure=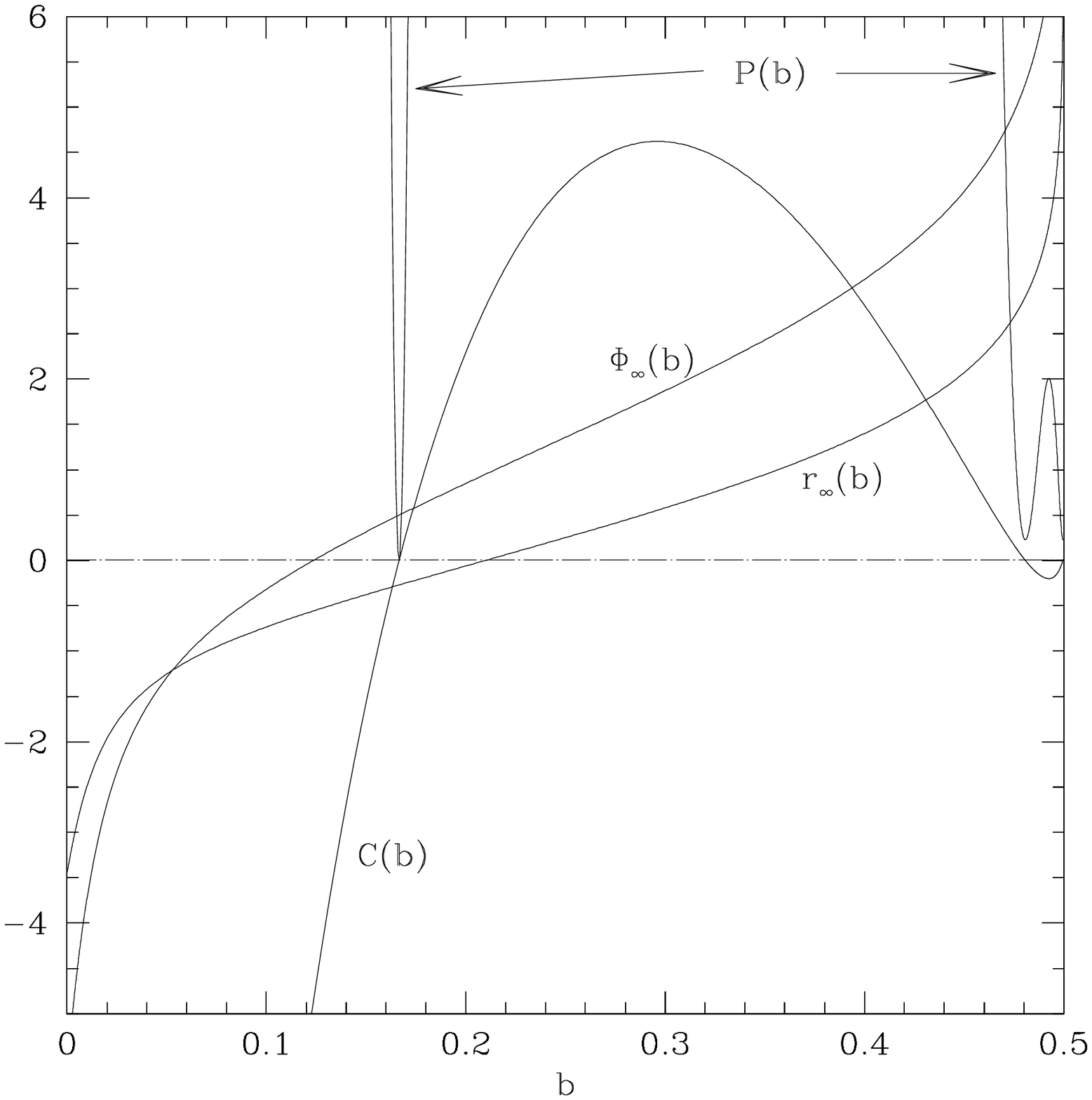,width=1.2\linewidth}
  \caption{{\fontsize{12}{12}\selectfont
 Parameters ${\cal P}(b)$, ${\cal C}(b)$, $r_\infty(b)$,
and $\Phi_\infty(b)$ for the globally regular solutions.
}}
  \label{figfPAR}
   \end{minipage} \hspace{15 mm}
\begin{minipage}[b]{0.40\linewidth}
 \centering\epsfig{figure=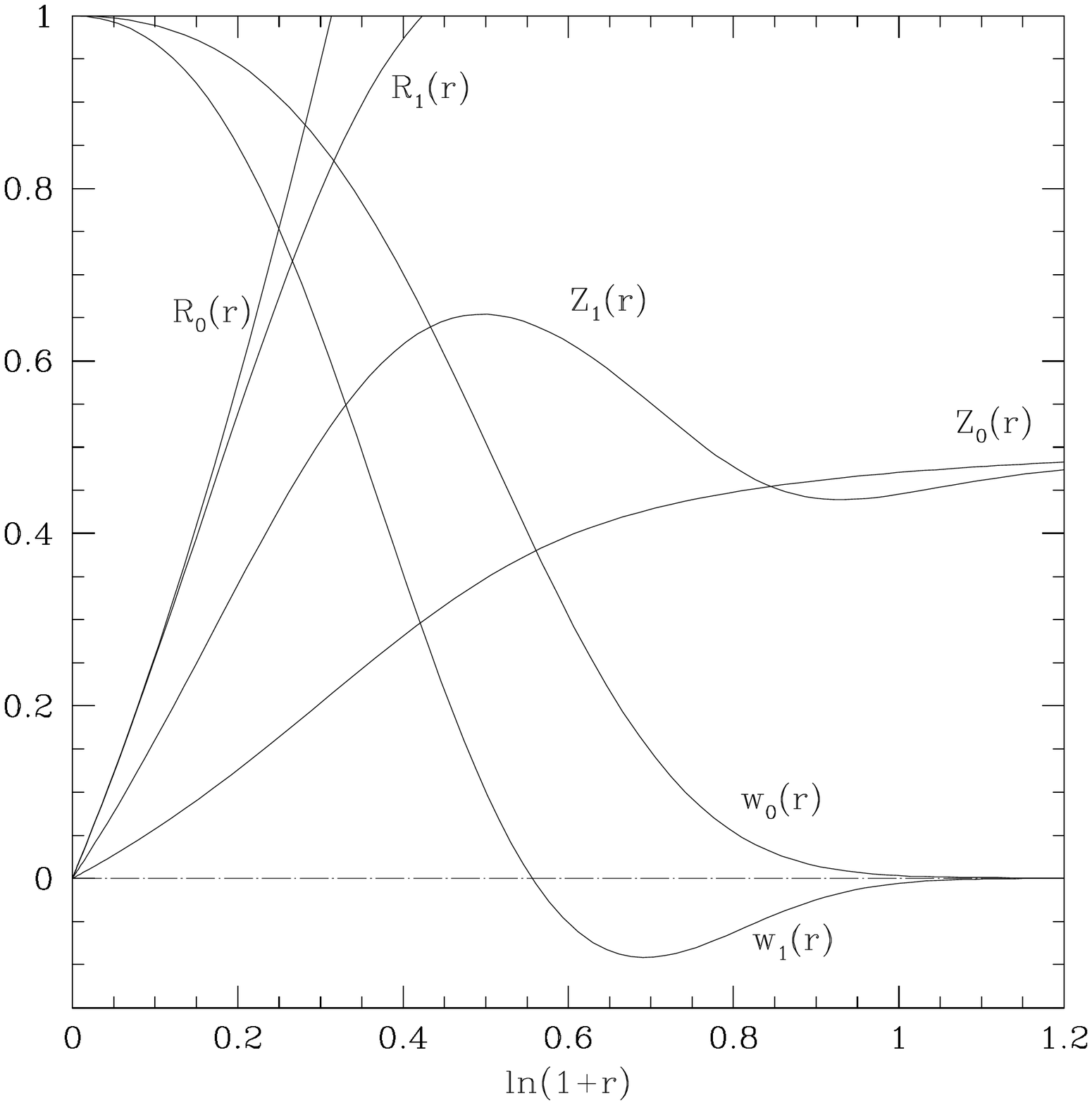,width=1.2\linewidth}
  \caption{{\fontsize{12}{12}\selectfont
The BPS $(b=1/6)$ solution and its first finite energy
excitation ($b=0.4807$).
}}
  \label{figffREG}
   \end{minipage}
\end{figure}

Let us first consider the  globally regular solutions.
These are parameterized by $b\in(0,1/2)$. If $b<1/6$, then
$w$ is everywhere positive, and therefore $\Upsilon>0$. For $1/6<b<0.48$
$w$ has a zero for some finite $r$, and therefore (see \rf{inf})
 $\Upsilon<0$.
As a result, there is a value of $b$ in between,
which is $b=1/6$,  for which $\Upsilon$ vanishes.
If we continue to increase $b$, we find
that for $b>0.48$ the function
 $w$ develops already two nodes
(see Fig.\ref{figREG})  such that
$\Upsilon$ is again positive.
This shows that $\Upsilon$ vanishes again
for $b\approx 0.48$. The number of nodes
of $w$ increases  as $b\to 1/2$, which shows that
there is a discrete sequence of values $b_n$, $n=0,1,\ldots,$
for which $\Upsilon(b_n)=0$. One has $b_0=1/6$, $b_1\approx 0.48$,
$\ldots$, $b_\infty=1/2$. The numerical plot for $\Upsilon(b)$
in Fig.\ref{figUps} shows the first three zeros of this function.
The remaining zeros accumulate near
$b=1/2$, where $\Upsilon(b)$
oscillates with a very small amplitude, which oscillations
are too small to be seen in the figure. The other
asymptotic parameters in (\ref{inf}) for the globally
regular solutions -- ${\cal P}(b)$,
$r_\infty(b)$, $\Phi_\infty(b)$, and (rescaled)
${\cal C}(b)$,  -- are shown in Fig.\ref{figfPAR}.
Notice that ${\cal P}(b)$ vanishes for $b=1/6$ and is positive for
other values of $b$.  $\Phi_\infty(b)\to\infty$ as $b\to 1/2$.

Summarizing, among all globally regular solutions
there is an infinite discrete subset of solutions
for which  $\Upsilon=0$ and the configurations
approach the BPS background exponentially fast. These solutions describe
the ``well-localized'' excitations over the BPS
background, and their energy, free energy, and action
turn out to be {\it finite}. The first such excitation is shown
in Fig.\ref{figffREG}.
Applying the same argument, one finds
also black holes with similar
properties. These finite energy black holes exist for arbitrary
values of $\R_h>0$, but only for some discrete values of $w_h$.
It is clear
that such finite energy configuration will be giving the {\it leading}
contribution to the path integral.

Let us explicitly compute the energy for
solutions with $\Upsilon=0$.
Asymptotics for large $r$ are obtained from (\ref{inf}):
\footnote{As was already mentioned earlier, both the
globally  regular  and the black hole solutions have 
the same large $r$ asymptotics given by  (\ref{inf}).
The constant  parameters there ($\Phi_\infty, {\cal P}, ...$)
are of course different 
in the two cases:  in the globally regular case they depend on 
the  two  constants $b$ and $\Phi(0)$ in \rf{6}, 
while  in the black hole case they depend on the three
constants $\R_h, w_h, \Phi_h$ in \rf{ini}.}
\bea              \label{f16}
\R&=&\sqrt{2r}+\sqrt{2}{\cal P}r{\rm e}^{-r}
(1+\frac{2}{r}+\ldots)\, ,  \ \ \ \ \ \ \
\nu=1-\frac{\alpha}{\sqrt{r}}{\rm e}^{-r-2\Phi_\infty}+\ldots,
\nonumber  \\
\Phi&=&\Phi_\infty+\frac{1}{2}r-\frac14\ln r -{\cal P}\sqrt{r}{\rm e}^{-r}(1+
\frac{1}{r}+\ldots)\, ,\ \ \ \ \ \ \ \
w={\cal C}r{\rm e}^{-r}+\ldots
\eea
where  we used the global symmetries (\ref{scaling}),(\ref{scaling2})
to set $r_\infty=0$ and $\mu=1$. Asymptotics of the regular BPS
solution (\ref{BPS}) can be obtained by putting
here ${\cal P}=\alpha=0$  (and ${\cal C}=2$) 
 and re-introducing  the two  free parameters 
in \rf{BPSa} by 
 arbitrary shifts of  $r$ and $\Phi$ 
($r_\ast= r_0 - { 1 \over 2}$, \  $\Phi_\ast= \Phi_0 +  { 1 \over 4}$)
\bea              \label{f17}
\R_{\rm BPS}&=&\sqrt{2(r+r_\ast)}+\dots,\ \ \ \ \ \
\Phi_{\rm BPS}=\Phi_\ast+\frac{1}{2} ( r+r_\ast)
-\frac14\ln(r+r_\ast)+\ldots,\ \ \
\nonumber \\
w_{\rm BPS}&=&(2r+2r_\ast+1)\,{\rm e}^{-r-r_\ast-\frac12}+\ldots  \ ,
\ \ \ \ \ \ \ \nu_{\rm BPS}={\rm const} \ .
\eea
We want to evaluate the expression for $E$ in (\ref{f13}) at
some large but finite
value of $r$ under the conditions (\ref{f12a}), (\ref{f12b}),
and (\ref{f14a}), which are equivalent to
\be                        \label{f170}
\nu=\nu_{\rm BPS},\ \ \ \ \ \
{\rm e}^{\Phi}\R={\rm e}^{\Phi_{\rm BPS}}\R_{\rm BPS},\ \ \ \ \ \
\R=\R_{\rm BPS}\, ,
\ee
and then take the limit $r\to\infty$.
The first of these conditions allows us to rewrite
the formula (\ref{f13}) for the energy as
\be                         \label{f17b}
E=-\frac12\,\lim_{r\to\infty}\sqrt{\nu}
\left\{\sqrt{\nu}(\R^2{\rm e}^{2\Phi})'
-(\R^2{\rm e}^{2\Phi})'_{\rm BPS}\right\}\, .
\ee
Since
\be                  \label{f17a}
\R^2{\rm e}^{2\Phi}=2\sqrt{r}{\rm e}^{r+2\Phi_\infty}
+4{\cal P}{\rm e}^{2\Phi_\infty}+\dots\, ,\ \ \ \
\R_{\rm BPS}^2{\rm e}^{2\Phi_{\rm BPS}}=
2\sqrt{r+r_\ast}{\rm e}^{r+r_\ast+2\Phi_\ast},
\ee
one has
\be                         \label{f17c}
E=-\lim_{r\to\infty}\sqrt{\nu}
\left\{
(1-\frac{\alpha}{2\sqrt{r}}{\rm e}^{-r-2\Phi_\infty}+\ldots)
(\sqrt{r}{\rm e}^{r+2\Phi_\infty} )'
-(\sqrt{r+r_\ast}{\rm e}^{r+r_\ast+2\Phi_\ast})'\right\}\, ,
\ee
which gives upon differentiation
\bea                  \label{f18}
E&=&\lim_{r\to\infty}\sqrt{\nu}
\left((\sqrt{r+r_\ast}{\rm e}^{r+r_\ast+2\Phi_\ast}
-\sqrt{r}{\rm e}^{r+2\Phi_\infty})
+\frac12(\frac{1}{\sqrt{r+r_\ast}}\,{\rm e}^{r+r_\ast+2\Phi_\ast}
-\frac{1}{\sqrt{r}}\,{\rm e}^{r+2\Phi_\infty})\right) \nonumber \\
&+&\lim_{r\to\infty}\sqrt{\nu}\,
\frac{\alpha}{2\sqrt{r}}{\rm e}^{-r-2\Phi_\infty}(\sqrt{r}{\rm e}^{r}
+\frac{1}{2\sqrt{r}}\,{\rm e}^{r})\,{\rm e}^{2\Phi_\infty}.
\eea
The second condition in (\ref{f170})  in view of (\ref{f17a}) reduces to
\be                    \label{f20}
\sqrt{r+r_\ast}{\rm e}^{r+r_\ast+2\Phi_\ast}=
\sqrt{r}{\rm e}^{r+2\Phi_\infty}+2{\cal P}{\rm e}^{2\Phi_\infty}\, .
\ee
Using it,  one can rewrite (\ref{f18}) as
\be                  \label{f21}
E=2{\cal P}\,{\rm e}^{2\Phi_\infty}+\frac12\,\alpha+
\frac12\lim_{r\to\infty}\sqrt{\nu}\,
(\frac{\sqrt{r}}{{r+r_\ast}}-\frac{1}{\sqrt{r}})\,{\rm e}^{r+2\Phi_\infty},
\ee
where we have set to zero those terms which clearly vanish in the limit.
The third  matching condition in (\ref{f170})
gives $r_\ast=2{\cal P}r^{3/2}{\rm e}^{-r}+\ldots$. 
In view of this,  the last term
on the right in (\ref{f21}) reduces in the limit to
$(-{\cal P}\,{\rm e}^{2\Phi_\infty})$, such that
\be                   \label{f24}
E={\cal P}\,{\rm e}^{2\Phi_\infty}+\frac12\,\alpha.
\ee
This is the final result for the conserved ADM energy for non-BPS --
either globally regular or black hole -- solutions with $\Upsilon=0$.
Since the energy is invariant under constant shifts of $r$, the same
expression holds for solutions with an arbitrary $r_\infty$
in the asymptotics.
If the dilaton is shifted by a constant, $\Phi\to\Phi+C$, then
$\alpha\to\alpha{\rm e}^{2C}$ (see (\ref{scaling1})),
while ${\cal P}$ remains intact,
and the energy therefore changes by the overall factor ${\rm e}^{2C}$.

The action for finite energy solutions is expressed in terms of
the energy and entropy as
\be                \label{ff25}
I=\beta E-S\, .
\ee
For the globally regular solution the entropy vanishes and $\alpha=0$, while
$\beta$ can be arbitrary, so that we get
\be                \label{f25}
I_{\rm regular}=\beta {\cal P}{\rm e}^{2\Phi_\infty}\, .
\ee
For the black holes,  the entropy is $S=\beta\alpha/2$,
while
$\beta=2\pi \alpha^{-1} \R_h^2{\rm e}^{2\Phi_h}$ (see (\ref{temp})), so that
\be                \label{f26}
I_{\rm BH}=\frac{2\pi}{\alpha}\,{\cal P} \R_h^2{\rm e}^{2\Phi_h+2\Phi_\infty}.
\ee
Under a  constant shift of the dilaton, $\Phi\to\Phi+C$,
${\cal P}$ and $\R_h$ are invariant, while $\alpha\to\alpha{\rm e}^{2C}$,
so that  the action acquires the overall factor ${\rm e}^{2C}$.

Summarizing the results obtained above, the  non-BPS solutions
described in the previous sections generically have infinite energy.
However, among these solutions there are special solutions with {finite}
 energy.
These form discrete sets;  they have fields
approaching their asymptotic values as $\exp(-r)$, and  thus describe
finite energy excitations over the BPS background.
In terms of  the geometrical Schwarzschild coordinate
$r_s=\R$e$^\Phi$,  the excitations decay is $1/r_s$, which is why
the energy is finite.

Let us now describe these finite energy
solutions in more detail.

\subsection{Globally regular solutions with finite energy}

In the globally regular case, the finite energy solutions comprise
a discrete one-parameter family. These solutions can be conveniently
labeled by the  integer $n=0,1,\ldots$, which is the number of nodes of the
gauge field function  $w(r)$ (solutions with $n=0,1$ are shown in
Fig.\ref{figffREG}).
Such solutions have asymptotics (\ref{6}) at the regular origin
(we set $\Phi(0)=0$).  At infinity the asymptotics are
those given in (\ref{inf}) with $\Upsilon=0$.
Such boundary conditions can be fulfilled
only for the discrete
values of the parameter $b=b_n$ in (\ref{6})
for which the function $\Upsilon(b)$ in Fig.\ref{figUps}
vanishes, $\Upsilon(b_n)=0$. The asymptotic parameters  in (\ref{inf})
then also assume only discrete values corresponding to
${\cal P}(b)$,
$r_\infty(b)$, $\Phi_\infty(b)$,
${\cal C}(b)$ shown in Fig.\ref{figfPAR} with $b=b_n$.

The ground state solution is the BPS one, with $b=1/6$ and $n=0$, since $w$
does not oscillate. Then
comes its first excitation for $b=0.4807$
with $n=1$, for which  $w$ has one zero
at some finite $r$. Then follow higher excitations.
We list  the parameters
of several such excitations in Tab.1. As one can see from this table,
for all excitations the coefficient ${\cal P}$ is approximately
the same,%
\footnote{Numerical values of the parameters of the 
solutions can be determined by the multiple shooting
method. The accurate determination of ${\cal P}$  
is, however, extremely involved, since ${\cal P}$ is the coefficient 
in front of the subleading terms which are exponentially small 
as compared to the other, leading terms. 
We used a simplified numerical procedure 
giving the value of ${\cal P}$ with $\sim 20\%$ uncertainty.  
As a result, the values of ${\cal P}$ and $E$ given in the table
are, in fact, approximate. Since it requires
considerable efforts to improve these numbers, 
we postpone this for a future publication. 
\label{foot}
} 
but $\Phi_\infty$ increases with $n$ as
approximately  $3n$. As a result, the energy grows
rapidly, $E\approx 0.2\times\exp(6n)$. The limit $n=\infty$
is reached for $b=1/2$. As was discussed above, the solutions then
change the topology, which costs infinite energy.

\begin{center}
Tab. 1. Parameters of the globally regular solutions with finite energy.
\vglue 0.4cm
\begin{tabular}{|c|c|c|c|c|c|} \hline
$n$ & $b$            & ${\cal P}$  & $\Phi_\infty$ &
$E={\cal P}\exp(2\Phi_\infty)$ & $r_\infty$ \\ \hline
$0$ & $1/6$         & $0$            & $(1-\ln 8)/4$ & $0$ & $-1/2$\\
$1$ & $0.4807$    & $0.2304$   & $2.902$         & $7.7\times 10^1$ & $5.5.258$\\
$2$ & $0.4996$    & $0.2295$   & $6.083$         & $4.4\times 10^4$ & $9.750$ \\
$3$ & $0.499991$ & $0.2294$   & $9.175$         & $2.1\times 10^7$ & $14.121$ \\
$\ldots$ & & & & & \\
$\infty$ & $0.5$ & $0.2294$   & $\infty$         & $\infty$ & $\infty$ \\
\hline
\end{tabular}
\end{center}

To summarize, the globally regular finite energy solutions
are characterized by the number $n=0,1,\ldots$ of nodes of $w$. The ground state
energy is zero, while for
all excitations the energy is positive and rapidly increases with $n$.
The action $I=\beta E$ also grows rapidly with $n$, where
the inverse temperature $\beta$ can be set to any value.
As a result, for any given $\beta$, the ground state solution gives
the leading contribution to the path integral. The contribution of the
excitations is highly suppressed.%
\footnote{Notice that the
normalization is important. One can
use (\ref{scaling1}) to rescale all solutions
to set $\Phi_\infty=0$, and then the energy will be
${\cal P}$, which is approximately the same for all excitations.}

\subsection{Black holes with finite energy}

\begin{figure}[h]
\begin{minipage}[b]{0.40\linewidth}
  \centering\epsfig{figure=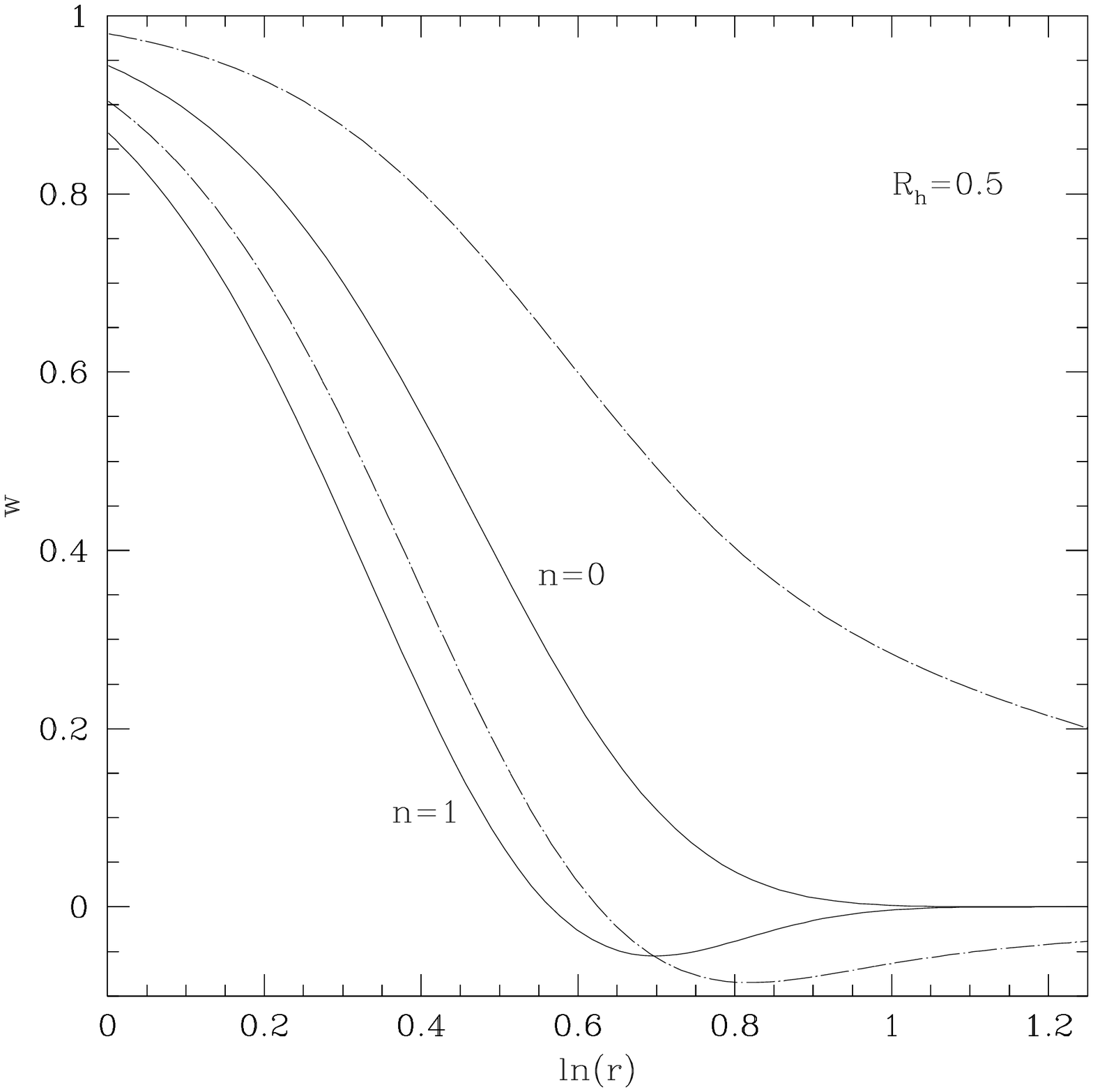,width=1.2\linewidth}
  \caption{{\fontsize{12}{12}\selectfont
Function  $w$ for $n=0,1$ finite energy black holes
with $\R_h=0.5$. For comparison, two other solutions
are shown, for which $w\sim1/\sqrt{r}$ for large $r$.
}}
  \label{figw}
   \end{minipage}\hspace{15 mm}
\begin{minipage}[b]{0.40\linewidth}
 \centering\epsfig{figure=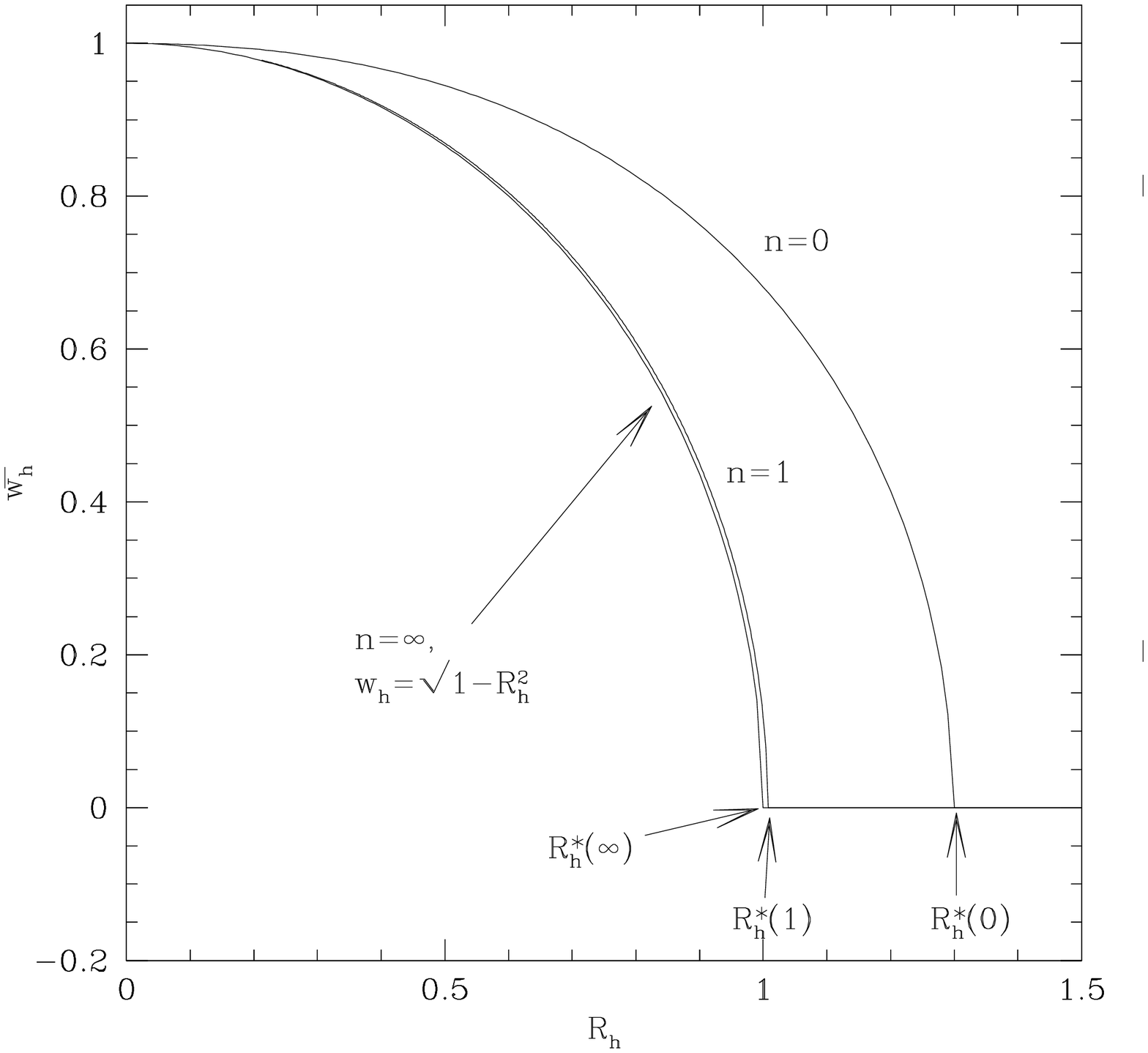,width=1.2\linewidth}
  \caption{{\fontsize{12}{12}\selectfont
 Parameters $w_h(n,\R_h)$ for finite energy black holes.
For other values of $w_h$, $w(r)$ tends to zero too slow
for the energy to be finite; see Fig.\ref{figw}.
}}
  \label{figfw}
   \end{minipage}
\end{figure}

Let us now consider  the  black holes with finite energy.
These  are obtained by selecting from the set of all
black holes considered in  section 5 only those
solutions for which $\Upsilon=0$ (we always assume
that $\nu(\infty)=1$).
For any given value of the event horizon size $\R_h$,
there are special values $w_h(n,\R_h)$
of the gauge field function $w$ at the horizon,
 shown in  Fig.\ref{figfw},
which give rise to solutions with $w\sim\exp(-r)$ for large $r$; see
Fig.\ref{figw}.
For all other values of $w_h$ one has $w\sim1/\sqrt{r}$ for large $r$
(Fig.\ref{figw})
and the energy is infinite.
The finite energy solutions therefore
comprise a discrete series of one-parameter families: particular
solutions are labeled by $(n,\R_h)$, where
$n=0,1,2,\ldots$ is the number of nodes of $w$
outside the black hole horizon, while $\R_h>0$.

For $n=0$ the set of such black holes
consists of two branches. First, there are the abelian black holes,
which exist for $1<\R_h<\infty$.

Second, for small $0<\R_h<1.3$
there are also  non-Abelian solutions.
For these $w$ starts from
some finite value at the horizon, and then exponentially quickly
tends to zero. In the limit $\R_h\to 0$ the field configurations
approach the BPS solution  pointwise (in the exterior black hole
region), and so in some sense they can be viewed as  black
hole generalizations of the BPS solution itself. As $\R_h$
increases, the value of $w_h$ for such solutions decreases,
and finally it vanishes for $\R_h\equiv \R_h^\ast(0)=1.3$,
at which point the abelian
and non-abelian branches merge. For $\R_h>1.3$ only the
abelian solutions exist.

There are also non-Abelian black holes with $n>0$.
For these $w$
starts from some finite value $w_h$ at the horizon, and then
after $n$ oscillations around zero exponentially fast tends to zero.
The function $w(r)$ for two such solutions with $n=0,1$ and $\R_h=0.5$
is shown in Fig.\ref{figw}.
In the limit $\R_h\to 0$ these solutions approach pointwise the
globally regular finite energy solutions described above.
As $\R_h$ increases, the value of $w_h$ decreases, and finally
for some finite $\R_h\equiv \R_h^\ast(n)$ the solutions
merge with the abelian black holes,
similarly to what happens to the $n=0$ non-abelian branch.

Summarizing, all non-abelian solutions exist only for small
values of $\R_h$, and all of them merge with  the abelian solution
for $\R_h=\R^\ast_h(n)$, where $\R^\ast_h(n)$
are $\R^\ast_h(0)=1.3$, $\R^\ast_h(1)=1.01$, $\ldots$,
$\R^\ast_h(\infty)=1$.
For $\R_h>1.3$ only the abelian solution exists.

\begin{figure}[h]
\begin{minipage}[b]{0.40\linewidth}
  \centering\epsfig{figure=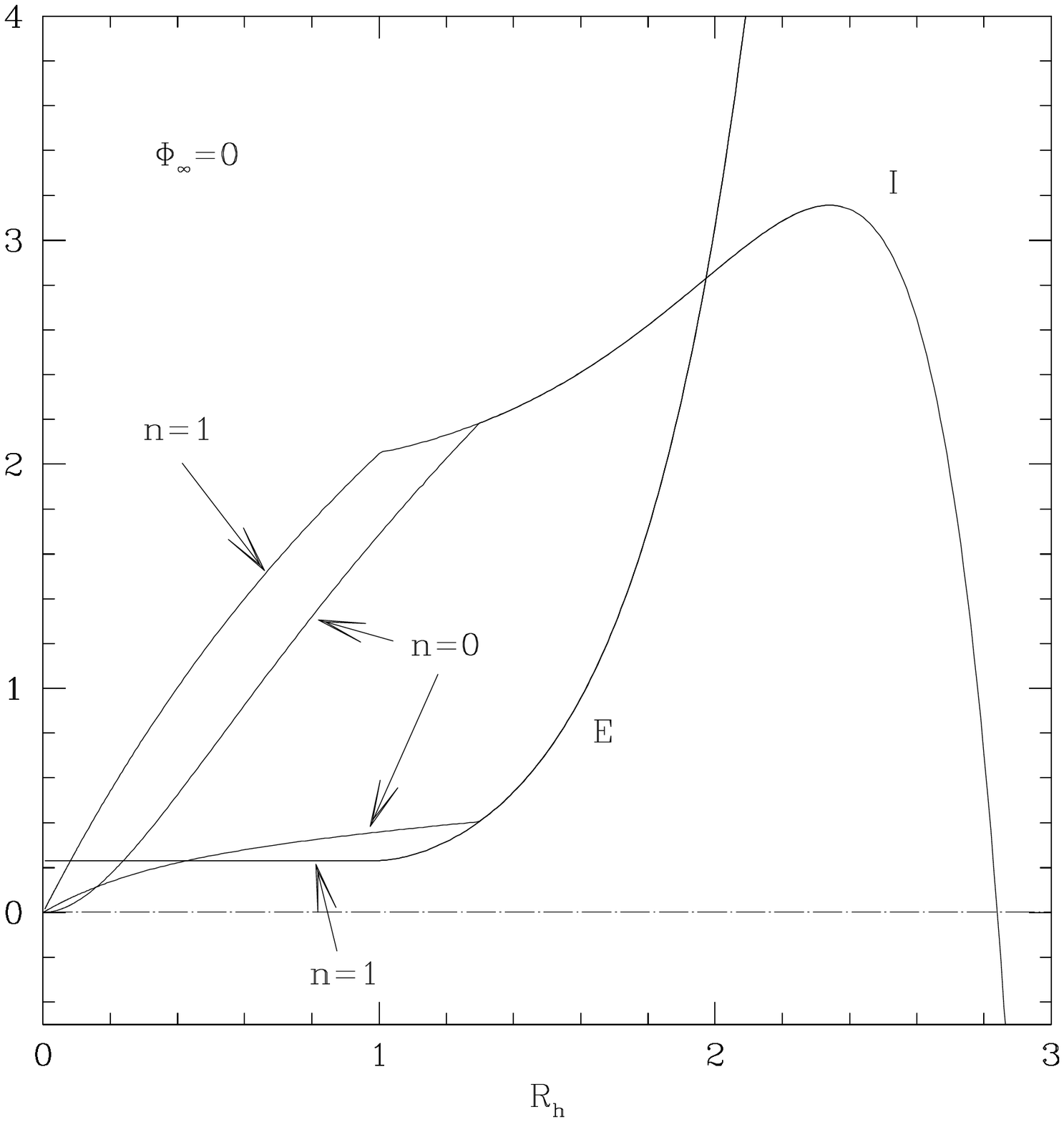,width=1.2\linewidth}
  \caption{{\fontsize{12}{12}\selectfont
Energy and action for the $n=0,1$ black holes.
}}
  \label{figEFb}
   \end{minipage} \hspace{15 mm}
\begin{minipage}[b]{0.40\linewidth}
 \centering\epsfig{figure=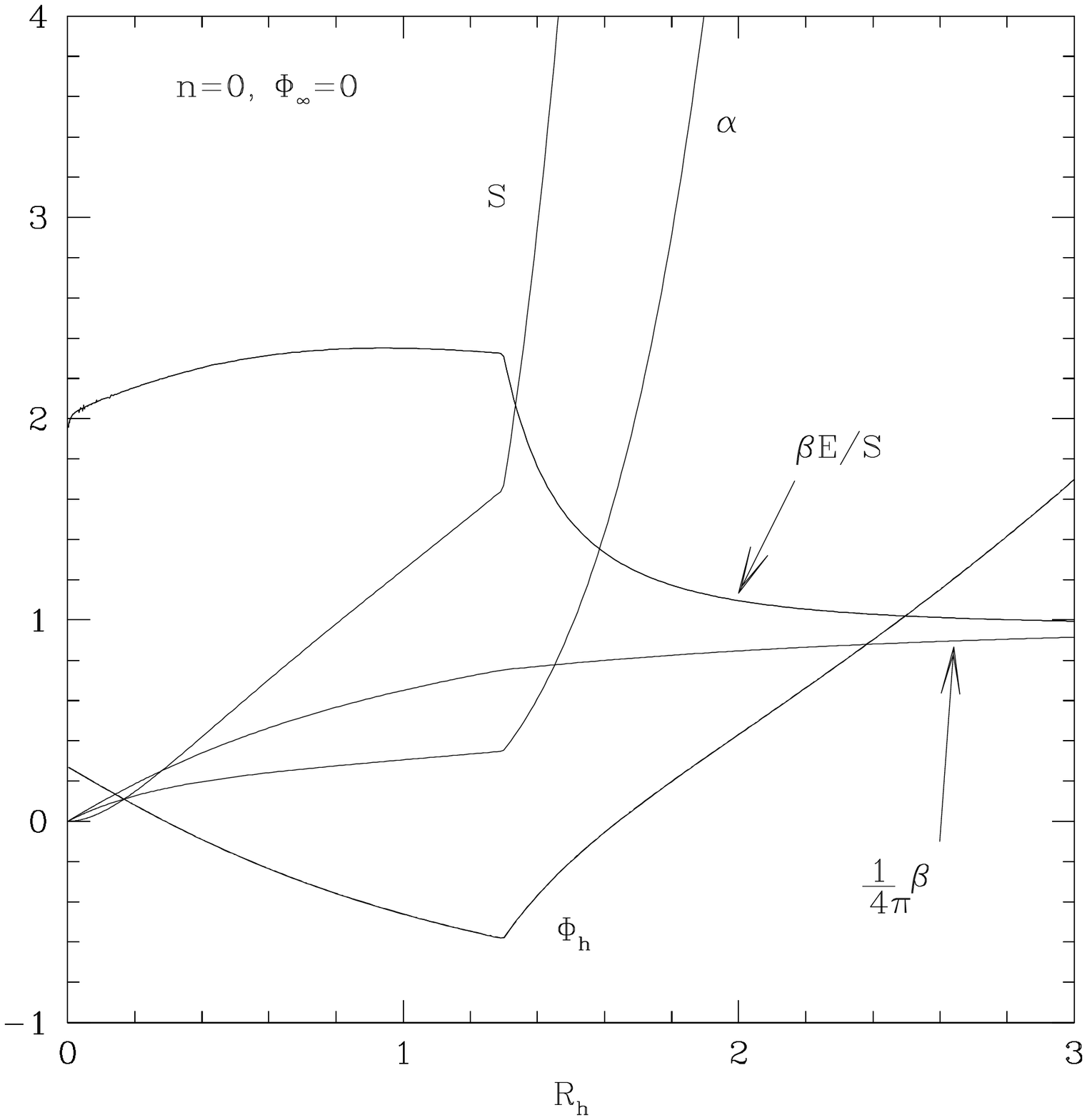,width=1.2\linewidth}
  \caption{{\fontsize{12}{12}\selectfont
Entropy $S$, $\alpha$, $\beta$, and $\Phi_h$ for the $n=0$ black
holes.
}}
  \label{figF0}
   \end{minipage}
\end{figure}

Having obtained the black hole solutions, we can compute their
thermodynamic parameters.
The energy $E$ (\ref{f24}) and the action $I$ (\ref{f26}) for the
$n=0,1$ black holes
are shown in Fig.\ref{figEFb}
with the normalization $\Phi_\infty=0$ for all solutions.%
\footnote{Since $E$ and $I$ depend on ${\cal P}$, their values 
are determined with some uncertainty; see footnote \ref{foot}. 
However, the qualitative behavior of the $E$ and $I$ curves 
seems to be independent on the numerical scheme used. 
}
For $\R_h\to 0$ the energy of the $n$-th non-Abelian black hole
coincides with that   of  the $n$-th regular solution.\footnote{Notice
that the energy of the regular solutions in Tab.1
is given in the different normalization: $\Phi(0)=0$.
Shifting the dilaton so  that $\Phi_\infty=0$, their
energy will be $E={\cal P}$, where the values of ${\cal P}$
are given in Tab.1.}
As $\R_h$ increases, the energy grows. For $\R_h=\R_h^\ast(n)$
the non-Abelian
solutions merge with the Abelian branch. The subsequent increase
in $\R_h$ along the Abelian branch is accompanied by  further
increase of the energy.

For all  black hole solutions the action $I(\R_h)$ is zero 
for $\R_h=0$,
positive for small values of $R_h$, and negative for all large enough $R_h$.%
\foot{The action vanishes for $\R_h\to 0$ because
$I\to\beta E$, where $E$ is the energy of the $n$-th regular solutions,
but $\beta=1/T\to 0$, since the black hole temperature diverges in the limit.}

\begin{figure}[h]
\begin{minipage}[b]{0.40\linewidth}
  \centering\epsfig{figure=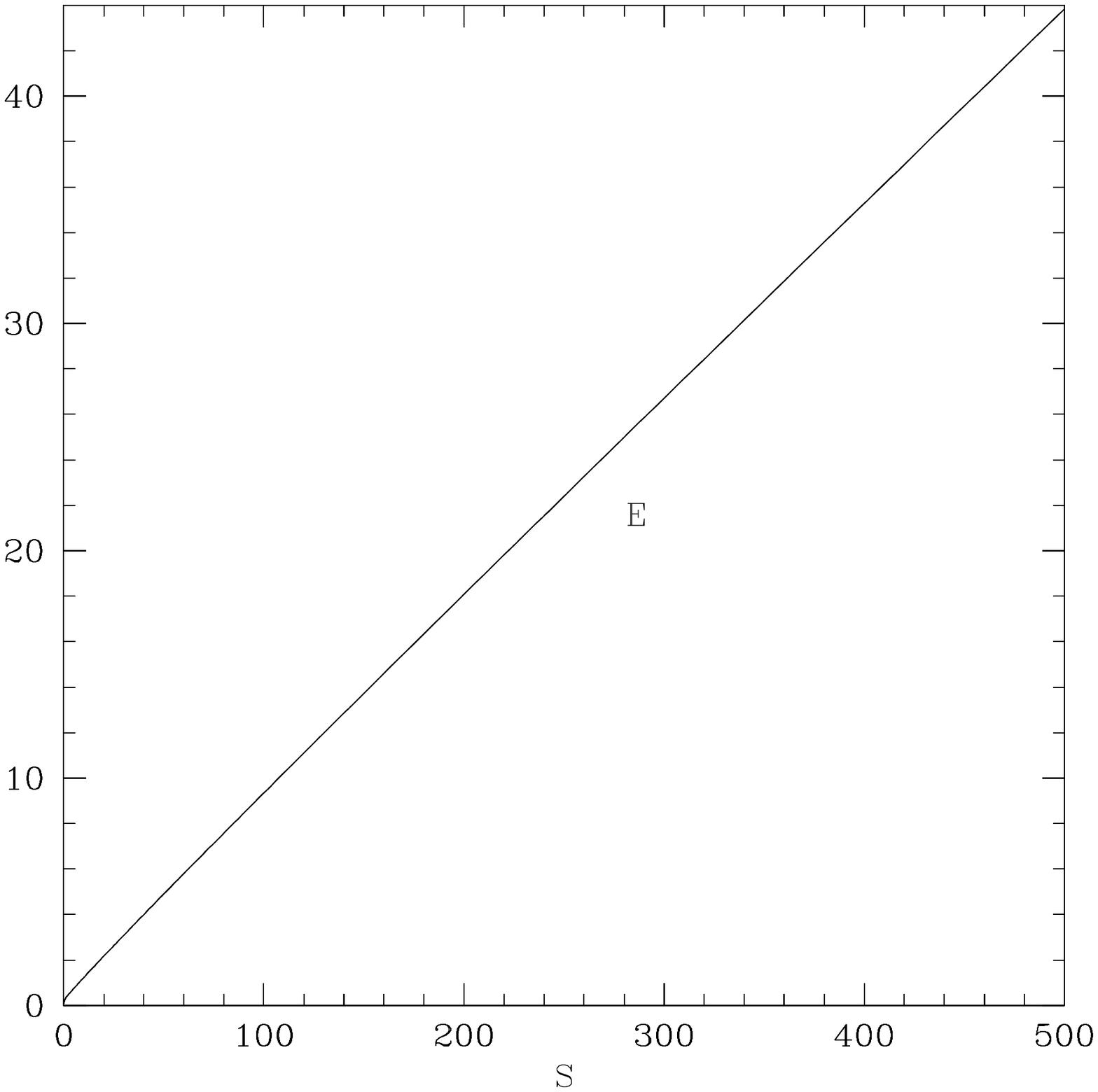,width=1.2\linewidth}
  \caption{{\fontsize{12}{12}\selectfont
Energy against entropy for the $n=0$ black holes.
}}
  \label{figEEE}
   \end{minipage} \hspace{15 mm}
\begin{minipage}[b]{0.40\linewidth}
 \centering\epsfig{figure=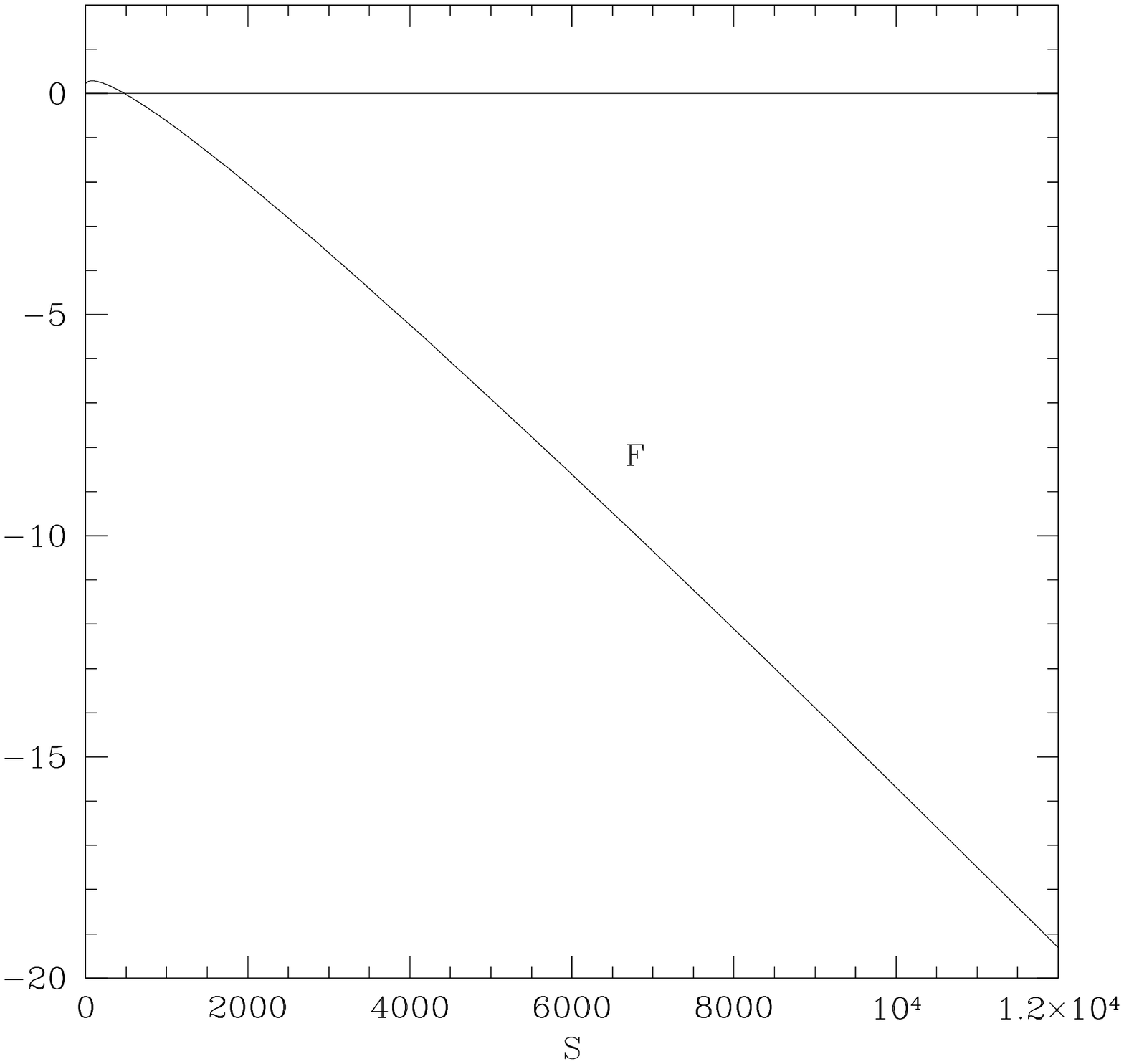,width=1.2\linewidth}
  \caption{{\fontsize{12}{12}\selectfont
Free energy versus entropy for the $n=0$ black holes.
}}
  \label{figFFF}
   \end{minipage}
\end{figure}

In Fig.\ref{figF0} we have  shown  the entropy $S(\R_h)$, the non-extremality
$\alpha(\R_h)$, the inverse temperature $\beta(\R_h)$, and the value of the
dilaton at the horizon $\Phi_h(\R_h)$ for the $n=0$ black holes.
In agreement with (\ref{temp2}), one has $\beta(\infty)=4\pi$.
In addition, the behavior of the ratio $\beta E/S$ shown in this figure
indicates that for large $\R_h$ the following equation of state holds:
\be                \label{ff}
E=TS\, .
\ee
This agrees with the first law of thermodynamics, $dE=TdS$, since
$T=1/\beta$ is constant for for large $\R_h$. We therefore recover in the UV
the standard NS5 brane  thermodynamics. 
In Figs.\ref{figEEE},\ref{figFFF} we also plot 
the energy and free energy against entropy 
for the $n=0$ black holes. As we can see, for large black holes 
$F$ also scales linearly with $S$. 

The value  $\Phi_h(\R_h)$ is an important parameter, since it determines
the  value of the string coupling constant.
It is therefore interesting to consider another normalization
for all solutions. 
For example, instead of fixing the value $\Phi_\infty=0$
one can fix $\Phi_h=0$. Using (\ref{scaling1}),
this can be achieved by translating
$\Phi(r)\to\Phi(r)-\Phi_h(\R_h)$, where $\Phi_h(\R_h)$ is shown in
Fig.\ref{figF0}. The energy, action, entropy, and
non-extremality $\alpha$ acquire then the factor $\exp(-2\Phi_h(\R_h))$,
while the temperature remains invariant.

To conclude this section, we have learned  the following about the value 
of the action for globally regular and black hole solutions.
 The action of all globally regular
solutions is non-negative, with the  minimal (zero)  value 
achieved for the BPS solution. 
For black holes, apart from those with small $R_h$, 
the action is negative.

\section{Restoration of chiral symmetry for $T>T_c$?}
\label{HawkingPage}

We have obtained the non-extremal generalizations of the globally
regular BPS solution
\cite{Maldacena:2000yy,Chamseddine:1997nm,Chamseddine:1998mc}, which
we have reproduced as \eno{BPS}.  The solutions which have no
singularities outside horizons are the original BPS solution, the
globally regular non-BPS solutions, and the black hole solutions
corresponding to the unshaded region in Fig.\ref{figA}.  Of this
two-parameter family of solutions, only a discrete series of
one-parameter families has finite energy.  These are the classical
saddle points which make important contributions to the path integral.
All the black hole solutions have temperature larger than the Hagedorn
temperature of the little string theory, as illustrated in
Figs.\ref{fig8},\ref{fig9}.  Thus, as remarked already in
section~\ref{Summary}, the solution that dominates the path integral
at temperatures lower than $T_c$ is the original BPS solution with
periodic Euclidean time.  (The contributions of the globally regular
non-BPS solutions are exponentially suppressed since their energy
density is finite and positive).  The energy, entropy, and free energy
of the periodized BPS solution are equal to zero  in the classical supergravity
approximation, which only indicates that they are less than $O(N^2)$.
Loop effects, due to the broken supersymmetry in the thermal boundary
conditions, would give rise to an $O(1)$ free energy.  This is
appropriate for the low-energy ${\cal N}=1$ gauge theory in its
confining phase.  Equally appropriate, chiral symmetry is broken in
this regime.  A deconfined phase might be expected to have restored
chiral symmetry, and energy, entropy, and free energy of order
$N^2$---like our abelian black hole solutions.

So far, the discussion is little different from that of
\cite{Witten:1998zw}, where it was argued that in global anti-de
Sitter space, a low-temperature phase corresponding to empty $AdS_5$
gives way to a high-temperature phase corresponding to
$AdS_5$-Schwarzschild through a Hawking-Page transition
\cite{HawkingPage} that corresponds to deconfinement in the gauge
theory.  The main differences in the current context are 1) the
putative high-temperature phase is thermodynamically unstable, and 2)
the little string theory is believed to have an exponential growth in
the number of states at high energy.  For both of these reasons, the
canonical ensemble is ill-defined above the Hagedorn temperature
$T_c$, and it doesn't make sense to speak of equilibrium processes at
controlled temperatures higher than $T_c$.  Thus, though it is
tempting to identify the abelian black hole solutions (which do have
$T > T_c$) as a high-temperature, deconfined phase, with restored
chiral symmetry, the truth is more complicated.

Suppose that a system such as the one we describe (that is, NS5-branes
on a shrinking $S^2$) were to come in thermal contact with a ``heat
bath'' at a temperature $T_{\rm bath} > T_c$.  Small black holes would
form and evaporate continually.  Eventually, through thermal
fluctuations, enough energy would be concentrated in one region to
make a larger black hole, with a temperature lower than $T_{\rm
bath}$.  The subsequent evolution would suck energy continually from
the heat bath until thermal contact ceased or the heat bath fell below
$T_c$.  
In regions of high energy density, where $R_h > 1.3$, 
chiral symmetry would be
restored because the only black hole solutions with high enough energy
are abelian.  In regions of low energy density, where $R_h < 1$,
chiral symmetry is broken because the only available black hole
solutions are non-abelian.
It is likely that the end state of the system would be
spatially non-uniform along the NS5-brane world-volume, since the
uniform state is thermodynamically unstable and this has been
associated \cite{Gubser:2000ec,Gubser:2000mm} with the presence of a
Gregory-Laflamme instability.

It may be noted from Figs.\ref{figFFF},\ref{figEFb} that the action,
$I = \beta F$, is negative for large $R_h$, but becomes positive for
small $\R_h$.  This might be regarded as the signal for a Hawking-Page
transition back to the periodized BPS solution at very high
temperatures; however this is not a coherent interpretation since the
canonical ensemble is still ill-defined.  More physically, it is
difficult to discuss a first order transition between two phases if
one is thermodynamically unstable, since the unstable phase may not
last long enough for the transition to take place.

For very large entropy/energy density (corresponding to
very large $\R_h$), 
Buchel has claimed $S = \beta_H E + a \log E$
plus subleading corrections, with $a < 0$ \ci{BUCH}, which   
result was obtained assuming that 
the thermodynamic description applies. This is
consistent with our result that the specific heat is negative.
However,  it also implies that $F > 0$, which 
is opposite to what we obtain in our analysis.  
Although we reproduce the energy--entropy relation in the 
leading order, the subleading terms are different, which probably
indicates the breakdown of the thermodynamic description.

\section{Conclusions}

Let us enumerate the solutions we have found.  In citing equation
numbers, we sometimes refer only to asymptotics if the solutions were
obtained numerically.  It helps to categorize solutions according
to whether they involve the non-abelian components of the $SU(2)$
gauge field when expressed in four-dimensional terms.  These
components are determined in terms of a single function $w(r)$, and
$U(1) \subset SU(2)$ is unbroken precisely if $w(r)=0$.  It happens
that $w$ vanishes for all $r$ if it vanishes at the horizon, if there
is a horizon, or if not, at the point where the radius of the $S^2$
vanishes.
  \begin{enumerate}
   \item The regular supersymmetric solution, \eno{BPS}.  This solution was
found in \cite{Chamseddine:1997nm,Chamseddine:1998mc}.  It preserves
four supercharges and has $w \neq 0$.  Its ten-dimensional lift was
shown in \cite{Maldacena:2000yy} to represent 5-branes wrapped on a
shrinking $S^2$, and it was therefore conjectured that the
supergravity geometries provided a holographic description for ${\cal
N}=1$, $D=4$ super-Yang-Mills theory.  The other solutions we obtain
can be viewed as excitations of this regular BPS one.

   \item Singular BPS solutions, \eno{BPSa}.  These solutions preserve
four supercharges, but they are unphysical because of a naked
singularity where the $S^2$ shrinks to zero size.  The abelian ``Dirac
monopole'' solution, \eno{chir}, is a special case of the
one-parameter family, \eno{BPSa}, which includes the BPS solution
\eno{BPS} as its only regular representative.

   \item The vanishing gauge field solution, \eno{10dw=1}.  This
solution breaks all supersymmetry, but it has $SU(2) \times SU(2)
\times SU(2)$ global symmetry, corresponding to an internal geometry
which is $S^2 \times S^3$.

   \item The factorized abelian solution, \eno{EEtube}.  All
supersymmetries are broken, but the geometry factorizes into a
five-dimensional compact coset manifold, $\tilde{T}^{1,1}$, and a
non-compact piece with a linear dilaton.  $\tilde{T}^{1,1}$ has a
bigger $U(1)$ fiber than the conventional $T^{1,1}$ metric, and the
interpretation is that NS5-branes have been wrapped on the 2-cycle and
then delocalized in the other directions.  We find an explicit sigma
model description of this geometry, valid in the weak coupling region.

   \item Globally regular non-BPS solutions, \eno{6}, \eno{inf}.
Superficially there is a one-parameter family of these solutions
labeled by $b$,
including the solution \eno{BPS} as its one BPS representative.  Of
these, only a discrete series has $w(r)$ falling off exponentially at
large radius, which we have found to be a necessary condition for
finite energy.  For solutions very far down the discrete series, there
is a long region which is nearly the factorized abelian solution, and
it is closed off on the inside by the $S^2$ shrinking, and on the
outside by asymptotics similar to the regular BPS solution.

  \item Abelian black hole solutions, \eno{ini} with $w_h = 0$.  These
solutions exist only if the entropy density is large enough: they are
parametrized by the horizon radius, $\R_h \geq 1$.  For $\R_h = 1$, we
have the analytic solution \eno{dilaton}, which is the factorized
abelian solution cut off on the inside by a black hole horizon: that
is, the standard 2-dimensional dilaton black hole times ${\bf R}^3$
times $\tilde{T}^{1,1}$.

  \item Non-abelian black hole solutions, \eno{ini} with $w_h \neq
0$.  Superficially there is a two-parameter family of solutions,
including all the other solutions listed as limiting cases (though in
some cases the relevant limit is only pointwise, not uniform in
$r$---allowing for instance the asymptotics to change).  However, only
a discrete series of one-parameter families has $w(r)$ falling off
exponentially at infinity.  Each of these one-parameter families
terminates at one end on the line of abelian solutions, and at the
other end at one of the globally regular solutions.
  \end{enumerate}

Many of the qualitative features of our results can be understood from
Fig.\ref{figA}.  Roughly speaking, the typical non-abelian black hole
solution has some oscillations of $w(r)$ in the region where it is
close to the factorized abelian solution.  This behavior is cut off at
one end by the horizon and at the other by expansion of the throat
into asymptotics similar to the BPS solution.

The globally regular non-BPS solutions, corresponding roughly to
excitations of a non-abelian gravitating monopole, are possibly
significant to {\it string theory cosmology}.  These solutions were
constructed with $3+1$-dimensional Poincar\'e invariance, but they
have finite positive energy density as compared to the supersymmetric
solution.  This translates to a positive contribution to the
four-dimensional cosmological constant.  To be more precise, suppose
we had constructed a compact solution where some local region was
well-approximated by one of our globally regular, non-BPS solutions.
And suppose the moduli, like the average value of the dilaton, were
fixed.  Then the non-compact four-dimensional part of the solution
would have to be de Sitter space, and the quantity $E$ in Tab.~1 would
translate into a cosmological constant.  The reason we were able to
construct solution with $3+1$-dimensional Poincar\'e invariance was
that the extra six dimensions were non-compact, so that gravity is
non-dynamical.  We can refine things a little further if we think in
terms of a toy model where the effects of compactification are
represented by cutting off our non-compact geometry at some large but
finite $r_C$.  Solutions with $\Upsilon=0$ in \eno{inf} have finite
energy as $r_C \to \infty$, but other solutions do not.
In short, we
expect that upon fixing finite $r_C$, the solutions in the discrete
series would ``broaden out'' into sharp, deep valleys in a
four-dimensional effective potential.  There would be only finitely
many minima, because for high excitation modes the nodes of $w$ would
fall outside the cutoff radius.  Thus the final picture is a
four-dimensional effective potential with many minima separated by
high walls.

So far we have assumed that moduli are stabilized, but so far in
string theory this seems very hard to do.  In the scenario of the
previous paragraph, the cosmological constant would have {\it very
weak} dependence on $r_C$, because in the $r_C \to \infty$ limit the
energy computed in Tab.~1 is finite.  However it would depend
exponentially on the dilaton, so each minimum would extend to a long,
low valley.  This is not much different from the conventional picture
of the effective potential in heterotic string compactifications with
broken supersymmetry.  The novelty is that the supersymmetry breaking
occurs as a non-BPS excitation of the internal geometry.

One may imagine a cosmological scenario where, at some stage in the
evolution of the universe, one finds local physics near the shrinking
$S^2$ described well by an abelian black hole.  As energy density
decreases due to expansion, the system would have to find its way onto
one of the non-abelian branches in Fig.\ref{figA}.  Only if the system
found the $n=0$ branch would it then relax into a supersymmetric
minimum; otherwise it would ``lock in'' some oscillations of $w(r)$,
and relax to a globally regular, extremal solution with a non-zero cosmological
constant and broken supersymmetry.  Thus we have given at least a
rough outline of how one might end up in a non-supersymmetric valley
of the four-dimensional effective potential and not be able to tunnel
into a supersymmetric solution.\footnote{One might in fact question
whether a four-dimensional effective potential is a valid notion.  We
use it for lack of a better language.}  This mechanism is intrinsically
non-field-theoretic because the Hawking temperature exceeds the
Hagedorn temperature of the little string theory.  We consider it
plausible that the contribution to the cosmological constant would be
small if the throat region, well-described by our non-compact
solutions, were long; however this is a point which deserves further
investigation.  Various drawbacks remain, notably the usual question
of why sparticle mass splittings are so much bigger than the
cosmological constant.  Also,  one may worry that the thermodynamic
instability will lead to unacceptably large spatial inhomogeneities.
But it nevertheless would be fascinating to see whether the {\it excited
monopole solutions} could be embedded into a global string
compactification---preferably one with other ingredients which fix the
dilaton.


\section*{Acknowledgements}
\noindent
We are grateful to  I. Klebanov
for participation at an initial stage
of this work and many useful discussions.
The work of S.S.G. is supported in part by the DOE under grant
DE-FG03-92ER40701.
The work of A.A.T. is partially supported by the DOE grant
  DE-FG02-91ER40690,
   PPARC SPG grant 00613,  INTAS
  project 991590 and
 CRDF Award RPI-2108.
 Part of this work was done while S.S.G.{} and
A.A.T. were  participating in the
M-theory program at ITP, Santa Barbara,
supported by the NSF grant PHY99-07949.
M.S.V. would like to acknowledge discussions with
G.W. Gibbons, and also with 
D. Maison, who was the first to numerically observe the
existence of the first integral (\ref{angle3}). 
The work of M.S.V. is supported
by the DFG grant Wi 777/4-3.

\end{document}